
%
\input amstex
%
%
%
%


\magnification=1200
\hsize=31pc
\vsize=55 truepc
\hfuzz=2pt
\vfuzz=4pt
\pretolerance=500
\tolerance=500
\parskip=0pt plus 1pt
\parindent=16pt
%

%
%
\font\fourteenrm=cmr10 scaled \magstep2
\font\fourteeni=cmmi10 scaled \magstep2
\font\fourteenbf=cmbx10 scaled \magstep2
\font\fourteenit=cmti10 scaled \magstep2
\font\fourteensy=cmsy10 scaled \magstep2

%
\font\large=cmbx10 scaled \magstep1

%

%

%

%
\font\eightrm=cmr8
\font\eighti=cmmi8
\font\eightbf=cmbx8
\font\eightit=cmti8

\font\eightsy=cmsy8
\font\sixrm=cmr6
\font\sixi=cmmi6
\font\sixsy=cmsy6

%
\def\tenpoint{\def\rm{\fam0\tenrm}%
  \textfont0=\tenrm \scriptfont0=\sevenrm
                      \scriptscriptfont0=\fiverm
  \textfont1=\teni  \scriptfont1=\seveni
                      \scriptscriptfont1=\fivei
  \textfont2=\tensy \scriptfont2=\sevensy
                      \scriptscriptfont2=\fivesy
  \textfont3=\tenex   \scriptfont3=\tenex
                      \scriptscriptfont3=\tenex
  \textfont\itfam=\tenit  \def\it{\fam\itfam\tenit}%
  \textfont\slfam=\tensl  \def\sl{\fam\slfam\tensl}%
  \textfont\bffam=\tenbf  \scriptfont\bffam=\sevenbf
                            \scriptscriptfont\bffam=\fivebf
                            \def\bf{\fam\bffam\tenbf}%
  \normalbaselineskip=20 truept
  \setbox\strutbox=\hbox{\vrule height14pt depth6pt
width0pt}%
  \let\sc=\eightrm \normalbaselines\rm}
\def\eightpoint{\def\rm{\fam0\eightrm}%
  \textfont0=\eightrm \scriptfont0=\sixrm
                      \scriptscriptfont0=\fiverm
  \textfont1=\eighti  \scriptfont1=\sixi
                      \scriptscriptfont1=\fivei
  \textfont2=\eightsy \scriptfont2=\sixsy
                      \scriptscriptfont2=\fivesy
  \textfont3=\tenex   \scriptfont3=\tenex
                      \scriptscriptfont3=\tenex
  \textfont\itfam=\eightit  \def\it{\fam\itfam\eightit}%
  \textfont\bffam=\eightbf  \def\bf{\fam\bffam\eightbf}%
  \normalbaselineskip=16 truept
  \setbox\strutbox=\hbox{\vrule height11pt depth5pt width0pt}}
\def\fourteenpoint{\def\rm{\fam0\fourteenrm}%
  \textfont0=\fourteenrm \scriptfont0=\tenrm
                      \scriptscriptfont0=\eightrm
  \textfont1=\fourteeni  \scriptfont1=\teni
                      \scriptscriptfont1=\eighti
  \textfont2=\fourteensy \scriptfont2=\tensy
                      \scriptscriptfont2=\eightsy
  \textfont3=\tenex   \scriptfont3=\tenex
                      \scriptscriptfont3=\tenex
  \textfont\itfam=\fourteenit  \def\it{\fam\itfam\fourteenit}%
  \textfont\bffam=\fourteenbf  \scriptfont\bffam=\tenbf
                             \scriptscriptfont\bffam=\eightbf
                             \def\bf{\fam\bffam\fourteenbf}%
  \normalbaselineskip=24 truept
  \setbox\strutbox=\hbox{\vrule height17pt depth7pt width0pt}%
  \let\sc=\tenrm \normalbaselines\rm}
\def\today{\number\day\ \ifcase\month\or
  January\or February\or March\or April\or May\or June\or
  July\or August\or September\or October\or November\or
December\fi
  \space \number\year}
\def\monthyear{\ifcase\month\or
  January\or February\or March\or April\or May\or June\or
  July\or August\or September\or October\or November\or
December\fi
  \space \number\year}

%
\newcount\secno      
\newcount\subno      
\newcount\subsubno   
\newcount\appno      
\newcount\tableno    
\newcount\figureno   
%

%
\normalbaselineskip=20 truept
\baselineskip=20 truept

%
%
\def\title#1
   {\vglue1truein
   {\baselineskip=24 truept
    \pretolerance=10000
    \raggedright
    \noindent \fourteenpoint\bf #1\par}
    \vskip1truein minus36pt}
%

%
\def\author#1
  {{\pretolerance=10000
    \raggedright
    \noindent {\large #1}\par}}

%
\def\address#1
   {\bigskip
    \noindent \rm #1\par}

%
\def\shorttitle#1
   {\vfill
    \noindent \rm Short title: {\sl #1}\par
    \medskip}

%
\def\pacs#1
   {\noindent \rm PACS number(s): #1\par
    \medskip}

%
\def\jnl#1
   {\noindent \rm Submitted to: {\sl #1}\par
    \medskip}

%
\def\date
   {\noindent Date: \today\par
    \medskip}

%

%
\def\keyword#1
   {\bigskip
    \noindent {\bf Keyword abstract: }\rm#1}

%

%
%
\def\contents
   {{\noindent
    \bf Contents
    \par}
    \rightline{Page}}

%
\def\entry#1#2#3
   {\noindent
    \hangindent=20pt
    \hangafter=1
    \hbox to20pt{#1 \hss}#2\hfill #3\par}

%
\def\subentry#1#2#3
   {\noindent
    \hangindent=40pt
    \hangafter=1
    \hskip20pt\hbox to20pt{#1 \hss}#2\hfill #3\par}
\def\checkforsub{\futurelet\nexttok\decide}
\def\ssf{\relax}
\def\decide{\if\nexttok\ssf\let\endspace=\nospace
                \else\let\endspace=\extraspace\fi\endspace}
\def\nospace{\nobreak\par\nobreak}
%
%
\def\section#1{%
    \goodbreak
    \vskip24pt plus12pt minus12pt
    \nobreak
    \gdef\extraspace{\nobreak\bigskip\noindent\ignorespaces}%
    \noindent
    \subno=0 \subsubno=0
    \global\advance\secno by 1
    \noindent {\bf \the\secno. #1}\par\checkforsub}

%
\def\subsection#1{%
     \goodbreak
     \vskip24pt plus12pt minus6pt
     \nobreak
     \gdef\extraspace{\nobreak\medskip\noindent\ignorespaces}%
     \noindent
     \subsubno=0
     \global\advance\subno by 1
     \noindent {\sl \the\secno.\the\subno. #1\par}\checkforsub}

%
\def\subsubsection#1{%
     \goodbreak
     \vskip15pt plus6pt minus6pt
     \nobreak\noindent
     \global\advance\subsubno by 1
     \noindent {\sl \the\secno.\the\subno.\the\subsubno. #1}\null.
     \ignorespaces}

%
\def\appendix#1
   {\vskip0pt plus.1\vsize\penalty-250
    \vskip0pt plus-.1\vsize\vskip24pt plus12pt minus6pt
    \subno=0
    \global\advance\appno by 1
    \noindent {\bf Appendix \the\appno. #1\par}
    \bigskip
    \noindent}

%
\def\subappendix#1
   {\vskip-\lastskip
    \vskip36pt plus12pt minus12pt
    \bigbreak
    \global\advance\subno by 1
    \noindent {\sl \the\appno.\the\subno. #1\par}
    \nobreak
    \medskip
    \noindent}

%


%

%
\def\tabcaption#1
   {\global\advance\tableno by 1
    \noindent {\bf Table \the\tableno.} \rm#1\par
    \bigskip}

%

%

%

%

%

%
\def\figcaption#1
   {\global\advance\figureno by 1
    \noindent {\bf Figure \the\figureno.} \rm#1\par
    \bigskip}

%

%

%
\def\refjl#1#2#3#4
   {\hangindent=16pt
    \hangafter=1
    \rm #1
   {\frenchspacing\sl #2
    \bf #3}
    #4\par}

%
\def\refbk#1#2#3
   {\hangindent=16pt
    \hangafter=1
    \rm #1
   {\frenchspacing\sl #2}
    #3\par}

%
\def\numrefjl#1#2#3#4#5
   {\parindent=40pt
    \hang
    \noindent
    \rm {\hbox to 30truept{\hss #1\quad}}#2
   {\frenchspacing\sl #3\/
    \bf #4}
    #5\par\parindent=16pt}

%
\def\numrefbk#1#2#3#4
   {\parindent=40pt
    \hang
    \noindent
    \rm {\hbox to 30truept{\hss #1\quad}}#2
   {\frenchspacing\sl #3\/}
    #4\par\parindent=16pt}

%

\def\ref#1{\par\noindent \hbox to 21pt{\hss
#1\quad}\frenchspacing\ignorespaces}

%
\def\frac#1#2{{#1 \over #2}}

%

%
\def\d{\hbox{\rm d}}

%
\def\e{\operatorname{e}}


\def\i{\operatorname{i}}
\chardef\ii="10

%

%

%

\catcode`\@=11
\def\vfootnote#1{\insert\footins\bgroup
    \interlinepenalty=\interfootnotelinepenalty
    \splittopskip=\ht\strutbox 
    \splitmaxdepth=\dp\strutbox \floatingpenalty=20000
    \leftskip=0pt \rightskip=0pt \spaceskip=0pt \xspaceskip=0pt
    \noindent\eightpoint\rm #1\ \ignorespaces\footstrut\futurelet\next\fo@t}

%
%
\def\eq(#1){\hfill\llap{(#1)}}
\catcode`\@=12
%
%



%
%





%
%

%
%

%
%

%

%

%

%
\def\gap{\;\lower3pt\hbox{$\buildrel > \over \sim$}\;}
%
%
\def\lap{\;\lower3pt\hbox{$\buildrel < \over \sim$}\;}
\def\tqs{\hbox to 25pt{\hfil}}

\def\cal#1{{\Cal #1}}

{\obeylines\gdef\startdisplay#1
  {\catcode`\^^M=5$$#1\halign\bgroup\indent##\hfil&&\qquad##\hfil\cr}}
\outer\def\enddisplay{\crcr\egroup$$}

\chardef\other=12
\def\ttverbatim{\begingroup \catcode`\\=\other \catcode`\{=\other
  \catcode`\}=\other \catcode`\$=\other \catcode`\&=\other
  \catcode`\#=\other \catcode`\%=\other \catcode`\~=\other
  \catcode`\_=\other \catcode`\^=\other
  \obeyspaces \obeylines \tt}
{\obeyspaces\gdef {\ }}  

\outer\def\begintt{$$\let\par=\endgraf \ttverbatim \parskip=0pt
  \catcode`\|=0 \rightskip=-5pc \ttfinish}
{\catcode`\|=0 |catcode`|\=\other 
  |obeylines 
  |gdef|ttfinish#1^^M#2\endtt{#1|vbox{#2}|endgroup$$}}

\catcode`\|=\active
{\obeylines\gdef|{\ttverbatim\spaceskip=.5em plus.25em minus.15em
                                            \let^^M=\ \let|=\endgroup}}%

\TagsOnRight


\tracingstats=1    

\normalbaselineskip=12pt
\baselineskip=12pt

\font\twelveit=cmti10 scaled 1200

\hyphenation{Eqs}

{\nopagenumbers
\pageno=0
\centerline{November 1992\hfill NTZ 29/92}
\centerline{\ }
\bigskip
\centerline{\fourteenbf AN INTRODUCTION INTO}
\medskip
\centerline{\fourteenbf THE FEYNMAN PATH INTEGRAL}
\bigskip
\centerline{\large CHRISTIAN GROSCHE}
\bigskip
\centerline{\twelveit International School for Advanced Studies}
\centerline{\twelveit Via Beirut 4}
\centerline{\twelveit 34014 Trieste, Miramare, Italy}
\vfill
\midinsert
\narrower
\noindent
{\it Lecture given at the graduate college ''Quantenfeldtheorie und
deren Anwendung in der Elementarteilchen- und Festk\"orperphysik'',
Universit\"at Leipzig, 16-26 November 1992.}
\endinsert
\vfill
\midinsert
\narrower
\noindent
{\bf Abstract.}
In this lecture a short introduction is given into the theory of the
Feynman path integral in quantum mechanics. The general formulation in
Riemann spaces will be given based on the Weyl- ordering prescription,
respectively product ordering prescription, in the quantum Hamiltonian.
Also, the theory of space-time transformations and separation of
variables will be outlined. As elementary examples I discuss the usual
harmonic oscillator, the radial harmonic oscillator, and the Coulomb
potential.
\endinsert
\eject
\pageno=0
\centerline{\ }\newpage}


\newcount\refno
\def\add{\advance\refno by 1}
\refno=1

\newcount\glno
\def\plus{\advance\glno by 1}
\def\num{\the\glno}

\newcount\chapno
\def\NUM{\the\chapno}

\def\makeheadline{\vbox to 0pt{\vskip-30.5pt
                  \line{\vbox to 8.5pt{}\the\headline}
                  \nointerlineskip\vskip1.5pt
                  \line{\hrulefill}\vss\nointerlineskip}}
\headline={\ifodd\pageno\rightheadline \else\leftheadline\fi}
\def\makefootline{\baselineskip=24pt
                  \line{\the\footline}}
\def\leftheadline{\eightpoint\eightrm\hfil\Chapter\hfil}
\def\rightheadline{\eightpoint\eightrm\Kapitel\hfil\Section}

\def\R{\hbox{\bf R}}

\def\N{\hbox{\bf N}}
\def\bqj{ {\bar q}^{(j)} }
\def\buj{ {\bar u}^{(j)} }
\def\bujq{ {\bar u}^{(j)\,2} }
\def\ih{{\i\over\hbar}}
\def\Norm{\left({m\over2\pi\i\epsilon\hbar}\right)}

\def\half{{1\over2}}
\def\bhalf{\hbox{${1\over2}$}}

\def\sqr#1#2{{\vcenter{\hrule height.#2pt
              \hbox{\vrule width.#2pt height#1pt
              \kern#1pt
              \vrule width.#2pt}
              \hrule height.#2pt}}}

\def\dfrac{\dsize\frac}
\def\cal{\Cal}

\def\CD{{\cal D}}

\def\CH{{\cal H}}
\def\CL{{\cal L}}


\def\diag{\operatorname{diag}}

\def\Tr{\operatorname{Tr}}

\def\cn{\operatorname{cn}}
\def\dn{\operatorname{dn}}
\def\sn{\operatorname{sn}}

\def\slim{\operatornamewithlimits{s-lim}}
\def\SO{\operatorname{SO}}
\def\SU{\operatorname{SU}}

\edef\ALB{\the\refno}\add
\edef\ABHKa{\the\refno}\add
\edef\ACHRS{\the\refno}\add
\edef\ARTb{\the\refno}\add
\edef\BJb{\the\refno}\add
\edef\CAST{\the\refno}\add
\edef\CARPa{\the\refno}\add
\edef\CARPb{\the\refno}\add
\edef\CBB{\the\refno}\add
\edef\CBBI{\the\refno}\add
\edef\CBI{\the\refno}\add
\edef\CGHa{\the\refno}\add
\edef\CGHc{\the\refno}\add
\edef\CGHd{\the\refno}\add
\edef\CHc{\the\refno}\add
\edef\DDS{\the\refno}\add
\edef\DEWb{\the\refno}\add
\edef\DEWMa{\the\refno}\add
\edef\DEWMNb{\the\refno}\add
\edef\DIRd{\the\refno}\add
\edef\DOTO{\the\refno}\add
\edef\DOWc{\the\refno}\add
\edef\DOWd{\the\refno}\add
\edef\DMa{\the\refno}\add
\edef\DURb{\the\refno}\add
\edef\DURd{\the\refno}\add
\edef\DKa{\the\refno}\add
\edef\DKb{\the\refno}\add
\edef\DKS{\the\refno}\add
\edef\EG{\the\refno}\add
\edef\EMOTa{\the\refno}\add
\edef\FEYa{\the\refno}\add
\edef\FEYb{\the\refno}\add
\edef\FH{\the\refno}\add
\edef\FLM{\the\refno}\add
\edef\FW{\the\refno}\add
\edef\GY{\the\refno}\add
\edef\GJ{\the\refno}\add
\edef\GOOb{\the\refno}\add
\edef\GRA{\the\refno}\add
\edef\GMV{\the\refno}\add
\edef\GMVc{\the\refno}\add
\edef\GROa{\the\refno}\add
\edef\GROe{\the\refno}\add
\edef\GROj{\the\refno}\add
\edef\GROm{\the\refno}\add
\edef\GROo{\the\refno}\add
\edef\GROp{\the\refno}\add
\edef\GRSb{\the\refno}\add
\edef\GRSf{\the\refno}\add
\edef\GRSg{\the\refno}\add
\edef\GROS{\the\refno}\add
\edef\GROGOb{\the\refno}\add
\edef\HIR{\the\refno}\add
\edef\HOI{\the\refno}\add
\edef\INH{\the\refno}\add
\edef\INOa{\the\refno}\add
\edef\INOb{\the\refno}\add
\edef\INOd{\the\refno}\add
\edef\INOWI{\the\refno}\add
\edef\BJa{\the\refno}\add
\edef\KLEh{\the\refno}\add
\edef\KLEk{\the\refno}\add
\edef\KLEm{\the\refno}\add
\edef\KLEMUS{\the\refno}\add
\edef\KUST{\the\refno}\add
\edef\LAI{\the\refno}\add
\edef\LRTd{\the\refno}\add
\edef\LEEb{\the\refno}\add
\edef\MCLS{\the\refno}\add
\edef\MSVW{\the\refno}\add
\edef\MIZa{\the\refno}\add
\edef\MARa{\the\refno}\add
\edef\MIZc{\the\refno}\add
\edef\MIZd{\the\refno}\add
\edef\MORE{\the\refno}\add
\edef\MDEW{\the\refno}\add
\edef\DEWMc{\the\refno}\add
\edef\DEWMEL{\the\refno}\add
\edef\NELb{\the\refno}\add
\edef\OMO{\the\refno}\add
\edef\PAKSc{\the\refno}\add
\edef\PAUb{\the\refno}\add
\edef\PI{\the\refno}\add
\edef\RS{\the\refno}\add
\edef\SCHUc{\the\refno}\add
\edef\SIMON{\the\refno}\add
\edef\SOKb{\the\refno}\add
\edef\SOKc{\the\refno}\add
\edef\STEa{\the\refno}\add
\edef\STEb{\the\refno}\add
\edef\STEc{\the\refno}\add
\edef\TROT{\the\refno}\add

\pageno=1

\def\Chapter{Contents}
\def\rightheadline{\eightpoint\hfil\eightrm Contents\hfil}
\contents
\entry{I}{Introduction}{2}
\entry{II}{General Theory}{5}
\subentry{II.1}{The Feynman Path Integral}{5}
\subentry{II.2}{Weyl-Ordering}{10}
\subentry{II.3}{Product-Ordering}{17}
\subentry{II.4}{Space-Time Transformations}{21}
\subentry{II.5}{Separation of Variables}{30}
\entry{III}{Important Examples}{33}
\subentry{III.1}{\ The Free Particle}{33}
\subentry{III.2}{\ The Harmonic Oscillator}{34}
\subentry{III.3}{\ The Radial Path Integral}{40}
\subentry{III.3.1}{\quad The General Radial Path Integral}{40}
\subentry{III.3.2}{\quad The Radial Harmonic Oscillator}{50}
\subentry{III.4}{\ Other Elementary Path Integrals}{57}
\subentry{III.5}{\ The Coulomb Potential}{59}
\subentry{III.5.1}{\quad The $1/r$-Potential in $\R^2$}{61}
\subentry{III.5.2}{\quad The $1/r$-Potential in $\R^3$ -
                   The Hydrogen Atom}{68}
\subentry{III.5.3}{\quad Coulomb Potential and $1/r$-Potential in $D$
                   Dimensions}{74}
\subentry{III.5.4}{\quad Axially Symmetric Coulomb-Like Potentials}{81}
\entry{ }{Bibliography}{89}

\vfill\eject
\baselineskip=11pt
\glno=0                          
\chapno=1                        

\def\Chapter{Introduction}

\def\rightheadline{\eightpoint\eightrm\Kapitel\hfil\Section}
\centerline{\fourteenbf I I\bf NTRODUCTION}

\bigskip\noindent
It was Feynman's (and Dirac's [\DIRd]) genius [\FEYa, \FEYb] to realize
that the integral kernel (propgator) of the time-evolution operator can
be expressed as a sum over all possible paths connecting the points $q'$
and $q''$ with weight factor $\exp\big[iS(q'',q';T)/\hbar\big]$,
where $S$ is the action, i.e.
\plus
$$K(q'',q';T)=\sum_{all\ paths} A\,\e^{iS(q'',q';T)/\hbar}
  \tag\NUM.\num$$\plus%
with some appropriate normalization $A$.

Surprisingly enough, the same calculus (``same'' in the  sense of a
na\"\ii ve analytical continuation) was already  know to mathematicians
due to Wiener in the study of stochastic processes. This {\it calculus}
in functional space (``Wiener measure'') attracted several
mathematicians, including Kac (who mentioned being influenced by
Feynman's work!), and was further developed by several authors, where
best known is the work of Cameron and Martin. The standard reference
concerning these achievements is the review paper of Gelfand and Yaglom
[\GY], where all these early work was first critically discussed.

Unfortunately, the discussion between physicists and mathematicians
remains near to nothing for quite a long time, except for [\GY]; the
situation changed with Nelson [\NELb], and nowadays there are many
attempts to understand the path integral mathematically despite its
pathalogies of ``infinite measure'', ``infinite sums of phases'' with
unit absolute values etc.

In particular, the work of Morette-DeWitt starting with her early paper
[\MORE] gave rise to a beautiful theory of the semiclassical expansion
in powers of $\hbar$ [\DEWMa, \DEWMNb, \MDEW-\DEWMEL]. As it is known
for quite a long time, the propagator can be expressed semi-classically
as $\e^{\i S_{Cl}/\hbar}$, with $S_{Cl}$ the classical action, times a
prefactor. This prefactor is remarkably simple, namely one has
$$\multline
  K_{WKB}(x'',x';t'',t')=[g(x')g(x'')]^{-{1\over4}}
  \bigg({1\over2\pi\i\hbar}\bigg)^{D/2}
  \\   \times
  \sqrt{\det\bigg(-{\partial^2 S_{Cl}[x',x'']\over
    \partial x_a'\partial x_b''}\bigg)}
  \exp\bigg(\ih S_{Cl}[x',x'']\bigg).
  \endmultline
  \tag\NUM.\num$$
\edef\numaa{\NUM.\num}\plus%
$g=\det(g_{ab})$ of some possible metric structure of a Riemannian
space, and the determinant
$$M:= \det\bigg(-{\partial^2 S_{Cl}[x',x'']\over
        \partial x_a'\partial x_b''}\bigg)
  \tag\NUM.\num$$\plus%
is known as the Pauli-van Vleck-Morette determinant. The semiclassical
(WKB-) solution of the Feynman kernel (we use the notions semiclassical
and WKB simultaneously) is based on the fact that the harmonic
oscillator, respectively, the general quadratic Lagrangian, is exactly
solvable and its solution is only determined by the classical path and
not on the summation over all paths. As it turns out, an arbitrary
kernel can be expanded in terms of the classical paths as an expansion
in powers of $\hbar$. The semiclassical kernel (at least its short time
representation) is known since van Vleck, derived by the correspondence
principle. Later on Pauli [\PAUb] has given a detailed discussion in
his well-known lecture notes. More rigorously the short time propagator
for the one-dimensional case was discussed by Morette [\MORE] in 1951
and a few years later by DeWitt [\DEWb] extending the previous work to
$D$-dimensional curved spaces. Concerning the semi-classical expansion,
every path integral with a Hamiltonian which is quadratic in its
momenta can be expanded about the semiclassical approximation (\numaa)
giving a consistent and converging theory, compare DeWitt-Morette
[\DEWMa, \DEWMNb, \MIZc, \MIZd, \MDEW].

Supersymmetric quantum mechanics provides a very convenient way of
classifying exactly solvable models in usual quantum mechanics, and a
systematic way of addressing the problem of finding all exactly solvable
potentials [\INH].

By e.g.\ Dutt et al.\ [\DDS, \DKS] it was shown that there are a total
of twelve different potentials. A glance on these potentials shows that
their corresponding Schr\"odinger equation leads either to the
differential equation of the confluent hypergeometric differential
equation with eigen-functions proportional to Laguerre polynomials
(bound states) and Whittaker functions (continuous states),
respectively to the differential equation of the hypergeometric
differential equation with eigen-functions proportional to Jacobi
polynomials (bound states) and hypergeometric functions (continuous
states). In Reference [\DDS] this topic was nicely addressed and it was
shown that in principle the radial harmonic oscillator and the
(modified) P\"oschl-Teller-potential solutions are sufficient to give
the solutions of all remaining ones, together with the technique of
space-time transformations as introduced by Duru and Kleinert [\DKa,
\DKb, \KLEh] (see also reference [\GRSb], Ho and Inomata [\HOI],
Steiner [\STEa], and Pak and S\"okmen [\PAKSc]) enables one to give the
explicit path integral solution of the Coulomb $V^{(C)}(r)=-e^2/r$
$(r>0)$ and the Morse potential $V^{(M)}(x)=(\hbar^2A^2/2m) (e^{2x}
-2\alpha e^x)$ $(x\in\R)$ (compare e.g.\ reference [\GROm] for a review
of some recent results). The same line of reasoning is true for the
path integral solution of the (modified) P\"oschl-Teller potential
[\BJb-\FLM] which give in turn the path integral solutions for the
Rosen-Morse $V^{(RM)}(x)=A\tanh x-B/\cosh^2x$ $(x\in\R)$, the
hyperbolic Manning-Rosen $V^{(MRa)}(r)=-A\coth r+B/\sinh^2r$ ($r>0)$,
and the trigonometric Manning-Rosen-like potential $V^{(MRb)}
(x)=-A\cot x+B/\sin^2x$ $(0<x<\pi)$, respectively [\GROe].

These lecture notes are far from being a comprehensive introduction
into the whole topic of path integrals, in particular if field theory is
concerned. As old as they be, the books of Feynman and Hibbs [\FH] and
Schulman [\SCHUc] as still a {\bf must} for becoming familiar with the
subject. A more recent contribution is due to Kleinert [\KLEm].
Myself and F.\ Steiner are presently preparing extended lecture notes
``Feynman Path Integrals''  and a ``Table of Feynman Path Integrals''
[\GRSf, \GRSg], which will appear next year.

Several reviews have been written about path integrals, let me note
Gelfand and Jaglom [\GY], Albeverio et al. [\ALB-\ACHRS],
DeWitt-Morette et al. [\DEWMNb, \DEWMEL], Marinov [\MARa], and e.g.\
for the topic of path integrals for Coulomb potentials [\GROm].

The contents of the lecture is as follows. In the next Chapter I
outline the basic theory of the Feynman path integral, i.e.\ the
lattice definition according to the Weyl-ordering prescription in the
Hamiltonian and a related prescription which is of use in several
applications in my own work. Furthermore, the technique of canonical
coordinate-, time- and space-time transformations will be presented.
The Chapter closes with the discussion of separation of variables
in path integrals.

In the third chapter I present some important examples of exact path
integral evaluations. This includes, of course, the harmonic oscillator
in its simplest form. I also discuss the radial path integral with its
application of the exact treatment of the radial (time-dependent)
harmonic oscillator. The chapter concludes with a comprehensive
discussion of the Coulomb potential, including the genuine Coulomb
problem, its $D$-dimensional generalization, and a generalization with
additional axially-symmetric terms.

\vfill\noindent
{\bf Acknowledgement}
\newline
Finally I want to thank the organizers of this graduate college for
their kind invitation to give this lecture. I also want to thank F.\
Steiner which whom part of the presented material was compiled.

\eject
\glno=0               
\chapno=1             
\def\Chapter{General Theory}

\noindent
\centerline{\fourteenbf II G\large ENERAL \fourteenbf T\large HEORY}

\section{The Feynman Path Integral}
In order to set up the requirements of the path integral formalism we
start with the generic case, where the time dependent Schr\"odinger
equation in some $D$-dimensional Riemannian manifold $M$ with metric
$g_{ab}$ and line element $ds^2=g_{ab}dq^a dq^b$ is given by
\plus
$$\bigg[-{\hbar^2\over2m}\Delta_{LB}+V(q)\bigg]
       \Psi(q;t)={\hbar\over\i}{\partial\over\partial t}\Psi(q;t).
  \tag\NUM.\num$$
$\Psi$ is some state function, defined in the Hilbert space $\CL^2$ -
the space of all square integrable functions in the sense of the
scalar product $(f_1,f_2) = \int_M \sqrt{g}\, f_1(q) f_2^*(q) dq$
\newline
[$g:=\det(g_{ab}),\ f_1,f_2\in\CL^2$]
and  $\Delta_{LB}$ is the Laplace-Beltrami operator
\plus
$$\Delta_{LB}:=g^{-{1\over2}}\partial_a g^{1\over2}g^{ab}\partial_b
 =g^{ab}\partial_a\partial_b+g^{ab}(\partial_a\ln\sqrt{g}\,)
 \partial_b+g^{ab}_{\ \ ,a}\partial_b
  \tag\NUM.\num$$
(implicit sums over repeated indices are understood).

The Hamiltonian $H:=-{\hbar^2\over2m}\Delta_{LB}+V(q)$ is usually
defined in some dense subset $D(H)\subseteq\CL^2$, so that $H$ is
selfadjoint. In contrast to the time independent Schr\"odinger
equation, $H\Psi=E\Psi$, which is an eigenvalue problem, and equation
(1.1) which are both defined on $D(H)$, the unitary operator
$U(T):=\e^{-\i TH/\hbar}$ describes the time evolution of arbitrary
states $\Psi\in\CL^2$ (time-evolution operator): $H$ is the
infinitessimal generator of $U$. The time evolution for some state
$\Psi$ reads: $\Psi(t'')=U(t'',t')\Psi(t')$ Rewriting the time
evolution with $U(T)$ as an integral operator we get
\plus
$$\Psi(q'';t'')=\int\sqrt{g(q')}K(q'',q';t'',t')\Psi(q';t')dq',
  \tag\NUM.\num$$
where $K(T)$ is the celebrated Feynman kernel. Equations (1.1) and
(1.3) are connected. Having an explicit expression for $K(T)$ in (1.3)
one can derive in the limit $T=\epsilon\to0$ equation (1.1). This, on
the other hand proves that $K(T)$ is indeed the correct integral kernel
corresponding to $U(T)$. A rigorous proof includes, of course, the
check of the selfadjointness of $H$, i.e. $H=H^*$. Due to the semigroup
property
\plus
$$U(t_2)U(t_1)=U(t_1+t_2)
  \tag\NUM.\num$$
($H$ time-independent) we have
\plus
$$K(q'',q';t''+t')=\int K(q'',q;t''+t)K(q,q';t+t')dq
  \tag\NUM.\num$$

In the case of an Euclidean space, where $g_{ab}=\delta_{ab}$, $S$
is just the classical action, $S_{Cl}=\int\big[{m\over2}\dot q^2-V(q)
\big] dt=\int\CL_{cl}(q,\dot q)dt$, and we get explicitly
[$\Delta q^{(j)}:=(q^{(j)}-q^{(j-1)})$,
$q^{(j)}=q(t_j)$, $t_j=t'+j\epsilon$, $\epsilon=(t''-t')/N$,
$N\to\infty$, $D$ = dimension of the Euclidean space] :
\hfuzz=6pt
\plus
$$\allowdisplaybreaks\align
  K(q'',q';T)
  &=\lim_{N\to\infty}
  \bigg({m\over2\pi\i\hbar T}\bigg)^{ND\over2}
  \prod_{j=1}^{N-1}\int_{-\infty}^\infty dq^{(j)}
  \exp\left\{\ih\sum_{j=1}^N\bigg[
  {m\over2\epsilon}\Delta^2q^{(j)}-\epsilon V(q^{(j)})\bigg]\right\}.
  \\  &\equiv
  \int\limits_{q(t')=q'}^{q(t'')=q''}\CD q(t)
  \exp\left\{\ih\int_{t'}^{t''}\bigg[{m\over2}\dot q^2-V(q)
  \bigg]dt\right\}
  \tag\NUM.\num\endalign$$
\hfuzz=3pt
For an arbitrary metric $g_{ab}$ things are unfortunately not so easy.
The first formulation for this case is due to DeWitt [\DEWb]. His
result reads:
\plus
$$\multline
  \!\!\!\!
  K(q'',q';T)
  \\  =
  \int\limits_{q(t')=q'}^{q(t'')=q''}\sqrt{g}\,\CD_{DeW} q(t)
  \exp\left\{\ih\int_{t'}^{t''}\bigg[{m\over2}g_{ab}(q)
   \dot q^a\dot q^b-V(q)+\hbar^2{R\over6m}\bigg] dt\right\}
  \hfill\\
  :=\lim_{N\to\infty}\Norm^{ND/2}
  \prod_{j=1}^{N-1}\int\sqrt{g(q^{(j)})}\,dq^{(j)}
  \hfill \\   \times
  \exp\left\{\ih\sum_{j=1}^N\bigg[{m\over2\epsilon}
  g_{ab}(q^{(j-1)})\Delta q^{a,(j)}\Delta q^{b,(j)}
  -\epsilon V(q^{(j-1)})+\epsilon{\hbar^2\over6m}R(q^{(j-1)})
  \bigg] \right\}
  \hfill\endmultline
  \tag\NUM.\num$$
($R=g^{ab}(\Gamma^c_{ab,c}-\Gamma^c_{cb,a}
    +\Gamma^d_{ab}\Gamma^c_{cd}-\Gamma^d_{cb}\Gamma^c_{ad})$
- scalar curvature; $\Gamma^a_{bc}=g^{ad}(g_{bd,c}+g_{dc,b}
-g_{bc,d})$ - Christoffel symbols).
Of course, we identify $q'=q^{(0)}$ and $q''=q^{(N)}$ in the limit
$N\to\infty$. Two comments are in order:
\item{1)} Equation (1.7) has the form of equation (1.6) but the
          corresponding $S=\int\CL dt$ is not the classical action,
          respectively the Lagrangian $\CL$ is not the classical
          Lagrangian $\CL(q,\dot q)={m\over2}g_{ab}\dot q^a \dot
          q^b-V(q)$, but rather an effective one:
\plus
          $$S_{eff}=\int\CL_{eff}dt\equiv
          \int(\CL_{cl}-\Delta V_{DeW})dt.
  \tag\NUM.\num$$
          The quantum correction $\Delta V_{DeW}=-{\hbar^2\over6m}R$ is
          indispensible in order to derive from the time evolution
          equation (1.3) the Schr\"odinger equation (1.1)
          (e.g.\ [\DMa]). The appearance of a quantum correction
          $\Delta V$ is a very general feature for path integrals
          defined on curved manifolds; but, of course, $\Delta
          V\sim\hbar^2$ depends on the lattice definition.
\item{2)} A specific lattice definition has been chosen.
          The metric terms in the action are evaluated at the
          ``prepoint'' $q^{(j-1)}$.
          Changing the lattice definition, i.e. evaluation of the
          metric terms at other points, e.g. the ``postpoint''
          $q^{(j)}$ or the ``mid-point'' $\bqj:={1\over2}
          (q^{(j)}+q^{(j-1)})$ changes $\Delta V$, because in a Taylor
          expansion of the relevant terms, {\bf all} terms of
          $O(\epsilon)$ contribute to the path integral.
          This fact is particularly important in the expansion
          of the kinetic term in the Lagrangian, where we have
          $\Delta^4 q^{(j)}/\epsilon\sim O(\epsilon)$.

\noindent
One of the basics requirements of the construction of the
path integral is Trotter's [\TROT] product formula.
Discussions and proofs can be found in many textbooks on
functional integration and functional analysis
(e.g.\ Reed and Simon [\RS], Simon [\SIMON]).
Let us shortly notice a simple proof [\SIMON] :

\noindent{\bf Theorem}:
Let $A$ and $B$ be selfadjoint operators on a separable Hilbert space
so that $A+B$, defined on $D(A)\dots D(B)$, is selfadjoint. Then
\plus
$$\e^{\i T(A+B)}=\slim_{N\to\infty}
  \Big(\e^{\i tA/N}\e^{\i tB/N}\Big)^N.
  \tag\NUM.\num$$
If furthermore, $A$ and $B$ are bounded from below, then
\plus
$$\e^{-t(A+B)}=\slim_{N\to\infty}\Big(\e^{-tA/N}\e^{-tB/N}\Big)^N.
  \tag\NUM.\num$$

Proof ([\NELb,\SIMON]):
Let $S_T=\e^{\i T(A+B)}$, $V_T=\e^{\i TA}$, $W_T=\e^{\i TB}$,
$U_T=V_TW_T$, and let $\Psi_T=S_T\Psi$ for some $\Psi\in\CH$,
with underlying Hilbert space $\CH$. Then
\plus
$$\allowdisplaybreaks\align
  \Vert (S_T-U_{T/N}^N)\Psi\Vert
  &=\Bigg\Vert
  \sum_{j=0}^{N-1}U_{T/N}^j(S_{T/N}-U_{T/N})^{N-j-1}\Psi\Bigg\Vert
  \\
  &\leq N\sup_{0\leq s\leq T}\Vert (S_T-U_{T/N})\Psi\Vert.
  \tag\NUM.\num\endalign$$
Let $\Phi\in D(A)\cap D(B)$. Then $s^{-1}(S_s-1)\Phi\to\i(A+B)\Phi$ for
$s\to0$ and
\plus
$$\allowdisplaybreaks\align
  {U_s-1\over s}\Phi
  &=V_s(\i B\Phi)+V_s\bigg({W_s-1\over s}-\i B\bigg)\Phi
   +{V_s-1\over s}\Phi
  \\
  &\to\i B\Phi+\i A\Phi+0
  \tag\NUM.\num
  \endalign$$
hence
\plus
$$\lim_{N\to\infty}\Big[
   N\Vert(S_{T/N}-U_{T/N})\Phi\Vert\Big] \to0,
  \qquad\hbox{for each }\Phi\in D(A)\cap D(B).
  \tag\NUM.\num$$
Now let $D$ denote $D(A)\cap D(B)$ with the norm $\Vert(A+B)\Phi\Vert+
\Vert\Phi\Vert\equiv\Vert\Phi\Vert_{A+B}$. By hypothesis, $D$ is a
Banach space. With the calculations of equations (1.11-1.13),
$\{N(S_{t/N}-U_{t/N})\}$ is a family of bounded operators from $D$ to
$\CH$ with $\sup_N\{N\Vert(S_{T/N}-U_{T/N})\Phi\Vert\}<\infty$ for each
$\Phi$. As a result, the uniform boundedness principle implies that
\plus
$$N\Vert(S_{T/N}-U_{T/N})\Phi\Vert\leq C\Vert\Phi\Vert_{A+B}
  \tag\NUM.\num$$
for some positive $C$. Therefore the limit (1.14) is uniform
over compact subsets of $D$. Now let $\Psi\in D$. Then
$s\to\Psi_s$ is a continuous map from
$[0,T] $ into $D$, so that
$\{\Psi_s\vert 0\leq s\leq T\}$ is compact in $D$. Thus the
right-hand side of equation (1.14) goes to zero as $N\to\infty$
The proof of equation (1.10) is similar.

\eject
\baselineskip=12pt

The conditions on the operators $A$ and $B$ can be weakened with the
requirement that they are fulfill some ``positiveness'' and possesses a
particular self-adjoint extension.

Thus we see that ordering prescriptions in the quantum Hamiltonian and
corresponding lattice formulations in the path integral are closely
related (compare also [\DOWc, \DOWd]). This can be formulated in a
systematic way (e.g.\ [\LRTd] ):
\newline
We consider a monomial in coordinates and momenta (classical quantities)
\plus
$$M(n,m)=q^{\mu_1}\dots q^{\mu_n}p^{\nu_1}\dots p^{\nu_m}
  \tag\NUM.\num$$
We want to have a correspondence rule according to
\plus
$$\exp\bigg[\ih(uq+vp)\bigg]
  \to D_\Omega(u,v;\b q,\b p)\equiv
  \Omega(u,v)\exp\bigg[\ih (u\b q+v\b p)\bigg]
  \tag\NUM.\num$$
to generate operators $\b q,\b p$ from coordinates $q,p$.
This produces the mapping
\plus
$$M(n,n)\to {1\over\i^{n+m}}\left\vert
  {\partial^{n+m}D_\Omega(u,v;\b q,\b p)\over
   \partial u_{\mu_1}\dots\partial u_{\mu_n}
   \partial v_{\nu_1}\dots\partial v_{\nu_m}}\right\vert_{u=v=0}.
  \tag\NUM.\num$$
Some better known examples are displayed in the following table
$$\aligned
\def\vsp{\vphantom{${\dsize\sum_{K=0}^N}$}}
\vbox{\offinterlineskip
\hrule
\halign{&\vrule#&
  \strut\quad\hfil#\quad\hfil\quad\cr
height2pt&\omit&&\omit&&\omit&\cr
&Correspondence Rule\hfil
               &&$\Omega(u,v)$
               &&Ordering Rule                              &\cr
height2pt&\omit&&\omit&&\omit&\cr
\noalign{\hrule}
height2pt&\omit&&\omit&&\omit&\cr
&Weyl          &&1                                          \vsp
               &&${\dsize
     {1\over 2^n}\sum_{l=0}^n
     \bigg(\matrix n\\ l\endmatrix\bigg)
     {\b q}^{n-l}{\b p}^m {\bar q}^l}$                      &\cr
&Symmetric     &&$\cos\dfrac{u\cdot v}2$                    \vsp
               &&${\dsize
     \half({\b q}^n{\b p}^m+{\b p}^m{\b q}^n)}$             &\cr
&Standard      &&$\exp\bigg(-\i\dfrac{u\cdot v}2\bigg)$      \vsp
               &&${\dsize
     {\b q}^n{\b p}^m}$                                     &\cr
&Anti-Standard &&$\exp\bigg(\i\dfrac{u\cdot v}2\bigg)$       \vsp
               &&${\dsize
     {\b p}^m{\b q}^n}$                                     &\cr
&Born-Jordan   &&$\sin\dfrac{u\cdot v}2\bigg/\dfrac{u\cdot v}2$
                                                            \vsp
               &&${\dsize
     {1\over m+1}\sum_{l=0}^n{\b p}^{m-l}{\b q}^n {\b p}^l}$&\cr
height2pt&\omit&&\omit&&\omit&\cr}
\hrule}
  \endaligned$$
We can make this correspondence explicit. Let us consider the Fourier
integral
\plus
$$\gathered
  A(p,q)
  =\int\widehat{A(u,v)}\exp\bigg[\ih(uq+ivp)\bigg] dudv
  \\
  \widehat{A(u,v)}
  ={1\over(2\pi\hbar)^D}\int A(p,q)\exp\bigg[-\ih(uq+ivp)\bigg] dpdq
  \endgathered
  \tag\NUM.\num$$
We define the operator $A^\Omega(\b p,\b q)$ via
\plus
$$A^\Omega(\b p,\b q)=\int\widehat{A(u,v)}\Omega(u,v)
  \exp\bigg[\ih(u\b q+iv\b p)\bigg] dudv,
  \tag\NUM.\num$$
which gives
\plus
$$\b A^\Omega(\b p,\b q)
  ={1\over(2\pi\hbar)^D}\int A(p,q)\Omega(u,v)
  \exp\bigg[-\ih u(q-\b q)-\ih v(p-\b p)\bigg] dudvdpdq.
  \tag\NUM.\num$$
The inverse transformation is denoted by $A^\Theta(p,q)$ and has the
form
\plus
$$A^\Theta(p,q)={1\over(2\pi\hbar)^D}\Tr\int
  \b A(\b p,\b q)[\Omega(u,v)]^{-1}
  \exp\bigg[\ih u(q-\b q)+i\ih v(p-\b p)\bigg] dudv.
  \tag\NUM.\num$$
In particular, the $\delta$-function has the correspondence operator
\plus
$$\Delta^\Omega(p-\b p,q-\b q)
  ={1\over(2\pi\hbar)^D}\int\Omega(u,v)
  \exp\bigg[-\ih u(q-\b q)-\ih v(p-\b p)\bigg] dudv.
  \tag\NUM.\num$$

Let us turn to the Hamiltonian. Given a classical Hamiltonian
$\CH_{cl}(p,q)$ the quantum Hamiltonian is calculated as
\plus
$$\gathered
  \b H(\b p,\b q)=\int\exp\bigg(\ih u\b p+\ih v\b q\bigg)
  \Omega(u,v)\widehat{\CH_{cl}(u,v)}dudv
  \\
  \widehat{\CH_{cl}(u,v)}={1\over(2\pi\hbar)^{2D}}
  \int\exp\bigg(-\ih up-\ih uq\bigg)\CH_{cl}(p,q)dpdq.
  \endgathered
  \tag\NUM.\num$$
Obviously, by proposing a particular classical Hamiltonian depending on
variables $q'$ and $q''$, respectively, produces a particular
Hamiltonian function,
\plus
$$H(p,q'',q')={1\over(2\pi\hbar)^D}
  \int \Omega(q''-q',y)\e^{ivy/\hbar}\CH_{cl}[p,\bhalf(q''+q')-v]dudv
  \tag\NUM.\num$$
For  $\Omega=1$ and $\Omega=\cos{uv\over2}$, respectively,
we obtain Hamiltonian functions according to
\plus
$$\gathered
  H(p,q'',q')=\CH(p,\bhalf(q'+q''))
  \\
  H(p,q'',q')=\bhalf[\CH_{cl}(p,q'')+\CH_{cl}(p,q')]
  \endgathered
  \tag\NUM.\num$$
which are the matrix elements
\plus
$$<q'\vert\b H(\b p,\b q)\vert q''>
  \tag\NUM.\num$$
of the quantum Hamiltonians which are Weyl- and symmetrically
ordered, respectively. Choosing $\Omega(u,v)=\exp[i(1-2u)uv]$ [\HIR]
yields
\plus
$$H(p,q;u)
  ={q\over2m}g^{ab}(q)p_ap_b
   +{\i\hbar\over m}(\bhalf-u)p_a{g^{ab}}_{,b}
   -{\hbar^2\over2m}(\bhalf-u)^2{g^{ab}}_{,ab}+\Delta V_{Weyl}.
  \tag\NUM.\num$$
with the well-defined quantum potential
\plus
$$\Delta V_{Weyl}={\hbar^2\over8m}(g^{ab}\Gamma^d_{ac}\Gamma^c_{bd}-R)
      ={\hbar^2\over8m}\Big[g^{ab}\Gamma_a\Gamma_b
       +2(g^{ab}\Gamma_a)_{,b}+g^{ab}_{\ \ ,ab}\Big]
  \tag\NUM.\num$$
$u=\half$ corresponds to the Weyl prescription and is clearly
emphazised.

In the next two Sections we discuss two particular ordering
prescriptions and the corresponding matrix elements which will appear
as the relevant quantities to be used in path integrals on curved
spaces, it will be the Weyl-ordering prescription (leading to the
midpoint rule in  the path integral) and a product ordering (leading to
a product rule in the path integral).

Another approach due to Kleinert [\KLEk, \KLEm] I will not discuss here.

\bigskip\bigskip
\glno=0               
\advance\chapno by 1  

\section{Weyl-Ordering}
A very convenient lattice prescription is the mid-point definition,
which is connected to the Weyl-ordering prescription in the Hamiltonian
$H$. Let us discuss this prescription in some detail.
First we have to construct momentum operators [\PAUb]:
\plus
$$p_a={\hbar\over\i}\left({\partial\over\partial q^a}
       +{\Gamma_a\over2}\right),\qquad
  \Gamma_a={\partial\ln\sqrt{g}\over\partial q^a}
  \tag\NUM.\num$$
which are hermitean with respect to the scalar product $(f_1,f_2)=\int
f_1f_2^*\sqrt{g}\,dq$. In terms of the momentum operators (2.1) we
rewrite the quantum Hamiltonian $H$ by using the Weyl-ordering
prescription [\GRSb,\LEEb,\MIZa]
\plus
$$H(p,q)={1\over8m}(g^{ab}\,p_a p_b+2p_a\,g^{ab}\,p_b
                     +p_a p_b\,g^{ab})+\Delta V_{Weyl}(q)+V(q).
  \tag\NUM.\num$$
Here a well-defined quantum correction appears which is given by
[\GRSb,\MIZa,\OMO]:
\plus
$$\Delta V_{Weyl}={\hbar^2\over8m}(g^{ab}\Gamma^d_{ac}\Gamma^c_{bd}-R)
      ={\hbar^2\over8m}\Big[g^{ab}\Gamma_a\Gamma_b
       +2(g^{ab}\Gamma_a)_{,b}+g^{ab}_{\ \ ,ab}\Big]
  \tag\NUM.\num$$
The Weyl-ordering prescription is the most discussed ordering
prescription in the literature. Let us start by defining it for powers
of position- and momentum- operators $\b q$ and $\b p$, respectively
[\LEEb]:
\plus
$$(\b q^m \b p^r)_{Weyl}=
  \bigg(\half\bigg)^m\sum_{l=0}^m
  {m!\over l!(m-l)!} \b q^{m-l}\b p^r \b q^l,
  \tag\NUM.\num$$
with the matrix elements (one-dimensional case)
\plus
$$\allowdisplaybreaks\align
  <q''\vert (\b q^m \b p^r)_{Weyl}\vert q'>
  &= \bigg(\half\bigg)^m\sum_{l=0}^m
  {m!\over l!(m-l)!} q^{m-l}<q''\vert\b p^r\vert q'> q^l
  \\
  &=\int {dp\over(2\pi\hbar)^D}p^r \e^{\i p(q''-q')/\hbar}
                                        \bigg({q'+q''\over2}\bigg)^m
  \tag\NUM.\num
  \endalign$$
and all coordinate dependent quantities turn out to be evaluated
at {\bf mid-points}. Here was used that the matrix elements of the
position $\vert q>$ and momentum eigen-states $\vert p>$ have the
property
\plus
$$<q''\vert q'>=(g'g'')^{-{1/4}}\delta(q''-q'),
  \qquad
  <q\vert p>=(2\pi\hbar)^{-D/2}\e^{\i pq/\hbar}.
  \tag\NUM.\num$$
This power rule is nothing but a special case of a more general
prescription. Of course, well behaved operator valued functions are
included.
\newline
We have the {\bf Weyl-transform} of an operator $A(\b p,\b q)$
as [\MIZa]:
\plus
$$A(p,q)\equiv\int dv\,\e^{\i pv}
  <q-\hbox{${v\over2}$}\vert\b A(\b p,\b q)\vert q+\hbox{${v\over2}$}>,
  \tag\NUM.\num$$
with the inverse transformation
\plus
$$\gathered
  \b A(\b p,\b q)={1\over(2\pi\hbar)^D}\int dpdq A(p,q)\Delta(p,q)
  \\
  \Delta(\b p,\b q)=\int du\,\e^{\i pu}
  \vert p-\tsize{u\over2}><p+\tsize{u\over2}\vert .
  \endgathered
  \tag\NUM.\num$$
Let us consider some simple examples. First of all for some operator
$f(\b p)$
\plus
$$\multline
  \int dv\,\e^{\i pv/\hbar}
  <q-\hbox{${v\over2}$}\vert f(\b p)\vert q+\hbox{${v\over2}$}>
  \\    \qquad
  =\int dvdp'\,\e^{\i pv/\hbar}
  <q-\hbox{${v\over2}$}\vert f(\b q)\vert p'>
  <p'\vert q+\hbox{${v\over2}$}>
  \hfill\\   \qquad
  ={1\over(2\pi\hbar)^D}\int dvdp'\e^{i(p-p')v/\hbar} f(p')=f(p).
  \hfill\endmultline
  \tag\NUM.\num$$
Hence (and similarly)
\plus
$$f(\b p)\iff f(p),\qquad f(\b q)\iff f(q).
  \tag2.10$$
Straightforwardly one shows the following correspondence
\plus
$$\multline
  F(\b q)\b p_i g^{ij}(\b q)\b p_j F(\b q)
  \\  \qquad
  \iff p_ip_jF^2(q)g^{ij}
  +\half\bigg(\half F^2(q){g^{ij}}_{,ij}(q)
  +F_{,i}(q)F_{,j}(q)g^{ij}(q)-F_{,ij}(q)F(q)g^{ij}(q)\bigg)
  \hfill\\ \ \endmultline
  \tag\NUM.\num$$
Special cases are
\plus
$$\aligned
  \b p_i\b p_j F(\b q)
  &\iff \bigg(p_i-{\i\hbar\over2}{\partial\over\partial q^i}\bigg)
        \bigg(p_j-{\i\hbar\over2}{\partial\over\partial q^j}\bigg)F(q)
  \\
  \b p_iF(\b q)\b p_j
  &\iff \bigg(p_i-{\i\hbar\over2}{\partial\over\partial q^i}\bigg)
        \bigg(p_j+{\i\hbar\over2}{\partial\over\partial q^j}\bigg)F(q)
  \\
  \b p_iF(\b q)\b p_j
  &\iff \bigg(p_i+{\i\hbar\over2}{\partial\over\partial q^i}\bigg)
        \bigg(p_j+{\i\hbar\over2}{\partial\over\partial q^j}\bigg)F(q).
  \endaligned
  \tag\NUM.\num$$
For the case that $F(q)$ is some symmetric $N\times N$ matrix we have
\plus
$${1\over4}\bigg[
  \b p_i\b p_j F^{ij}(\b q) +2\b p_iF^{ij}(\b q)p_j
  +F^{ij}(\b q)\b p_i\b p_j\bigg]
  \iff p_ip_jF^{ij}(q).
  \tag\NUM.\num$$
There is, of course, a one-to-one correspondence between the function
$A(p,q)$ and the operator $A(\b p,\b q)$, called {\bf
Weyl-correspondence}. It gives a prescription how {\bf Weyl-ordered}
operators can be constructed by the classical counterpart. Starting now
with an operator $A$, the Weyl-correspondence gives an unique
prescription for the construction of the path integral. We have for the
Feynman kernel for an arbitrary $N\in\N$ [which is due to the
semi-group property of $U(T)$, i.e. $U(t_1+t_2)=U(t_1)U(t_2)$]:
\plus
$$\multline
  K(q'',q';T)=\bigg<q''\bigg\vert
  \exp\bigg[-\ih\b H(\b p,\b q)\bigg]\bigg\vert q'\bigg>
  \\
  =\left(\prod_{j=1}^{N-1}\int\sqrt{g^{(j)}}\,dq^{(j)}\right)
  \prod_{j=1}^N\bigg<q^{(j)}\bigg\vert
  \exp\bigg[-\ih{T\over N}\b H(\b p,\b q)\bigg]
  \bigg\vert q^{(j-1)}\bigg>.
  \endmultline
  \tag\NUM.\num$$
Observing
\plus
$$\allowdisplaybreaks\align
  <q'\vert\Delta(\b p,\b q)\vert q''>
  &=\int du\,\e^{\i uq}
  <q'\vert p-\tsize{u\over2}><p+\tsize{u\over2}\vert q''>
  \\
  &={1\over(2\pi\hbar)^D}\int du\,\exp\bigg[\ih p(q'-q'')
  +\ih u\bigg(q-{q'+q''\over2}\bigg)\bigg]
  \\
  &=\e^{\i p(q'-q'')/\hbar}\delta\bigg(q-{q'+q''\over2}\bigg)
  \tag\NUM.\num
  \endalign$$
and making use  of the Trotter formula $\e^{-\i t(A+B)}=
s\!-\!\lim_{N\to\infty}\left(\e^{-\i tA/N}\e^{-\i tB/N}\right)^N$
and the short-time approximation for the matrix element
$<q''\vert\e^{-\i\epsilon H}\vert q'>$:
\plus
$$\multline
  <q^{(j)}\vert\exp\Big[
  -\i\epsilon\b H(\b p,\b q)/\hbar\Big]\vert q^{(j-1)}>
  \\  \qquad
  ={1\over(2\pi\hbar)^D}<q^{(j)}
  \bigg\vert\int dpdq\,\e^{-\i h(p,q)/\hbar}
   \int dudv\,\exp\big[
  i(q-\b q)u+i(p-\b p)v\big]\bigg\vert q^{(j-1)}>
  \hfill\\   \qquad
  ={1\over(2\pi\hbar)^D}\int dp\,
  \exp\bigg[{\i\epsilon\over\hbar}p(q^{(j)}-q^{(j-1)})
               -{\i\epsilon\over\hbar}H(p,\bar q^{(j)})\bigg],
  \hfill\endmultline
  \tag\NUM.\num$$
where $\bar q^{(j)}=\half(q^{(j)}+q^{(j-1)})$ is the mid-point
coordinate.
\newline
Let us consider the classical Hamilton-function
\plus
$$\CH_{cl}(p,q)={1\over2m}g^{ab}(q)
  \big[p_a-A_a(q)\big]
  \big[p_b-A_b(q)\big]+V(q).
  \tag\NUM.\num$$
The corresponding quantum mechanical operator is not clearly defined
due to the factor ordering ambiguity. Let us try the manifestly
hermitean operator
\plus
$$H(\b p,\b q)={1\over2m}g^{-{1\over4}}
  \big[\b p_a-A_a(\b q)\big]g^\half(\b q)g^{ab}(\b q)
  \big[\b p_b-A_a(\b q)\big]g^{-{1\over4}}(\b q)+V(\b q).
  \tag\NUM.\num$$
The Weyl-transformed of $H$ reads
\plus
$$\allowdisplaybreaks\align
  H(p,q)&=\CH_{cl}(p,q)+{\hbar^2\over2m}
        \bigg[{1\over4}g^{-\half}\big(g^\half g^{ab}\big)_{,ab}
  \\
  &\qquad
   +\half\big(g^{-{1\over4}}\big)_{,a}
              \big(g^{-{1\over4}}\big)_{,b}g^\half g^{ab}
   -\half\big(g^{-{1\over4}}\big)_{,ab}g^{1\over4}g^{ab}\bigg]
  \\
  &=\CH_{cl}(p,q)+{\hbar^2\over8m}
   \big[\Gamma^m_{la}\Gamma^l_{mb}g^{ab}-R\big]
  \\
  &=\CH_{cl}(p,q)+{\hbar^2\over8m}\Big[g^{ab}\Gamma_a\Gamma_b
       +2(g^{ab}\Gamma_a)_{,b}+g^{ab}_{\ \ ,ab}\Big]
  \tag\NUM.\num
  \endalign$$
and, of course, $\Delta V_{Weyl}$ appears again. For the path integral
all quantities have to be evaluated at $\bar q^{(j)}$. Note: The
Weyl-transformed of the the Weyl-ordered operator (2.16) is just the
classical Hamiltonian (2.17). Inserting all quantities in equation
(2.13) we obtain the {\bf Hamiltonian path integral}:
\plus
$$\multline
  K(q'',q';T)
  =[g(q')g(q'')]^{-{1\over4}}
   \lim_{N\to\infty}\prod_{j=1}^{N-1}\int dq^{(j)}
   \cdot\prod_{j=1}^N\int{dp^{(j)}\over(2\pi\hbar)^D}
   \hfill\\    \qquad\times
   \exp\left\{\ih\sum_{j=1}^N \Big[
   \Delta q^{(j)}\cdot p^{(j)}-\epsilon\CH_{eff}(p^{(j)},{\bar q}^{(j)})
             \Big]\right\}.
  \hfill\endmultline
  \tag\NUM.\num$$
The effective Hamiltonian to be used in the path integral (2.20) reads,
\plus
$$H(p,\bar q)=\CH_{eff}(p^{(j)},{\bar q}^{(j)})=
  {1\over2m}g^{ab}(\bqj)p^{(j)}_a p^{(j)}_b
  +V(\bqj)+\Delta V_{Weyl}(\bqj).
  \tag\NUM.\num$$
With the help of the famous Gaussian integral
\plus
$$\int_{-\infty}^\infty \e^{-p^2x^2\pm qx}\,dx
  ={\sqrt{\pi}\over p}\exp\bigg({q^2\over4p^2}\bigg)
  \tag\NUM.\num$$
and, respectively, by its $D$-dimensional generalization
\plus
$$\int_{\R^n}dp\,\e^{iq\cdot p-\half g^{ab}p_ap_b}
  =(2\pi)^{D/2}\sqrt{\det(g_{ab})}\,
  \exp\bigg(-\half g_{ab}q^aq^b\bigg),
  \tag\NUM.\num$$
we get by integrating out the momenta the {\bf Lagrangian path
integral} which reads ($MP$ = {\bf M}id-{\bf P}oint):
\plus
$$\multline
  K(q'',q';T)=[g(q')g(q'')]^{-{1\over4}}
  \int\limits_{q(t')=q'}^{q(t'')=q''}\sqrt{g}\,\CD_{MP} q(t)
  \exp\bigg[\ih\int_{t'}^{t''}\CL_{eff}(q,\dot q)dt\bigg]
  \hfill\\  \quad
  :=[g(q')g(q'')]^{-{1\over4}}\lim_{N\to\infty}
  \Norm^{ND\over2}
  \left(\prod_{j=1}^{N-1}\int dq^{(j)}\right)
  \hfill\\ \qquad\times
  \prod_{j=1}^N\sqrt{g(\bqj)}
  \exp\left\{\ih\bigg[{m\over2\epsilon}g_{ab}(\bqj)
  \Delta q^{a,(j)}\Delta q^{b,(j)}-\epsilon V(\bqj)
  -\epsilon\Delta V_{Weyl}(\bqj)\bigg]\right\}.
  \hfill\\ \quad\endmultline
  \tag\NUM.\num$$
Equation (2.24) is, of course, equivalent with equation (1.7).
The mid-point prescription arises here in a very natural way,
as a consequence of the Weyl-ordering prescription. This is a general
feature that ordering prescriptions lead to {\bf specific} lattices and
that different lattices define {\bf different} $\Delta V$.

To proof that the path integral (2.24) is indeed the correct one,
one has to show that with the corresponding short time kernel
\plus
$$\multline
  K(q^{(j)},q^{(j-1)};\epsilon)=
  \Norm^{D/2}[g(q^{(j-1)})g(q^{(j)})]^{-{1\over4}}\sqrt{g(\bqj)}
  \hfill\\   \hfill\times
  \exp\left\{\ih\bigg[{m\over2\epsilon}g_{ab}(\bqj)
  \Delta q^{a,(j)}\Delta q^{b,(j)}-\epsilon V(\bqj)
  -\epsilon\Delta V_{Weyl}(\bqj)\bigg]\right\}.
  \endmultline
  \tag\NUM.\num$$
and the time evolution equation  (1.3) the Schr\"odinger equation (1.1)
follows. For this purpose a Taylor expansion has to performed in
equation (1.3) yielding
\plus
$$\Psi(q'',t)+\epsilon{\partial\Psi(q'';t)\over\partial t}
  =B_0\Psi(q'';t)+B_{q^b}{\partial\Psi(q'';t)\over\partial{q'}^b}
  +B_{q^aq^b}{\partial^2\Psi(q'';t)\over\partial{q'}^b{q'}^a}+\dots,
  \tag\NUM.\num$$
where the coefficients in the expansion are given by
\advance\glno by -1
$$\allowdisplaybreaks\align
  B_0&=\int dq'\sqrt{g(q')}K(q'',q';\epsilon)
  \\     &\simeq\Norm^{D/2}
   g^{-{1\over4}}(q'')
   \e^{-\i\epsilon [V(q'')+\Delta V(q'')]/\hbar}
  \\     &\qquad\times
  \int dq'g^\half(\bar q)g^{1\over4}(q')
  \exp\bigg({\i m\over2\epsilon\hbar}\Delta q^a\,{_a}g_b\Delta q^b\bigg)
  \tag\NUM.\num\\    \global\plus
  B_{q^b}&=\int dq'\sqrt{g(q')}K(q'',q';\epsilon)\Delta q^b
  \\     &\simeq\Norm^{D/2}
  g^{-{1\over4}}(q'')\int dq'g^\half(\bar q)g^{1\over4}(q')
  \exp\bigg({\i m\over2\epsilon\hbar}\Delta q^a\,{_a}g_b\Delta q^b\bigg)
  \Delta q^b
  \tag\NUM.\num\\   \global\plus
  B_{q^aq^b}
  &=\int dq'\sqrt{g(q')}K(q'',q';\epsilon)\Delta q^a\Delta q^b
  \\
  &\simeq\left({m\over2\pi\i\epsilon}\right)^{D/2}
   g^{-{1\over4}}(q'')\int dq'g^\half(\bar q)g^{1\over4}(q')
  \exp\bigg({\i m\over2\epsilon\hbar}\Delta q^a\,{_a}g_b\Delta g^b\bigg)
   \Delta q^a\Delta q^b
  \tag\NUM.\num
  \endalign$$
{}From these representations it is clear that we need equivalence
relations (in the sense of path integrals) for $\Delta q^a\Delta q^b$
etc. They are given by
$$\allowdisplaybreaks\align
  \Delta q^a\Delta q^b&{\dot=}{\i\epsilon\hbar\over m}g^{ab}
  \tag\NUM.\num\\    \global\plus
  \Delta q^a\Delta q^b\Delta q^c\Delta q^d&{\dot=}
  \bigg({\i\epsilon\hbar\over m}\bigg)^2
  \big[g^{ab}\,g^{cd}+g^{ac}\,g^{bd}+g^{ad}\,g^{bc}\big].
  \tag\NUM.\num
  \endalign$$
\plus
$$\multline
  \Delta q^a\Delta q^b\Delta q^c\Delta q^d\Delta q^e\Delta q^f
  \\  \quad
  {\dot=}\bigg({\i\epsilon\hbar\over m}\bigg)^3\Big[
   g^{ab}\,g^{cd}\,g^{cd}+g^{ac}\,g^{bd}\,g^{ef}+g^{ad}\,g^{bc}\,g^{ef}
  \hfill\\   \qquad
  +g^{ab}\,g^{ce}\,g^{df}+g^{ab}\,g^{cf}\,g^{de}+g^{cd}\,g^{ae}\,g^{bf}
  +g^{cd}\,g^{af}\,g^{be}+g^{ac}\,g^{be}\,g^{df}+g^{ac}\,g^{bf}\,g^{de}
  \hfill\\    \qquad
  +g^{bd}\,g^{ae}\,g^{cf}+g^{bd}\,g^{af}\,g^{ce}+g^{ad}\,g^{be}\,g^{cf}
  +g^{ad}\,g^{bf}\,g^{ce}+g^{bc}\,g^{ae}\,g^{df}+g^{bc}\,g^{af}\,g^{de}
  \Big].\hfill\\ \ \endmultline
  \tag\NUM.\num$$
We just show the identity (2.30), the proof of the remaining ones is
similarly. Let us consider the integral
\plus
$$I(g_{cd})=\int dq\exp\bigg({\i m\over2\epsilon\hbar}
                  \Delta q^c\,g_{cd}\,\Delta q^d\bigg)
           =\sqrt{m\over2\pi\i\epsilon\hbar g}
  \tag\NUM.\num$$
with the $g_{cd}$ as free parameters.
Differentiation with respect to one of the parameters gives on the
one hand:
\plus
$${\partial\over\partial g_{ab}}I(g_{cd})=
  {\i m\over2\epsilon\hbar}\int dq\exp\bigg({\i m\over2\epsilon\hbar}
           \Delta q^c\,g_{cd}\,\Delta q^d\bigg)\Delta q^a\Delta q^b
  \tag\NUM.\num$$
and on the other
\plus
$${\partial\over\partial g_{ab}}I(g_{cd})=
  \sqrt{m\over2\pi\i\epsilon\hbar}
  {\partial\over\partial g_{ab}}g^{-1/2}=
  -\half\sqrt{m\over2\pi\i\epsilon\hbar}
  g^{-1/2}\,g^{ab}=-\half g^{ab}I(g_{cd}).
  \tag\NUM.\num$$
Here we have used the formula for the differentiation of determinants:
$\partial g/\partial g_{ab}=g\,g^{ab}$. Combining the last two
equations yield (2.30).
\newline
Let us denote by $\xi=q''-q'$ and $q=q''$. The various Taylor expanded
contributions yield:
$$\allowdisplaybreaks\align
  g^{1/4}(q-\xi)g^{1/2}(q-\xi)
   &\simeq g^{3/4}(q)\Bigg[1-\Gamma_a\xi^a
   +{1\over8}(4\Gamma_a\Gamma_b+3\Gamma_{a,b})\xi^a\xi^b\Bigg].
  \tag\NUM.\num\\    \global\plus
  \exp\bigg[
  {\i m\over2\epsilon\hbar}g_{ab}(q-\xi)\xi^a\xi^b\Bigg]
  &\simeq\exp\bigg[
  {\i m\over2\epsilon\hbar}g_{ab}(q)\xi^a\xi^b\Bigg]
  \\    \times
  \bigg[1+{m\over2\i\epsilon\hbar}g_{ab}\Gamma^b_{cd}\xi^a\xi^c\xi^d
  & -{m\over8\i\epsilon\hbar}
      \Big(g_{av}\Gamma^v_{bc,d}+g_{au}\Gamma^a_{vd}\Gamma^v_{bc}
          +g_{uv}\Gamma^u_{ad}\Gamma^v_{bc}\Big)\xi^a\xi^b\xi^c\xi^d
  \\
  &+\half\bigg({m\over2\i\epsilon\hbar}\bigg)^2
          g_{av}\,g_{du}\Gamma^v_{bc}\Gamma^u_{ef}
                            \xi^a\xi^b\xi^c\xi^d\xi^e\xi^f\bigg].
  \tag\NUM.\num\endalign$$
Here the various derivatives of the metric tensor $g_{ab}$ have been
expressed by the Christoffel symbols. Thus combining the last two
equations yield:
\plus
$$\multline
  g^{1/4}(q-\xi)g^{1/2}(q-\xi)
  \exp\bigg[
  {\i m\over2\epsilon\hbar}g_{ab}(q-\xi)\xi^a\xi^b\Bigg]
  \simeq
  g^{3/4}(q)\exp\bigg[{\i m\over2\epsilon\hbar}
     g_{ab}(q)\xi^a\xi^b\Bigg]
  \hfill\\  \qquad\times
  \bigg[1-\bigg(\Gamma_a\xi^a+{m\over2\i\epsilon\hbar}
     g_ad\Gamma^d_{bc}\bigg) \xi^a\xi^b\xi^c
  +\half\bigg({m\over2\i\epsilon\hbar}\bigg)^2
  g_{av}\,g_{du}\Gamma^v_{bc}\Gamma^u_{ef}
  \xi^a\xi^b\xi^c\xi^d\xi^e\xi^f
  \hfill\\  \hfill
  -{m\over8\i\epsilon\hbar}\Big(g_{av}\Gamma^v_{bc,d }
  +g_{au}\Gamma^a_{vd}\Gamma^v_{bc}
  +g_{uv}\Gamma^u_{ad}\Gamma^v_{bc}\Big )
  \xi^a\xi^b\xi^c\xi^d
  +{1\over8}(4\Gamma_a\Gamma_b+3\Gamma_{a,b})\xi^a\xi^b\bigg].
  \\  \quad\endmultline
  \tag\NUM.\num$$
Let us start with the $B_{ab}$-terms. We get immediately by equation
(2.30):
\plus
$$B_{ab}{\dot=}-{\i\epsilon\hbar\over2m}g^{ab}.
  \tag\NUM.\num$$
Similarly:
\plus
$$B_a{\dot=}-{\i\epsilon\hbar\over2m}[\Gamma_a\,g^{ab}
                                    +(\partial_a\,g^{ab})]
  \tag\NUM.\num$$
For the $\xi^2$- and $\xi^4$-terms in $B_0$ we get:
$$\allowdisplaybreaks\align
  {1\over8}(4\Gamma_a\Gamma_b+3\Gamma_{a,b})\xi^a\xi^b
  &{\dot=}
  {\i\epsilon\hbar\over8m}g^{ab}(4\Gamma_a\Gamma_b+3\Gamma_{a,b}),
  \tag\NUM.\num\\    \global\plus
  -{m\over8\i\epsilon\hbar}\Big(g_{av}\Gamma^v_{bc,d}
  &+g_{au}\Gamma^a_{vd}\Gamma^v_{bc}
  +g_{uv}\Gamma^u_{ad}\Gamma^v_{bc}\Big)\xi^a\xi^b\xi^c\xi^d
  \\  &
  {\dot=}-{\i\epsilon\hbar\over8m}g^{ab}
  \Big[8\Gamma_a\Gamma_b+\Gamma^c_{ab,c}+2\Gamma_{a,b}
  \\  &\
  +g_{uv}g^{cd}(2\Gamma^u_{ad}\Gamma^v_{bc}+\Gamma^u_{ab}\Gamma^v_{cd})
  +5\Gamma^c_{ab}\Gamma^c_{bd}+2\Gamma^d_{ac}\Gamma^c_{bd}\Big].
  \tag\NUM.\num
  \endalign$$
For the $\xi^6$-terms equation (2.32) yields
\plus
$$\multline
  \half\bigg({m\over2\i\epsilon\hbar}\bigg)^2
  g_{av}\,g_{du}\Gamma^v_{bc}\Gamma^u_{ef}
  \xi^a\xi^b\xi^c\xi^d\xi^e\xi^ f
  \hfill\\   \hfill
  {\dot=}{\i\epsilon\hbar\over8m}\,g^{ab}\Big[
  4\Gamma_a\Gamma_b+4\Gamma^c_{ab}\Gamma_c+4\Gamma^d_{ac}\Gamma^c_{bd}
  +g_{uv}g^{cd}(2\Gamma^u_{ac}\Gamma^v_{bd}+\Gamma^u_{ab}\Gamma^v_{cd})
  \Big].\endmultline
  \tag\NUM.\num$$
Therefore combining the relevant terms yields finally:
\plus
$$\multline
  \int g^{1/4}(q-\xi)g^\half(q-\bar\xi)
  \exp\bigg[{\i m\over2\epsilon\hbar}
  g_{ab}(q-\bar\xi)\xi^a\xi^b\bigg]d\xi
  \hfill\\   \qquad
  {\dot=}\Norm^{-{D/2}}
  \bigg[1+{\i\epsilon\hbar\over8m}g^{ab}\big(
  \Gamma_{a,b}-\Gamma^c_{ab}\Gamma_c+2\Gamma^d_{ac}\Gamma^c_{bd}
  -\Gamma^c_{ab,c}\Big)\bigg]
  \hfill\\  \qquad
  {\dot=}\Norm^{-{D/2}}
  \exp\bigg({\i\epsilon\over\hbar}\Delta V_{Weyl}\bigg).
  \hfill\endmultline
  \tag\NUM.\num$$
Inserting all the contributions into equation (2.67) yields the
Schr\"odinger equation (1.1).

Let us emphasize that the above procedure is nothing but a formal proof
of the path integral. A rigorous proof {\bf must} include at least two
more ingredients
\item{1)}
One must show that in fact
\plus
$$\lim_{N\to\infty}\vert [\e^{-\i TH/\hbar}-K(T)]\Psi\vert\to0
  \tag\NUM.\num$$
for all $\Psi\in\CH$ ($\CH$: relevant Hilbert space).
\item{2)}
One must show that the domain $\CD$ of the infinitessimal
generator of the kernel $K(T)$ is in fact identical
with the domain of the Hamiltonian corresponding to the Schr\"odinger
equation (1.1), i.e.\ the infinitessimal generator is the
(selfadjoint) Hamiltonian.

\bigskip\noindent
It is quite obvious that these strong mathematical requirements hold
only under certain assumptions on the potential involved in the
Hamiltonian.
Here Nelson [\NELb] has shown the validity of the path
integral for one dimensional path integrals.
Some instructive proof can be furthermore found in the books
of Simon [\SIMON]\ and Reed and Simon [\RS].
Also due to Albeverio et al.\ [\ALB-\ACHRS] is
a wide range of discussion to formulate the Feynman path integral
without the delay of go back to the definition of Wiener integrals.

\bigskip\bigskip
\glno=0               
\advance\chapno by 1  

\section{Product-Ordering}
In order to develop another useful lattice formulation for path
integrals we consider again the generic case [\GROa].
We assume that the metric tensor $g_{ab}$ is real and symmetric
and has rank$(g_{ab})=D$, i.e. we have no constraints on the
coordinates.
Thus one can always find a linear transformation $C:q_a=C_{ab}y_b$
such that $\CL_{Cl}={m\over2}\Lambda_{ab}\dot y^a\dot y^b$
with $\Lambda_{ab}=C_{ac}g_{cd}C_{db}$ and where $\Lambda$ is diagonal.
$C$ has the form $C_{ab}=u_a^{(b)}$ where the ${\vec u}^{(b)}$
$(b\in\{1,\dots,d\})$ are the eigenvectors of $g_{ab}$ and
$\Lambda_{ab}=f_c^2\delta_{ac}\delta_{bc}$ where
$f_a^2\not=0$ ($a\in\{1,\dots,d\}$) are the eigenvalues of $g_{ab}$.
Without loss of generality we assume $f_a^2>0$ for all
$a\in\{1,\dots,d\}$.
(For a time like coordinate $q_a$ one might have e.g.
$f_a^2<0$, but cases like this we want to exclude).
Thus one can always find a representation for $g_{ab}$ which reads,
\plus
$$g_{ab}(q)= h_{ac}(q)h_{bc}(q).
  \tag\NUM.\num$$
Here the $h_{ab}=C_{ac}f_c C_{cb}=u_c^{(a)}f_c u_c^{(b)}$ are real
symmetric $D\times D$ matrices and satisfy $h_{ab}h^{bc}=\delta_a^c$.
Because there exists the orthogonal transformation $C$ equation
(\NUM.1) yields for the y-coordinate system (denoted by $M_y$):
\plus
$$\Lambda_{ab}(y)=f_c^2(y)\delta_{ac}\delta_{bc}.
  \tag\NUM.\num$$
equation (\NUM.2) includes the special case $g_{ab}=\Lambda_{ab}$.
The square-root of the determinant of $g_{ab}, \sqrt{g}$ and the
Christoffels $\Gamma_a$ read in the q-coordinate system (denoted by
$M_q$):
\plus
$$\sqrt{g}=\det(h_{ab})=:h,\quad
  \Gamma_a={h_{,a}\over h},\quad
  p_a={\hbar\over\i}
  \left({\partial\over\partial q_a}+{h_{,a}\over2h}\right).
  \tag\NUM.\num$$
The Laplace-Beltrami-operator expressed in the $h^{ab}$ reads
on $M_q$:
\plus
$$\Delta_{LB}^{M_q}=\left\{h^{ac}h^{bc}
  {\partial^2\over\partial q^a\partial q^b}
  +\bigg[{\partial h^{ac}\over\partial q^a}h^{bc}
  +h^{ac}{\partial h^{bc}\over\partial q^a}
  +{h_{,a}\over h}h^{ac}h^{bc}\bigg]
  {\partial\over\partial q^b}\right\}
  \tag\NUM.\num$$
and on $M_y$:
\plus
$$\Delta_{LB}^{M_y}=
  {1\over f_a^2}\bigg[{\partial^2\over\partial y_a^2}
  +\bigg({f_{b,a}\over f_b}-2f_{a,a}\bigg)
  {\partial\over\partial y_a}\bigg].
  \tag\NUM.\num$$
With the help of the momentum operators (\NUM.3) we rewrite the
Hamiltonian in the ''pro\-duct-ordering'' form
\plus
$$H=-{\hbar^2\over2m}\Delta_{LB}^{M_q}+V(q)
   ={1\over2m} h^{ac}(q)p_a p_b h^{bc}(q)+V(q)+\Delta V_{Prod}(q),
  \tag\NUM.\num$$
with the well-defined quantum correction
\plus
$$\Delta V_{Prod}={\hbar^2\over8m}\bigg[ 4h^{ac}h^{bc}_{\ \ ,ab}
  +2h^{ac}h^{bc}{h_{,ab}\over h}
  +2h^{ac}\left(h^{bc}_{\ \ ,b}{h_{,a}\over h}
  +h^{bc}_{\ \ ,a}{h_{,b}\over h}\right)
  -h^{ac}h^{bc}{h_{,a}h_{,b}\over h^2}\bigg].
  \tag\NUM.\num$$
On $M_y$ the corresponding $\Delta V_{Prod}(y)$ is given by
\plus
$$\Delta V_{Prod}(y)={\hbar^2\over8m}{1\over f_a^2}
  \bigg[\bigg({f_{b,a}\over f_b}\bigg)^2-4{f_{a,aa}\over f_a}
  +4{f_{a,a}\over f_a}\bigg(2{f_{a,a}\over f_a}
  -{f_{b,a}\over f_b}\bigg)
  +2\bigg({f_{b,a}\over f_b}\bigg)_{\!,a}\bigg].
  \tag\NUM.\num$$
The expressions (\NUM.7) and (\NUM.8) look somewhat circumstantial,
so we display a special case and the connection to the quantum
correction $\Delta V_{Weyl}$ which corresponds to the Weyl-ordering
prescription.
\item{1)} Let us assume that $\Lambda_{ab}$ is proportional to the
         unit tensor, i.e. $\Lambda_{ab}=f^2\delta_{ab}$.
         Then $\Delta V_{Prod}(y)$ simplifies into
\plus
$$\Delta V_{Prod}(y)=\hbar^2{D-2\over8m}
         {(4-D)f_{,a}^2+2f\cdot f_{,aa}\over f^4}.
  \tag\NUM.\num$$
     This implies: Assume that the metric has or can be transformed into
     the special form $\Lambda_{ab}=f^2\delta_{ab}$. If the dimension
     of the space is $D=2$, then the quantum correction $\Delta
     V_{Prod}$ vanishes.
\item{2)}
     A comparison between (\NUM.7) and (2.3) gives the connection with
     the quantum correction corresponding to the Weyl-ordering
     prescription:
\plus
$$\Delta V_{Prod}=\Delta V_{Weyl}+{\hbar^2\over8m}
          \left( 2h^{ac}h^{bc}_{\ \ ,ab}-h^{ac}_{\ \ ,a}h^{bc}_{\ \ ,b}
                  -h^{ac}_{\ \ ,b}h^{bc}_{\ \ ,a}\right).
  \tag\NUM.\num$$
In the case of equation (\NUM.2) this yields:
\plus
$$\Delta V_{Prod}(y)=\Delta V_{Weyl}
   +{\hbar^2\over4m}{f_{a,a}^2-f_a f_{a,aa}\over f_a^4}
  \tag\NUM.\num$$
These equations often simplify practical applications.

Next we have to consider the short-time matrix element
$<q''\vert\e^{-\i TH}\vert q'>$
in order to derive the path integral formulation
corresponding to our ordering prescription (\NUM.6).
\newline
We proceed similar as in the previous section.
We consider the short-time approximation to the matrix element
$[\epsilon=T/N,\ g^{(j)}=g(q^{(j)})]$:
\plus
$$\multline
  <q^{(j)}\vert\e^{-\i\epsilon H/\hbar}\vert q^{(j-1)}>\simeq
        <q^{(j)}\vert 1-\hbox{$\i\epsilon$} H/\hbar\vert q^{(j-1)}>
  \hfill\\   \qquad
  ={[g^{(j)}\,g^{(j-1)}]^{-{1\over4}}\over(2\pi\hbar)^D}
   \int \e^{i p\Delta q^{(j)}}dp
  \hfill\\     \hfill
   -{\i\epsilon\hbar\over2m}
   <q^{(j)}\vert h^{ac}p_a p_b h^{bc}\vert q^{(j-1)}>
   -{\i\epsilon\over\hbar}
   <q^{(j)}\vert V+\Delta V_{Prod}\vert q^{(j-1)}>.
   \endmultline
  \tag\NUM.\num$$
Therefore we get for the short-time matrix element ($\epsilon\ll1$):
\plus
$$\multline
  <q^{(j)}\vert\e^{-\i\epsilon H/\hbar}\vert q^{(j-1)}>\simeq
   {[g^{(j)}\,g^{(j-1)}
   ]^{-{1\over4}}\over(2\pi\hbar)^D}\int dp
  \hfill\\    \hfill \times
  \exp\bigg[\ih p\Delta q^{(j)}-{\i\epsilon\over2m\hbar}
  h^{ac}(q^{(j)})h^{bc}(q^{(j-1)})p_a p_b
  -{\i\epsilon\over\hbar}V(q^{(j)})
  -{\i\epsilon\over\hbar}\Delta V_{Prod}(q^{(j)})\bigg].
  \\  \ \endmultline
  \tag\NUM.\num$$
The choice of the ``post point'' $q^{(j)}$ in the potential terms
is not unique.
A ``pre-point'', ``mid-point'' or a ``product-form''-expansion is also
legitimate.
However, changing from one to another formulation does not alter the
path integral, because differences in the potential terms are of
$O(\epsilon)$, i.e.\ of $O(\epsilon^2)$ in the short-time Feynman
kernel and therefore do not contribute. The Trotter formula $\e^{-\i
T(A+B)}:=s-\lim_{N\to\infty}(\e^{-\i TA/N} \e^{-\i TB/N})^N$ states
that all approximations in equations (\NUM.12) to (\NUM.13) are valid
in the limit $N\to\infty$ and we get for the {\bf Hamiltonian path
integral in the ``product form''-definition}
[$h_{ab}^{(j)}=h_{ab}(q^{(j)})$]:
\plus
$$\multline
  K(q'',q';T)=[g(q')g(q'')]^{-{1\over4}} \lim_{N\to\infty}
  \left( \prod_{j=1}^{N-1}\int dq^{(j)}\times\prod_{j=1}^N
                  {dp^{(j)}\over(2\pi\hbar)^D}\right)
  \hfill\\  \hfill\times
  \exp\left\{\ih\sum_{j=1}^N\bigg
  [p\Delta q^{(j)}-{\epsilon\over2m}
  h^{ac,(j)}h^{bc,(j-1)}p_a^{(j)} p_b^{(j)}
  -\epsilon V(q^{(j)})-\epsilon\Delta V_{Prod}(q^{(j)})
  \bigg]\right\}.
  \\  \quad\endmultline
  \tag\NUM.\num$$
Performing the momentum integrations we get for the {\bf Lagrangian
path integral in the ``product form''-definition}:
($PF$={\bf P}roduct-{\bf F}orm)
\plus
$$\multline
  K(q'',q';T)
  \\   \qquad
  =\int\limits_{q(t')=q'}^{q(t'')=q''}\sqrt{g}\,\CD_{PF} q(t)
  \exp\bigg\{\ih\int_{t'}^{t''}\bigg[{m\over2}
  h_{ac}h_{bc}\dot q^a\dot q^b
  -V(q)-\Delta V_{Prod}(q)\bigg]dt\bigg\}
  \hfill\\   \qquad
  :=\lim_{N\to\infty}\Norm^{ND\over2}
  \prod_{j=1}^{N-1}\int\sqrt{g(q^{(j)})}\,dq^{(j)}
  \hfill\\   \hfill\times
  \exp\left\{\ih\sum_{j=1}^N\bigg[{m\over2\epsilon}
  h_{ac}^{(j)}h_{bc}^{(j-1)}\Delta q^{a,(j)}\Delta q^{b,(j)}
  -\epsilon V(q^{(j)})
  -\epsilon\Delta V_{Prod}(q^{(j)})\bigg]\right\}.
  \endmultline
  \tag\NUM.\num$$
In the last step we have to check that the Schr\"odinger equation (1.1)
can be deduced from the short-time kernel of equation (\NUM.15).
Because one can always transform from the  q-coordinates to the
y-coordinates, which is a linear orthogonal transformation and thus
does not produce any quantum correction in the path integral (\NUM.14)
defined on $M_y$, we shall use in the following the representation of
equation (\NUM.2). Looking at the Hamiltonian formulation of the path
integral (\NUM.14) we see that the unitary transformation $y=C^{-1}q$
changes the metric term $p_ah^{ac}h^{cb}p_b\to p_a\Lambda^{ab}p_b$,
i.e.\ in the correct feature, the measure $dq^{(j)}dp^{(j)}\to
dy^{(j)}dp^{(j)}$ due to the Jacobean $J=1$ so that the Feynman kernel
is transformed just in the right manner.

We restrict ourselves to the proof that the short-time kernels of
equations (2.25) and (\NUM.15) are equivalent, i.e.\ we have to show
($\bar y=(y''+y')/2$):
\plus
$$\multline
  [g(y')g(y'')]^{-{1\over4}}\sqrt{g(\bar y)}
  \exp\bigg[{\i m\over2\epsilon\hbar}
  \Lambda_{ab}(\bar y)\Delta y^a\Delta y^b
  -{\i\epsilon\over\hbar}V(\bar y)
  -\i\epsilon\hbar\Delta V_{W}(\bar y)\bigg]
  \hfill\\     \hfill\dot=
  \exp\bigg\{{\i m\over2\epsilon\hbar} f_a(y')f_a(y'')
  \Delta^2y^a-{\i\epsilon\over\hbar} V(y'')
  -\i\epsilon\hbar\Delta V_{Prod}(y'')\big]\bigg\}.
  \endmultline
  \tag\NUM.\num$$
Clearly, $\e^{-\i\epsilon V(\bar y)/\hbar}\dot
=\e^{-\i\epsilon V(y'')/\hbar}$ for the potential term. It suffices to
show that a Taylor expansion of the $g$ and the kinetic energy terms on
the left-hand side of equation (\NUM.15) yield an additional potential
$\Delta\tilde V$ given by
\plus
$$\Delta\tilde V(y)=\Delta V_{Prod}(y)-\Delta V_{Weyl}(y)
  ={\hbar^2\over4m}{f_{a,a}^2(y)-f_a(y) f_{a,aa}(y)\over f_a^4(y)}.
  \tag\NUM.\num$$
We consider the $g$-terms on the left-hand side of equation (\NUM.16)
and expand them in a Taylor-series around $y'$. This gives
($\xi_a=(y_a''-y_a')$, $f_a(y')\equiv f_a$):
\plus
$$[g(y')g(y'')]^{-{1\over4}}\sqrt{g(\bar y)}
  \simeq\bigg[1-{1\over8}
  {f_c f_{c,ab}-f_{c,a}f_{c,b}\over f_c^2}\xi^a\xi^b\bigg].
  \tag\NUM.\num$$
Exploiting the path integral identity (2.30) we get by exponentiating
the $O(\epsilon)$-terms,
\plus
$$[g(y') g(y'')]^{-{1\over4}}\sqrt{g(\bar y)}
  \simeq\exp\bigg[-{\i\epsilon\hbar\over8m}
  {f_a f_{a,bb}-f_{a,b}^2\over f_a^2 f_b^2}\bigg].
  \tag\NUM.\num$$
Repeating the same procedure for the exponential term gives:
\plus
$$\multline
 \exp\bigg[
  {\i m\over2\epsilon\hbar}\Lambda_{ab}(\bar y)\xi^a\xi^b\bigg]
  \simeq\exp\bigg[{\i m\over2\epsilon\hbar}
                  f_a(y')f_a(y'')\xi^a\xi^a\bigg]
  \hfill\\    \hfill\times
  \Bigg[1-{\i\epsilon\over8m}\left(f_c f_{c,ab}-f_{c,a}f_{c,b}\right)
  \xi^a\xi^b\xi^c\xi^c\Bigg].
  \endmultline
  \tag\NUM.\num$$
We use the identity (2.30) to get
\plus
$$\multline
 \exp\bigg[
  {\i m\over2\epsilon\hbar}\Lambda_{ab}(\bar y)\xi^a\xi^b\bigg]
  \simeq\exp\bigg[{\i m\over2\epsilon\hbar}
                   f_a(y')f_a(y'')\xi^a\xi^a\bigg]
   \hfill\\    \hfill\times
  \exp\Bigg[{\i\epsilon\hbar\over8m}
  {f_a f_{a,bb}-f_{a,b}^2\over f_a^2 f_b^2}
  +{\i\epsilon\hbar\over4m}{f_a f_{a,aa}-f_{a,a}^2\over f_a^4}\Bigg].
  \endmultline
  \tag\NUM.\num$$
Combining equations (\NUM.20) and (\NUM.21) yields the additional
potential $\Delta\tilde V$ and equation (\NUM.16) is proven.
Thus we conclude that the path integral (\NUM.15) is well-defined and is
the correct path integral corresponding to the Schr\"odinger equation
(1.1).

A combination of lattice prescriptions were in the metric $g_{ab}$ in
the kinetic term in the exponential and in the determinant $g$
different lattices are used exist also in the literature.
Let us note the result of McLaughlin and Schulman [\MCLS]:
\plus
$$\multline
  K(q'',q';T)=\lim_{N\to\infty}\Norm^{ND/2}
  \prod_{j=1}^{N-1}\int\sqrt{g(q^{(j)}}\,dq^{(j)}
  \\   \times
  \exp\left\{\ih\sum_{j=1}^N\bigg[{m\over2\epsilon}
   g_{ab}(\bar q^{(j)})\Delta q^{(j)\,a}q^{(j)\,b}
   -\epsilon U(\bar q^{(j)})\bigg]\right\}
  \endmultline
  \tag\NUM.\num$$
with the effective potential
\plus
$$U(q)=V(q)-{e\over c}A(q)\cdot \dot q-F(q),
  \tag\NUM.\num$$
$A(q)$ an additional vector potential term and $F(q)$ given by
\plus
$$\aligned
  F(q)&=-\hbar^2F_{abcd}\Big(g^{ab}g^{cd}+g^{ac}g^{bd}+g^{ad}g^{bc}\Big)
  \\
  F_{abcd}&={1\over48m}\Big(g_{ab,cd}-2g^{ef}
     \Gamma_{abe}\Gamma_{cdf}\Big)
  \endaligned
  \tag\NUM.\num$$
[$\Gamma_{abc}=\half(g_{ab,c}+g_{ac,b}-g_{bc,a})$].
According to D'Olivio and Torres [\DOTO] the effective potential can be
 rewritten as
\plus
$$U(q)=V(q)-{e\over c}A(q)\cdot\dot q+{\hbar^2\over8m}
  \Big[R+g^{ab}(\Gamma^c_{ad}\Gamma^d_{bc}
  +\partial_b\Gamma^c_{ac})\Big].
  \tag\NUM.\num$$
which is up the derivative term the result of Mizrahi [\MIZa].

\bigskip\bigskip
\glno=0               
\advance\chapno by 1  

\section{Space-Time Transformation}
Let us consider a one-dimensional path integral
\plus
$$K(x'',x';T)=
  \int\limits_{x(t')=x'}^{x(t'')=x''}\CD x(t)
  \exp\left[\ih\int_{t'}^{t''}\bigg({m\over2}\dot x^2
  -V(x)\bigg)dt\right].
  \tag\NUM.\num$$
It is now assumed that the potential $V$ is so complicated that a
direct evaluation of the path integral is not possible. We want to
describe a method how a transformed path integral can be achieved to
calculate $K(T)$ or $G(E)$, respectively, nevertheless. This method is
called ``space-time'' transformation. This technique was originally
developed by Duru and Kleinert [\DKa,\DKb]. It was further evolved by
Steiner [\STEa], Pak and S\"okmen [\PAKSc], Inomata [\INOd], Kleinert
[\KLEh, \KLEm] and Grosche and Steiner [\GRSb]. In the rigorous lattice
derivation of the following formul\ae\ we follow ourselves [\GRSb] and
Pak and S\"okmen [\PAKSc].
\newline
Let us start by considering the Legendre-transformed of the general
one-dimensional Hamiltonian:
\plus
$$H_E=-{\hbar^2\over2m}\bigg({d^2\over dx^2}
  +\Gamma(x){\d\over\d x}\bigg)+V(x)-E
  \tag\NUM.\num$$
which is hermitean with respect to the scalar product
\plus
$$(f_1,f_2)=\int f_1(x) f_2^*(x)J(x)dx,\qquad
  J(x)=\e^{\int^x\Gamma(x')dx'}.
  \tag\NUM.\num$$
Introducing the momentum operator
\plus
$$p_x={\hbar\over\i}\bigg({\d\over\d x}+\half\Gamma(x)\bigg),\qquad
  \Gamma(x)={d\ln J(x)\over dx}
  \tag\NUM.\num$$
$H_E$ can be rewritten as
\plus
$$H_E={p_x^2\over2m}+V(x)
  +{\hbar^2\over8m}[\Gamma^2(x)+2\Gamma'(x)]-E
  \tag\NUM.\num$$
with the corresponding path integral
\plus
$$K_E(x'',x';T)=\e^{\i TE/\hbar}K(x'',x';T),
  \tag\NUM.\num$$
where $K(T)$ denotes the path integral of equation (\NUM.1).
\newline
Let us consider the space-time transformation
\plus
$$x=F(q), \qquad  dt=f(x)ds
  \tag\NUM.\num$$
with new coordinate $q$ and new ``time'' $s$.
Let $G(q)=\Gamma[F(q)]$, then
\plus
$$\hat H_E=-{\hbar^2\over2m}{1\over {F'}^2(q)}
  \bigg[{d^2\over dq^2}+\bigg(G(q)F'(q)-{F''(q)\over F'(q)}\bigg)
  {\d\over\d q}\bigg]+V[F(q)]-E.
  \tag\NUM.\num$$
With the constraint $f[F(q)]={F'}^2(q)$ we get for the new
Hamiltonian $\tilde H=f\hat H_E$:
\plus
$$\aligned
  \tilde H&=-{\hbar^2\over2m}
  \bigg[{d^2\over dq^2}+\bigg(G(q)F'(q)-{F''(q)\over F'(q)}\bigg)
  {\d\over\d q}\bigg]+f[F(q)][V(F(q))-E]
  \\
  &=-{\hbar^2\over2m}\bigg[{d^2\over dq^2}
  +\tilde\Gamma(q){\d\over\d q}\bigg]+f[F(q)][V(F(q))-E],
  \endaligned
  \tag\NUM.\num$$
where $\tilde\Gamma(q)=G(q)F'(q)-F''(q)/F'(q)$. The corresponding
measure in the scalar product and the hermitean momentum are
\plus
$$p_q={\hbar\over\i}\bigg[
  {\d\over\d q}+\half\tilde\Gamma(q)\bigg],\qquad
  J(q)=\sqrt{g(q)}=\e^{\int^q\tilde\Gamma(q')dq'}.
  \tag\NUM.\num$$
The Hamiltonian $\tilde H$ expressed in the position- $q$ and momentum
operator $p_q$ is
\plus
$$\tilde H={p_q\over2m}+f[F(q)][V(F(q))_E]+\Delta V(q)
  \tag\NUM.\num$$
with the well-defined quantum potential
\plus
$$\Delta V(q)={\hbar^2\over8m}\left[
  3\bigg({F''(q)\over F'(q)}\bigg)^2-2{F'''(q)\over F'(q)}
  +\big(G(q)F'(q)\big)^2+2G'(q)F'(q)\right].
  \tag\NUM.\num$$
Note that for $G\equiv0$, $\Delta V$ is proportional to the Schwartz
derivative of the transformation $F$.
The path integral corresponding to the Hamiltonian $\tilde H$ is
\plus
$$\multline
  \tilde K(q'',q';s'')=
  \int\limits_{q(t')=q'}^{q(t'')=q''}\CD q(s)
  \\   \times
  \exp\left\{\ih\int_0^{s''}\bigg[{m\over2}\dot q^2
       -f[F(q)][V(F(q))-E]-\Delta V(q)\bigg]ds\right\}.
  \endmultline
  \tag\NUM.\num$$
As it is easily checked it is possible to derive from the
short time kernel of equation (\NUM.13) via the time evolution equation
\plus
$$\tilde\Psi(q'';s)=\int\tilde K(q'',q';s'')\tilde\Psi(q';s)dq'
  \tag\NUM.\num$$
the time dependent Schr\"odinger equation
\plus
$$H\tilde\Psi(q;s)=\i\hbar{\partial\over\partial s}\tilde\Psi(q;s).
  \tag\NUM.\num$$
The crucial point is now, of course, the rigorous lattice derivation
of $\tilde K(s'')$ and the relation between $\tilde K(s'')$ and $K(T)$.
It turns out that $\tilde K$ is given in terms of $K(T)$ by the
equations
\plus
$$\gathered
  K(x'',x';T)
  ={1\over 2\pi\i\hbar}\int_{-\infty}^\infty \e^{-\i TE/\hbar}
  G(x'',x';E)dE
  \\
  G(x'',x';E)=\i[f(x')f(x'')]^{1\over4}\int_0^\infty
              \tilde K(q'',q';s'')ds''.
  \endgathered
  \tag\NUM.\num$$
This we want to justify!

Let us consider the path integral $K(T)$ in its lattice definition
\plus
$$\multline
  K(x'',x';T)=\lim_{N\to\infty}\Norm^{ND/2}
  \prod_{j=1}^{N-1}\int dx^{(j)}
  \\  \times
  \exp\left\{\ih\sum_{j=1}^N\bigg[{m\over2\epsilon}
  (x^{(j)}-x^{(j-1)})^2    -\epsilon V(\bar x^{(j)})
  \bigg]\right\} .
  \endmultline
  \tag\NUM.\num$$
We consider a $D$-dimensional path integral. To transform the
coordinates $x$ into the coordinates $q$ we use the midpoint
prescription expansion method. It reads that one has to expand any
dynamical quantity in question $F(q)$ which is defined on the points
$q^{(j)}$ and $q^{(j-1)}$ of the $j^{th}$-interval in the lattice
version about the midpoints $\bqj=\half(q^{(j)}+q^{(j-1)})$ maintaining
terms up to order $(q^{(j)}-q^{(j-1)})^3$. This gives:
\plus
$$\allowdisplaybreaks\align
  \Delta F(q^{(j)})&\equiv F(q^{(j)})-F(q^{(j-1)})
  =F\bigg(\bqj+{\Delta q^{(j)}\over2}\bigg)-
   F\bigg(\bqj-{\Delta q^{(j)}\over2}\bigg)
  \\
  &=\Delta q^{(j)}_m{\partial F(q)\over\partial q_m}\bigg\vert_{q=\bqj}
  +{1\over24}\Delta q^{(j)}_m \Delta q^{(j)}_n \Delta q^{(j)}_k
  {\partial^3 F(q)\over\partial q_m \partial q_n \partial q_k}
                                             \bigg\vert_{q=\bqj}+\dots.
  \tag\NUM.\num
  \endalign$$
Introducing the abbreviations
\plus
$$F^{(j)}_{,m}\equiv{\partial F(q)\over\partial q_m}\bigg\vert_{q=\bqj}
  ,\qquad
  F^{(j)}_{,mnk}\equiv{\partial^3 F(q)\over\partial q_m \partial q_n
                   \partial q_k}\bigg\vert_{q=\bqj}
  \tag\NUM.\num$$
we get therefore
\plus
$$\allowdisplaybreaks\align
  \Delta^2F^{(j)}&=\bigg[\Delta q^{(j)}_m F^{(j)}_{,m}
     +{1\over24}\Delta q^{(j)}_m\Delta q^{(j)}_n\Delta^{(j)}_k
      F^{(j)}_{,mnk}\bigg]^2
  \\
  &\simeq\Delta q^{(j)}_m\Delta q^{(j)}_n F^{(j)}_{,m}F^{(j)}_{,n}
  +{1\over12}
    \Delta q^{(j)}_m\Delta q^{(j)}_n\Delta q^{(j)}_k\Delta q^{(j)}_l
    F^{(j)}_{,m}F^{(j)}_{,nkl}
   \\
   &\hbox{$\dot=$}
   \Delta q^{(j)}_m\Delta q^{(j)}_n F^{(j)}_{,m}F^{(j)}_{,n}
   +{1\over12}\bigg({\i\epsilon\hbar\over m}\bigg)^2
    F^{(j)}_{,m}F^{(j)}_{,nkl}F_{mnkl}^{-1}(q^{(j)})
  \tag\NUM.\num
  \endalign$$
according to equation (2.30) with the abbreviation
\plus
$$\multline
  F_{mnkl}^{-1}(q^{(j)})=
  (F^{(j)}_{,m}F^{(j)}_{,n})^{-1}
       (F^{(j)}_{,k}F^{(j)}_{,l})^{-1}
  \\
      +(F^{(j)}_{,m}F^{(j)}_{,k})^{-1}(F^{(j)}_{,l}F^{(j)}_{,n})^{-1}
      +(F^{(j)}_{,m}F^{(j)}_{,l})^{-1}(F^{(j)}_{,k}F^{(j)}_{,n})^{-1}.
  \endmultline
  \tag\NUM.\num$$
Furthermore we have to transform the measure. Because
\plus
$$\prod_{j=1}^{N-1}dx^{(j)}=\prod_{j=1}^{N-1}
  \bigg\vert
  {\partial F(q^{(j)})\over\partial q^{(j)}}\bigg\vert dq^{(j)}
  \equiv\prod_{j=1}^{N-1} F_{;q}(q^{(j)})dq^{(j)}
  \tag\NUM.\num$$
does not have a symmetric look, we must expand about $\bqj$. We get:
\plus
$$\multline
  \prod_{j=1}^{N-1}dx^{(j)}
  =\big[F_{;q}(q')F_{;q}(q'')\big]^{-1/2}
  \prod_{j=1}^{N-1}dq^{(j)}
  \prod_{j=1}^N\big[F_{;q}(q^{(j)})F_{;q}(q^{(j-1})
  \big]^\half
  \\  \qquad\simeq
  \big[F_{;q}(q')F_{;q}(q'')\big]^{-1/2}
  \prod_{j=1}^{N-1}dq^{(j)}\prod_{j=1}^N F_{;q}(\bqj)
  \hfill\\  \qquad\qquad\times
  \bigg[1-{1\over8}\bigg({F_{;q,m}(\bqj)F_{;q,n}(\bqj)
        \over F_{;q}^2(\bqj) }
         -{F_{;q,mn}(\bqj)\over F_{;q}(\bqj)}\bigg)
         \Delta q^{(j)}_m\Delta q^{(j)}_n\bigg]
  \\  \qquad\dot=
  \big[F_{;q}(q')F_{;q}(q'')\big]^{-\half}
  \prod_{j=1}^{N-1}dq^{(j)}\prod_{j=1}^N F_{;q}(\bqj)
  \hfill\\  \qquad\qquad\times
  \exp\bigg[-{\i\epsilon\hbar\over8m}
  (F^{(j)}_{,m}F^{(j)}_{,n})^{-1}
      \bigg({F_{;q,m}(\bqj)F_{;q,n}(\bqj)\over F_{;q}^2(\bqj)}
         -{F_{;q,mn}(\bqj)\over F_{;q}(\bqj)}\bigg)\bigg]
  \hfill\endmultline
  \tag\NUM.\num$$
Thus we have the coordinate transformed path integral
\plus
$$\multline
  \hat K(q'',q';T)
  \\   \qquad
  =[F_{;q}(q')F_{;q}(q'')]^{-1/2}
  \lim_{N\to\infty}\Norm^{ND/2}\prod_{j=1}^{N-1}\int dq^{(j)}
  \hfill\\   \qquad\qquad\times
  \prod_{j=1}^N F_{;q}(\bqj)
  \exp\bigg\{\ih\bigg[{m\over2\epsilon}
                         \Delta q_m^{(j)}\Delta q_n^{(j)}
           F_{,m}(\bqj)F_{,n}(\bqj)-\epsilon V(\bqj)
  \hfill\\  \qquad\qquad\qquad
  -{\epsilon\hbar^2\over8m}
   (F^{(j)}_{,m}F^{(j)}_{,n})^{-1}
   \bigg({F_{;q,m}(\bqj)F_{;q,n}(\bqj)\over F_{;q}^2(\bqj)}
         -{F_{;q,mn}(\bqj)\over F_{;q}(\bqj)}\bigg)
  \hfill\\  \qquad\qquad\qquad
  -{\epsilon\hbar^2\over8m}
   F^{(j)}_{,m}F^{(j)}_{,nkl}
  F_{mnkl}^{-1}(q^{(j)})\bigg)\bigg]\bigg\}
  \hfill\endmultline
  \tag\NUM.\num$$
This path integral has the canonical form (2.24).
In order to see this note that the factor $F_{,m}(\bqj)F_{,n}(\bqj)$
in the dynamical term can be interpreted as a metric $g_{nm}$
appearing in an effective Lagrangian.
Thus we can rewrite the transformed Lagrangian in terms of this
metric (Gervais and Jevicki [\GJ]):
Firstly, we have $F_{;q}(q)=\sqrt{\det(g_{nm}(q))}\,\equiv\sqrt{g(q)}$.
In the expansion of the determinant about the midpoints we then get
\plus
$$\multline
  [F_{;q}(q^{(j)})F_{;q}(q^{(j-1})\big]^{1/2}
  \\
  \simeq\sqrt{q(\bqj)}\bigg\{1+{1\over16}\bigg[
  g^{mn}(\bqj)g_{mn,kl}+{g^{mn}}_{,k}(\bqj)g_{mn,l}(\bqj)\bigg]
  \Delta q_k^{(j)}\Delta q_k^{(j)}\bigg\}.
  \endmultline
  \tag\NUM.\num$$
Using now successively the identities
\plus
$${g_{,l}(q)\over g(q)}=g^{mn}(q)g_{mn,l}(q),\qquad
  F^m_{,kl}(q)=\Gamma_{kl}^m(q)F^m_{,l}(q),\qquad
  \Gamma^m_{ml,k}(q)=\half\bigg({g_{,l}(q)\over g(q)}\bigg)_{,k}
  \tag\NUM.\num$$
we can cast the quantum potential into the form
\plus
$$\Delta V(q)={\hbar^2\over8m}g^{mn}(q)
  \Gamma^k_{lm}(q)\Gamma^l_{km}(q)
  \tag\NUM.\num$$
which is just the Weyl-ordering quantum potential without curvature
term.  In the present case the curvature vanishes since we started in
flat Euclidean space which remains flat during the transformation.

In particular equation (\NUM.24) takes in the one-dimensional case the
form [$F'\equiv dF/dq$]:
\plus
$$\multline
  \hat K(q'',q';T)
  =[F'(q')F'(q'')]^{-\half}
  \lim_{N\to\infty}\Norm^{N/2}\prod_{j=1}^{N-1}\int dq^{(j)}
  \\  \qquad\times
  \prod_{j=1}^N F'(\bqj)
  \exp\left\{\ih\left[{m\over2\epsilon\hbar}
                                     \Delta q_m^{(j)}\Delta q_n^{(j)}
   {F'}^2(\bqj)-\epsilon V(\bqj)
  -{\epsilon\hbar^2\over8m}{{F'}^2(\bqj)\over{F'}^4(\bqj)}
  \right]\right\}
  \hfill\\  \ \endmultline
  \tag\NUM.\num$$
which is the same result as taking for $D=1$ in $\Delta V_{Weyl}$
the metric tensor $g_{ab}={F'}^2$.
The infinitesimal generator of the corresponding time-evolution
operator is $\hat H_E$.
To proceed further we perform a time transformation according to
\plus
$$dt=f(s)ds,\qquad  s'=s(t')=0,\qquad s''=s(t'')
  \tag\NUM.\num$$
and we make the choice $f={F'}^2$.
Translating this transformation into the discrete notation
requires special care. In accordance with the midpoint prescription we
must first symmetrize over the interval $(j,j-1)$ to prefer neither of
the endpoints over the other, i.e.
\plus
$$\Delta t^{(j)}=\epsilon=\Delta s^{(j)} F'(q^{(j)})F'(q^{(j-1)}),
  \qquad
  \Delta s^{(j)}=s^{(j)}-s^{(j-1)}\equiv\delta^{(j)}.
  \tag\NUM.\num$$
Expanding about midpoints yields
\plus
$$\epsilon\simeq\delta^{(j)}{F'}^2(\bqj)
  \left\{1+{\Delta^2 q^{(j)}\over4}
  \left[{F'''(\bqj)\over F'(\bqj)}
       -\bigg({F''(\bqj)\over F'(\bqj)}\bigg)^2\right]\right\}.
  \tag\NUM.\num$$
Insertion yields for each j$^{th}$-term
[identify $V(x^{(j)})\to V[F(q^{(j)})]\equiv V(q^{(j)})$]
\plus
$$\allowdisplaybreaks\align
  &\Norm^{N\over2}\prod_{j=1}^{N-1}dx^{(j)}\times\prod_{j=1}^N
  \exp\bigg\{\ih\bigg[{m\over2\epsilon}\Delta^2x^{(j)}
    -\epsilon V(x^{(j)})+\epsilon E\bigg]\bigg\}
  \\   &
  =\prod_{j=1}^{N-1}F'(q^{(j)})dq^{(j)}\prod_{j=1}^N\Norm^\half
  \\   &\qquad\times
  \exp\bigg(\ih\bigg\{{m\over2\epsilon}\bigg[
  {F'}^2(\bar q^{(j)})\Delta^2q^{(j)}
  +{1\over12}F'(\bar q^{(j)})F'''(\bar q^{(j)})
  \Delta^4q^{(j)}\bigg]
  -\epsilon V(x^{(j)})+\epsilon E\bigg\}\bigg)
  \\   &
  =\Big[F'(q')F'(q'')\Big]^{-\half}\prod_{j=1}^{N-1}dq^{(j)}
  \prod_{j=1}^N\bigg({m\over2\pi\i\delta^{(j)}\hbar}\bigg)^\half
  \\   &\qquad\times
  \prod_{j=1}^N
  \exp\left[\!\!\left[\ih\left\{{m\over2\delta^{(j)}}
  \bigg[
  1+{\Delta^2 q^{(j)}\over4}\bigg({F'''(\bqj)\over F'(\bqj)}
       -\bigg({F''(\bqj)\over F'(\bqj)}\bigg)^2\bigg)\bigg]^{-1}
  \right.\right.\right.\\   &\qquad\times\left.\left.\left.
  \vphantom{\bigg({F''(\bqj)\over F'(\bqj)}\bigg)^2\bigg)
                                               \bigg]^{-1}}
  \bigg[ \Delta^2q^{(j)}+{\Delta^4q^{(j)}\over12}
  {F'''(\bar q^{(j)})\over F'(\bar q^{(j)})}\bigg]
  -\delta^{(j)}{F'}^2(\bar q^{(j)})\Big[V(q^{(j)})
  -E\Big]\right\}\right]\!\!\right].
  \\  &
  \dot =\Big[F'(q')F'(q'')\Big]^{-\half}
  \bigg({m\over2\pi\i\delta\hbar}\bigg)^{N\over2}
  \prod_{j=1}^{N-1}dq^{(j)}
  \\   &\qquad\times
  \exp\left[\!\!\left[\ih\left\{{m\over2\delta}\Delta^2q^{(j)}
  -\delta^{(j)}{F'}^2(q^{(j)})\Big[V(q^{(j)})-E\Big]
  \vphantom{\bigg({F''(q^{(j)})\over F'(q^{(j)})}\bigg)^2}
  \right.\right.\right.\\
  &\qquad\qquad\qquad\qquad\left.\left.\left.
  -{i\delta\hbar\over8m}
  \left[3\bigg({F''(q^{(j)})\over F'(q^{(j)})}\bigg)^2
  -2{F'''(q^{(j)})\over F'(q^{(j)})}\right]\right\}\right]\!\!\right].
  \tag\NUM.\num
  \endalign$$
Let us assume that the constraint
\plus
$$\int_0^{s''}ds f[F(q(s))]=T
  \tag\NUM.\num$$
has for all admissible path a unique solution $s''>0$. Of course, since
$T$ is fixed, the ``time'' $s''$ will be path dependent. To incorporate
the constraint let us consider the Green function $G(E)$:
\plus
$$\gathered
  G(x'',x';E)=\i\int_0^\infty dT\, \e^{\i TE/\hbar}K(x'',x';T)
  \\
  K(x'',x';T)={1\over2\pi\i\hbar}\int_{-\infty}^\infty dE\,
              \e^{-\i TE/\hbar}G(x'',x';E).
  \endgathered
  \tag\NUM.\num$$
This incorporation is far from being trivial, in particular the fact
that the new time-slicing $\delta^{(j)}$ is treated in the same way as
the old time-slicing $\epsilon$, i.e.\ $\delta^{(j)}\equiv\delta$ is
considered as being fixed, but can however be justified in a more
comprehensive way, see e.g.\ [\CAST, \FLM, \INOd]. We now observe that
$G(E)$ can be written e.g.\ in the following ways [\KLEh]:
\plus
$$\allowdisplaybreaks\align
  G(x'',x';E)
  &=f(x'')\int_0^\infty<x''\vert\e^{-\i s''f(x)(H-E)/\hbar}\vert x'>ds''
  \tag\NUM.\num a\\
  &=f(x')\int_0^\infty<x''\vert\e^{-\i s''(H-E)f(x)/\hbar}\vert x'>ds''
  \tag\NUM.\num b\\
  &=\sqrt{f(x')f(x'')}\int_0^\infty<x''\vert
   \e^{-\i s''\sqrt{f(x)}(H-E)\sqrt{f(x)}/\hbar}\vert x'>ds''.
  \tag\NUM.\num c
  \endalign$$
\edef\numBExa{\NUM.\num}%
The last line give for the short-time matrix element in the usual
manner
\hfuzz=10pt
\plus
$$\allowdisplaybreaks\align
  <&x^{(j)}\vert
  \e^{-\i\delta^{(j)}f^\half(x)(H-E)f^\half(x)/\hbar}\vert x^{(j-1)}>
  \\  &\simeq<x^{(j)}\vert 1-\i\delta^{(j)}f^\half(x)
                 (H-E)f^\half(x)/\hbar\vert x^{(j-1)}>
  \\  &=<x^{(j)}\vert x^{(j-1)}>
  -{\i\delta^{(j)}\over\hbar}
      \sqrt{f(x^{(j)})f(x^{(j)})}\,<x^{(j)}\vert H-E\vert x^{(j-1)}>
  \\  &=<F(q^{(j)})\vert F(q^{(j-1)})>
  -{\i\delta^{(j)}\over\hbar}
      \sqrt{f[F(q^{(j)})]f[F(q^{(j)})]}\,
  <F(q^{(j)})\vert\hat H_E\vert F(q^{(j-1)})>
  \\  &=<q^{(j)}\vert q^{(j-1)}>-{\i\delta\over\hbar}
  <q^{(j)}\vert\tilde H\vert q^{(j-1)}>
  \\  &={1\over2\pi}\int dp_q
  \exp\bigg[\ih p_q\Delta q^{(j)}-{\i\delta\over2m\hbar}p_q^2
      -{\i\delta\over\hbar}{F'}^2(\bqj)(V(\bqj)-E)
  -{\i\delta\over\hbar}\Delta V(\bqj)\bigg]
  \\  &=\bigg({m\over2\pi\i\delta\hbar}\bigg)^\half
  \exp\bigg[{\i m\over2\delta\hbar}\Delta^2 q^{(j)}
      -{\delta\over\hbar}{F'}^2(\bqj)(V(\bqj)-E)
  -{\delta\over\hbar}\Delta V(\bqj)\bigg],
  \tag\NUM.\num
  \endalign$$
\hfuzz=3pt
where the well-defined quantum potential $\Delta V$ is given by
\plus
$$\Delta V(q)={\hbar^2\over8m}\left[3{F'''(q)\over F'(q)}
          -2\bigg({F''(q)\over F'(q)}\bigg)^2\right].
  \tag\NUM.\num$$
Note that this quantum potential has the form of a Schwartz-derivative.
This procedure justifies our symmetrization prescription of equations
(\NUM.30). Thus we arrive finally at the {\bf transformation formul\ae}:
\plus
$$\gathered
  K(x'',x';T)={1\over2\pi\i\hbar}
  \int_{-\infty}^\infty dE\,\e^{-\i TE/\hbar} G[F(q''),F(q');E]
  \\
  G[F(q''),F(q');E]
  =\i[F'(q'')F'(q')]^{-\half}\int_0^\infty ds''\tilde K(q'',q';s'')
  \endgathered
  \tag\NUM.\num$$
where the path integral $\tilde K(s'')$ is given by
\plus
$$\multline
  \tilde K(q'',q';s'')=\lim_{N\to\infty}
  \bigg({m\over2\pi\i\delta\hbar}\bigg)^\half
  \prod_{j=1}^{N-1}\int dq^{(j)}
  \\  \times
  \exp\left\{\ih\sum_{j=1}^N\bigg[{m\over2\delta}\Delta^2q^{(j)}
   -\delta{F'}^2(\bqj)(V(\bqj)-E)-\delta\Delta V(\bqj)\bigg]\right\}.
  \endmultline
  \tag\NUM.\num$$
This is exactly the path integral (\NUM.13) and thus we have proven it.
Note that only for $D=1$ we have the simple transformation property
\plus
$${\i m\over2\epsilon}\Delta^2x^{(j)}\to
  {\i m\over2\delta}\Delta^2q^{(j)}.
  \tag\NUM.\num$$
For higher dimensional problem this may only possible for one
coordinate kinetic term. Anyway, the requirement for higher dimensional
problems are
\plus
$$F_{;q}=F_{,m}F_{,n}
  \tag\NUM.\num$$
for at least one pair $(m,n)\in(1,\dots,D)$.
\newline
Finally we consider a pure time transformation in a path
integral. We consider equation (\numBExa c), i.e.
\plus
$$G(x'',x';E)=\sqrt{f(x')f(x'')}\int_0^\infty ds''
  <q''\vert\e^{-\i s''\sqrt{f}\,(H-E)\sqrt{f}/\hbar}\vert q'>,
  \tag\NUM.\num$$
where we assume that the Hamiltonian $H$ is product ordered.
Then
\plus
$$\multline
  G(x'',x';E)
  \\   \qquad
  =\sqrt{f(x')f(x'')}\int_0^\infty ds''
  \prod_{j=1}^{N-1}\int dq^{(j)}\prod_{j=1}^N
  <q^{(j)}\vert\e^{-\i\delta\sqrt{f}\,(H-E)\sqrt{f}/\hbar}\vert q^{(j-1)}>
  \hfill\\   \qquad
  ={\sqrt{f(x')f(x'')}\over(g'g'')^{1\over4}}
  \int_0^\infty ds''\lim_{N\to\infty}\prod_{j=1}^{N-1}\int dq^{(j)}
  \hfill\\   \qquad\qquad\times
  \prod_{j=1}^N\int{dp^{(j)}\over(2\pi\hbar)^D}
  \exp\left(\ih\left\{p^{(j)}\Delta q^{(j)}
  \vphantom{{h^{ac}(q^{(j-1)})h^{bc}(q^{(j)})\over2m}}
  -\delta\sqrt{f(q^{(j)})f(q^{(j-1)})}
  \right.\right.\hfill\\   \hfill\left.\left.\times
  \left[{h^{ac}(q^{(j-1)})h^{bc}(q^{(j)})\over2m}
  p_a^{(j)}p_b^{(j)}+V(q^{(j)})+\Delta V_{Prod}(q^{(j)})-E\right]
  \right\}\right)
  \\   \qquad
  =(f'f'')^{\half(1-D/2)}\lim_{N\to\infty}\Norm^{ND/2}
  \prod_{j=1}^{N-1}\int\sqrt{g(q^{(j)})\over f^D(q^{(j)})}\,dq^{(j)}
  \hfill\\  \qquad\qquad\times
  \exp\left\{\ih\sum_{j=1}^N\left[{m\over2\epsilon}
  {h_{ac}(q^{(j-1)})h_{bc}(q^{(j)})\over\sqrt{f(q^{(j-1)})f(q^{(j)})}}
  \Delta q^{a,(j)}\Delta q^{b,(j)}
  \right.\right.\hfill\\   \hfill\left.\left.
  \vphantom{{h_{ac}(q^{(j-1)})h_{bc}(q^{(j)})
                \over\sqrt{f(q^{(j-1)})f(q^{(j)})}}}
  -\i\epsilon f(q^{(j)})\Big(V(q^{(j)})+\Delta V_{Prod}(q^{(j)})-E\Big)
  \right]\right\}
  \hfill\\   \qquad
  =(f'f'')^{\half(1-D/2)}\int_0^\infty\,\tilde K(q'',q';s'')ds''
  \hfill\endmultline
  \tag\NUM.\num$$
with the path integral
\plus
$$\multline
  \tilde K(q'',q';s'')
  =\int\limits_{q(t')=q'}^{q(t'')=q''}\sqrt{\tilde g}\,\CD_{PF} q(t)
  \\  \times
  \exp\left\{\ih\int_0^{s''}\bigg[{m\over2}
  \tilde h_{ac}\tilde h_{cb}\dot q^a\dot q^b
  -f\Big(V(q)+\Delta V_{Prod}(q)-E\Big)\bigg]ds\right\}.
  \endmultline
  \tag\NUM.\num$$
Here denote $\tilde h_{ac}=h_{ac}/\sqrt{f}$ and $\sqrt{\tilde g}
=\det(\tilde h_{ac})$ and equation (\NUM.44) is of the canonical
product form. Note that for $D=2$ the prefactor is ``one''.

\bigskip\bigskip
\glno=0               
\advance\chapno by 1  

\section{Separation of Variables}
Let us assume that the potential problem $V(x)$ has an exact solution
according to
\plus
$$\int\limits_{x(t')=x'}^{x(t'')=x''} \CD x(t)\exp\left[
   \ih\int_{t'}^{t''}\bigg({m\over2}\dot x^2-V(x)\bigg)dt\right]
  =\int dE_\lambda\,\e^{-\i E_\lambda T/\hbar}
  \Psi_\lambda^*(x')\Psi_\lambda(x'').
  \tag\NUM.\num$$
Here $\int dE_\lambda$ denotes a Lebesque-Stieltjes integral to include
discrete as well as continuous states.  Now we consider the path
integral
\plus
$$\multline
  \!\!\!\!\!\!
  K(z'',z',x'',x';T)
  \\
  =\int\limits_{z(t')=z'}^{z(t'')=z''} f^d(z)\prod_{i=1}^{d'}g_i(z)\CD
  z_i(t)\int\limits_{x(t')=x'}^{x(t'')=x''}\prod_{k=1}^d\CD x_k(t)
  \hfill\\   \times
  \exp\left\{\ih\int_{t'}^{t''}\left[{m\over2}\left(
  \sum_{i=1}^{d'}g_i(z)\dot z^2_i+f^2(z)\sum_{k=1}^d\dot x_k^2\right)
  -\left({V(x)\over f^2(z)}+W(z)+\Delta W(z)\right)\right]dt\right\}
  \hfill\\
  =\lim_{N\to\infty} \Norm^{ND\over2}\prod_{j=1}^{N-1}
  \int f^d(z^{(j)})\prod_{i=1}^{d'}g_i(z^{(j)})dz_i^{(j)}
  \int\prod_{k=1}^d dx_k^{(j)}
  \hfill\\   \times
  \exp\left\{\ih\sum_{j=1}^N\left[{m\over2\epsilon}
  \left(\sum_{i=1}^{d'}g_i(z^{(j-1)})g_i(z^{(j)})\Delta^2z_i^{(j)}
     +f(z^{(j-1)})f(z^{(j)})\sum_{k=1}^d\Delta^2x_k^{(j)}\right)
  \right.\right.\\   \left.\left.\vphantom{\sum_j^N}
  -\epsilon\bigg({V(x^{(j)})\over f^2(z^{(j)})}
  +W(z^{(j)})+\Delta W(z^{(j)})\bigg)\right]\right\}.
  \endmultline
  \tag\NUM.\num$$
Here $(z,x)\equiv(z_i,x_k)$ ($i=1,\dots,d';k=1,\dots,d$, $d'+d=D$)
denote a $D$-dimensional coordinate system, $g_i$ and $f$ the
corresponding metric terms, and $\Delta W$ the quantum potential.
For simplicity I assume that the metric tensor $g_{ab}$
involved has only diagonal elements, i.e.\ $g_{ab}=
\diag(g_1^2(z),g_2^2(z),\dots,g_{d'}^2(z),f^2(z),\dots,f^2(z)]$.
Of course, $\det(g_{ab})=f^{2d}\prod_{i=1}^{d'} g_i^2\equiv
f^{2d}G(z)$. The indices $i$ and $k$ will be omitted in the following.
We perform the time transformation
\plus
$$s=\int_{t'}^t{d\sigma\over f^2[z(\sigma)]},
  \qquad s''=s(t''),\qquad s (t')=0,
  \tag\NUM.\num$$
where the lattice interpretation reads $\epsilon/ [f(z^{(j-1)})
f(z^{(j)})]=\delta^{(j)}\equiv\delta$. Of course, we identify
$z(t)\equiv z[s(t)]$ and $x(t)\equiv x[s(t)]$. The transformation
formul\ae\ for a pure time transformation are now
$$\allowdisplaybreaks\align
  K(z'',z',x'',x';T)&={1\over2\pi\i\hbar}\int_{-\infty}^\infty
  dE\,\e^{-\i ET/\hbar}G(z'',z',x'',x';E)
   \tag\NUM.\num\\   \global\plus
  G(z'',z',x'',x';E)&=\i[f(z')f(z'')]^{1-D/2}\int_0^\infty ds''
  \tilde K(z'',z',x'',x';s'' ),
  \tag\NUM.\num
  \endalign$$
where the transformed path integral $\tilde K(s'')$ is given by
\plus
$$\multline
  \!\!\!\!\!\!
  \tilde  K(z'',z',x'',x';s'')
  \\  =
  \int\limits_{z(0)=z'}^{z(s'')=z''}
  {\sqrt{G(z)}\over f^{2-d}(z)} \CD z(s)
  \int\limits_{x(0)=x'}^{x(s'')=x''}\CD x(s)
  \hfill\\   \hfill\times
  \exp\left\{\ih\int_0^{s''}\bigg[{m\over2}
  \bigg({g^2(z)\over f^2(z)}\dot z^2+\dot x^2\bigg)
  -V(x)-f^2(z)\Big(W(z)+\Delta W(z)\Big)+f^2(z)E\bigg]ds\right\}
  \\  =
  \int dE_\lambda\,\e^{-\i E_\lambda s''/\hbar}
  \Psi_\lambda^*(x')\Psi_\lambda(x'')\hat K(z'',z';s''),
  \hfill\endmultline
  \tag\NUM.\num$$
with the remaining path integration
\plus
$$\multline
  \hat K(z'',z';s'')=\int\limits_{z(0)=z'}^{z(s'')=z''}
  {\sqrt{G(z)}\over f^{2-d}(z)}\CD z(s)
  \hfill\\   \hfill\times
  \exp\left\{\ih\int_0^{s''}\bigg[{m\over2}{g^2(z)\over f^2(z)}\dot z^2
      -f^2(z)\Big(W(z)+\Delta W(z)\Big)+f^2(z)E\bigg]ds\right\}.
  \endmultline
  \tag\NUM.\num$$
Of course, in the path integrals (\NUM.6,\NUM.7) the same lattice
formulation is assumed as in the path integral (\NUM.2). Note the
difference in comparison with a combined space-time transformation
where a factor $[f(z')f(z'')]^{1/4}$ would instead appear.
We also see that for $D=2$ the prefactor is identically ``one''.
We perform a second time transformation in $\hat K(s'')$ effectively
reversing the first:
\plus
$$\sigma=\int_0^sf^2[z(\omega)]d\omega,\qquad \sigma''=s''
  \tag\NUM.\num$$
with the transformation on the lattice interpreted as
$\sigma^{(j)}=\delta^{(j)}f(z^{(j-1)})f(z^{(j)})$.
Therefore we obtain the transformation formul\ae
$$\allowdisplaybreaks\align
  \hat K(z'',z';s'')&={1\over2\pi\i\hbar}\int_{-\infty}^\infty dE'\,
                  \e^{-\i E's''/\hbar}\hat G(z'',z';E')
  \tag\NUM.\num\\   \global\plus
  \hat G(z'',z';E')&=\i[f(z')f(z'')]^{{D-d\over2}-1}
  \int_0^\infty d\sigma''\,\e^{\i E\sigma''/\hbar}
  \tilde{\hat K}(z'',z';\sigma'')
  \tag\NUM.\num
  \endalign$$
with the transformed path integral given by
\plus
$$\multline
  \tilde{\hat K}(z'',z';\sigma'')
  =\int\limits_{z(0)=z'}^{z(\sigma'')=z''}
  \sqrt{G(z)}\,\CD z(\sigma)
  \\  \times
  \exp\left\{\ih\int_0^{\sigma''}\bigg[{m\over2}g^2(z)\dot z^2
       -W(z)-\Delta W(z)+{E'\over f^2(z)}\bigg]d\sigma\right\}.
  \endmultline
  \tag\NUM.\num$$
Plugging all the relevant formul\ae\ into equation (\NUM.5) yields
\hfuzz=10pt
\plus
$$\multline
  K(z'',z',x'',x';T)=[f(z')f(z'')]^{-D/2}\int dE_\lambda
  \Psi_\lambda^*(x')\Psi_\lambda(x'')
  \hfill\\   \qquad\times
  {1\over2\pi\hbar}\int_0^\infty d\sigma''\int_{-\infty}^\infty dE\,
                                            \e^{-\i E(\sigma''-T)/\hbar}
  {1\over2\pi\hbar}\int_{-\infty}^\infty dE'\int_0^\infty ds''\,
   \e^{-\i s''(E_\lambda+E')/\hbar}
  \hfill\\  \qquad\times
  \int\limits_{z(0)=z'}^{z(\sigma'')=z''}\sqrt{G(z)}\,\CD z(\sigma)
  \exp\left\{\ih\int_0^{\sigma''}\bigg[{m\over2}g^2(z)\dot z^2
       -W(z)-\Delta W(z)+{E'\over f^2(z)}\bigg]d\sigma\right\}.
  \hfill\\   \quad
  \endmultline
  \tag\NUM.\num$$
\hfuzz=3pt
The $d\sigma''dE$-integration produces just $\sigma''\equiv T$,
whereas  the $dE'ds''$-integration can be evaluated by giving
$E_\lambda+E$ a small negative imaginary part and applying the residuum
theorem yielding $E_\lambda\equiv-E'$. Therefore we arrive finally at
the identity [\GROj]
\plus
$$\multline
  K(z'',z',x'',x';T)=[f(z')f(z'')]^{-D/2}\int dE_\lambda
  \Psi_\lambda^*(x')\Psi_\lambda(x'')
  \hfill\\  \hfill\times
  \int\limits_{z(t')=z'}^{z(t'')=z''}\sqrt{G(z)}\,\CD z(t)
  \exp\left\{\ih\int_{t'}^{t''}\bigg[{m\over2}g^2(z)\dot z^2
       -W(z)-\Delta W(z)-{E_\lambda\over f^2(z)}\bigg]dt\right\}.
  \\ \ \endmultline
  \tag\NUM.\num$$
Note that this result can be short handed interpreted by inserting
\plus
$$\multline
  \bigg({m\over2\pi\i\delta^{(j)}\hbar}\bigg)^{D/2}
  \exp\bigg[\ih\bigg({m\over2\delta^{(j)}}\Delta^2x^{(j)}
       -\delta^{(j)}V(x^{(j)})\bigg)\bigg]
  \\
  =\int dE_{\lambda^{(j)}}\,\e^{-\i E_{\lambda^{(j)}}\delta^{(j)}/\hbar}
  \Psi_{\lambda^{(j)}}^*(x^{(j-1)})\Psi_{\lambda^{(j)}}(x^{(j)})
  \endmultline
  \tag\NUM.\num$$
with $\delta^{(j)}=\epsilon/[f(z^{(j-1)})f(z^{(j)})]$ for all $j^{th}$
and applying the orthonormality of the $\Psi_\lambda$ in each
$j^{th}$-path integration. Equation (\NUM.14) describes therefore a
``short-cut'' to establish equation (\NUM.12) instead of performing a time
transformation back and forth.


\newpage
\secno=0
\glno=0      
\chapno=1    
\def\Chapter{Important Examples}

\centerline{\fourteenbf III I\large mportant \fourteenbf E\large xamples}
\bigskip

\noindent
In this Chapter I present some of the most important
path integral solutions which are
\item{1)} The free particle as a simple example.
\item{2)} The harmonic oscillator in its basic form, where we
          allow in addition a time dependent frequency.
\item{3)} Path integration in polar coordinates. We shall discuss
          the various features of properly defined path integrals
          including the ``Besselian functional measure''.
          In particular the (time-dependent) radial harmonic oscillator
          will be exactly evaluated.
\item{4)} The Coulomb potential.

\bigskip\bigskip
\section{The Free Particle}
For warming up we calculate the path integral for the free particle
in an $D$-dimensional Euclidean space.
{}From the representation
\plus
$$\allowdisplaybreaks\align
  K&(x'',x';T)
  \\
  &=\int\limits_{x(t')=x'}^{x(t'')=x''}\CD x(t)
  \exp\left({\i m\over2\hbar}\int_{t'}^{t''}\dot x^2dt\right)
  \\
  &=\lim_{N\to\infty}\Norm^{ND\over2}
  \prod_{j=1}^{N-1}\int_{-\infty}^\infty dx^{(j)}
  \exp\left[{\i m\over2\epsilon\hbar}\sum_{j=1}^N
  \big(x^{(j)}-x^{(j-1)}\big)^2\right].
  \tag\NUM.\num\endalign$$
it is obvious that the various integrations separate into a
$D$-dimensional product. All integrals are Gaussian. Starting with
($\mu=1,\dots,D$):
\plus
$$\multline
  K^{N=2}(x_\mu^{(2)},x_\mu^{(0)};2\epsilon)
  \\   \qquad
  =\Norm\int_{-\infty}^\infty dx^{(1)}
  \exp\bigg[-{m\over2\i\epsilon\hbar}(x_\mu^{(2)}-x_\mu^{(1)})^2
            -{m\over2\i\epsilon\hbar}(x_\mu^{(1)}-x_\mu^{(0)})^2\bigg]
  \hfill\\    \qquad
  =\sqrt{m\over4\pi\i\hbar\epsilon}
  \exp\bigg[
  -{m\over4\i\epsilon\hbar}(x_\mu^{(2)}-x_\mu^{(0)})^2\bigg],
  \hfill\endmultline
  \tag\NUM.\num$$
it is easy to show that
\plus
$$K^N(x^{(N)}_\mu,x^{(0)}_\mu;\epsilon N)
  =\sqrt{m\over2\pi\i\hbar\epsilon N}\exp\bigg[
  -{m\over2\i\hbar\epsilon N}(x_\mu^{(N)}-x_\mu^{(0)})^2\bigg].
  \tag\NUM.\num$$
Thus we get in the limit $N\to\infty$ (note $\epsilon N=T$):
\plus
$$\aligned
  K(x'',x';T)&=\bigg({m\over2\pi\i\hbar T}\bigg)^{D/2}
  \exp\bigg[{\i m\over2\hbar T}(x''-x')^2\bigg]\\
        &={1\over(2\pi)^D}\int_{-\infty}^\infty
  \exp\bigg[\i p(x''-x')-\i T{\hbar p^2\over2m}\bigg]dp.
  \endaligned
  \tag\NUM.\num$$
{}From this representation the normalized wave-functions and the energy
spectrum can be read off ($p\in\R^D$):
\plus
$$\Psi(x)={\e^{\i p\cdot x}\over(2\pi)^{D/2}},\qquad
  E_p={p^2\hbar^2\over2m}\qquad(x\in\R^D)
  \tag\NUM.\num$$
which is the known result.
\newline
The corresponding energy-dependent Green function is given by
($D=1$):
\plus
$$G^{(1)}(x'',x';E)=\i\int_0^\infty K(x'',x';T) \e^{\i ET/\hbar}dT
   =\sqrt{-{m\over2E}}\exp\bigg(\i{\vert x''-x'\vert\over\hbar}
        \sqrt{2mE}\bigg).
  \tag\NUM.\num$$
For the $D$-dimensional case we obtain
$$\allowdisplaybreaks\align
  G^{(D)}&(x'',x';E)
  \\
  &=2\i\bigg({m\over2\pi\i\hbar}\bigg)^{D/2}
   \bigg({m\over2E}\vert x''-x'\vert^2\bigg)^{\half(1-{D/2})}
   K_{1-{D/2}}\bigg(-\i{\vert x''-x'\vert^2\over\hbar}
      \sqrt{2mE}\bigg)\quad
  \tag\NUM.\num\\   \global\plus
  &=\i\bigg({m\over2\hbar}\bigg)^{D/2}
   \bigg({m\pi^2\over2E}\vert x''-x'\vert^2\bigg)^{\half(1-{D/2})}
   H_{1-{D/2}}^{(1)}
  \bigg({\vert x''-x'\vert\over\hbar}\sqrt{2mE}\bigg),
  \tag\NUM.\num\endalign$$
\hfuzz=3pt
where use has been made of the integral representation
[\GRA, p.340]
\plus
$$\int_0^\infty dt\,t^{\nu-1}\exp\bigg(-{a\over4t}-pt\bigg)
  =2\bigg({a\over 4p}\bigg)^{\nu\over2}K_\nu(\sqrt{ap}\,),
  \tag\NUM.\num$$
and $K_\nu(z)={\i\pi\over2}\e^{\i\pi\nu/2}H_\nu^{(1)}(\i z)$,
$K_{\pm\half}(z)=\sqrt{\pi/2z}\,\e^{-z}$.

\bigskip\bigskip
\glno=0               
\advance\chapno by 1  

\section{The Harmonic Oscillator}
After this easy task we proceed to the path integral calculation of
the harmonic oscillator. We consider the simple one-dimensional case
with the Lagrangian
\plus
$$\CL[x,\dot x]
  ={m\over2}\dot x^2-{c(t)\over2}x^2+b(t)x\dot x-e(t)x
  \tag\NUM.\num$$
Here we assume that the various coefficients may be time-dependent.
The path integral has the form
\plus
$$K(x'',x';t'',t')=
  \int\limits_{x(t')=x'}^{x(t'')=x''}\CD x(t) \e^{\ih S[x]}=
  \int\limits_{x(t')=x'}^{x(t'')=x''}\CD x(t)
  \e^{\ih\int_{t'}^{t''}\CL[x,\dot x]dt}
  \tag\NUM.\num$$
In this quadratic Lagrangian, of course, an ordering problem appears.
This is due to the stochastic nature of the path integral,
the ``time''-integral in the exponent is not a Riemann-integral
but a so called ``It\^o-integral''. This we have already discussed in
chapter II. The ordering ambiguity appears in the ``b(t)''-term, where
we have in the corresponding Hamiltonian a $px$-term. The consequence
is that we have to use the mid-point formulation, i.e.\ ($\bar x^{(j)}=
\half(x^{(j)}+x^{(j-1)})$, and $c^{(j)}=c(t^{(j)})$, etc., see also
Schulman's book [\SCHUc]):
\hfuzz=20pt
\plus
$$\multline
  \!\!\!\!
  K(x'',x';t'',t')
  \\
  =\int\limits_{x(t')=x'}^{x(t'')=x''}\CD x(t)
  \exp\left\{\ih\int_{t'}^{t''}\bigg[
  {m\over2}\dot x^2-{c(t)\over2}x^2+b(t)x\dot x-e(t)x\bigg]dt\right\}
  \hfill\\
  =\lim_{N\to\infty}\Norm^{N\over2}
  \prod_{j=1}^{N-1}\int_{-\infty}^\infty dx^{(j)}
  \hfill\\  \hfill\times
  \exp\left\{\ih\sum_{j=1}^N\left[{m\over2\epsilon}
  \Big(x^{(j)}-x^{(j-1)}\Big)^2
  -\epsilon{c^{(j)}\over2}({\bar x^{(j)}})^2
  +b^{(j)}\bar x^{(j)}\Big(x^{(j)}-x^{(j-1)}\Big)
  -\e^{(j)}\bar x^{(j)}\right]\right\}.
  \\  \quad\endmultline
  \tag\NUM.\num$$
\hfuzz=3pt
In the following we use the notations in equations (2.2) and (2.3)
as synonymous. Let us expand the ``path'' $x(t)$ about the classical
path $x_{Cl}(t)$, i.e.:
$$x(t)=x_{Cl}(t)+y(t),$$
where $y(t)$ denotes a fluctuating path about the classical one.
The classical path obeys, of course, the Euler Lagrange equations:
\plus
$${\delta \CL[x_{Cl},\dot x_{Cl}]\over\delta x_{Cl}}=
  {d \over dt}{\partial\CL
  [x_{Cl},\dot x_{Cl}]\over\partial\dot x_{Cl}}
  -{\partial\CL[x_{Cl},\dot x_{Cl}]\over\partial x_{Cl}}=0,\quad
 x_{Cl}(t')=x',\quad x_{Cl}(t'')=x_{Cl}''.
  \tag\NUM.\num$$
Expanding we obtain for the action
\plus
$$S[x]=S[x_{Cl}]
  +\half{\delta^2S[x]\over\delta x^2}
  \bigg\vert_{x=x_{Cl}}y^2
  =S[x_{Cl}(t''),x_{Cl}(t')]+\int_{t'}^{t''}
   \bigg[{m\over2}\dot y^2-{c\over2}y^2+by\dot y\bigg]dt.
  \tag\NUM.\num$$
The linear functional variation vanishes due to the Euler-Lagrange
equations (2.4). For the path integral we now find
\plus
$$\aligned
  K(x''x';t'',t')
  &=\exp\bigg\{\ih S[x_{Cl}(t''),x_{Cl}(t')]\bigg\}F(t'',t'),
  \\
  F(t'',t')&=\int\limits_{y(t')=0}^{y(t'')=0}\CD y(t)
  \exp\left[{\i m\over2\hbar}\int_{t'}^{t''}
  \big(\dot y^2-\omega^2(t)y^2\big)dt\right].
  \endaligned
  \tag\NUM.\num$$
Here we have used the abbreviations $m\omega^2(t)=c(t)+\dot b(t)$,
$y'=y(t')$ and $y''=y(t'')$. Now
\plus
$$\allowdisplaybreaks\align
  F(t'',t')&=\lim_{N\to\infty}F_N
  \\
  &=\lim_{N\to\infty}\Norm^{N\over2}
  \int_{-\infty}^\infty dy^{(1)}\dots\int_{-\infty}^\infty dy^{(N-1)}
  \\    &\qquad\times
  \exp\left\{-{m\over2\i\epsilon\hbar}\sum_{j=1}^N
  \left[(y^{(j)}-y^{(j-1)})^2
  -\epsilon^2\omega^{(j)\,2}y^{(j)\,2}\right]
  \right\}
  \tag\NUM.\num
  \endalign$$
Let us introduce a $N-1$-dimensional vector $z=(x_1,\dots,x_{N-1})^T$
and the $(N-1)\times(N-1)$ matrix $B$:
\hfuzz=20pt
\plus
$$B=\pmatrix 2-\epsilon^2\omega^{(1)\,2} &-1 &$\dots$&0&0\\
             -1 &2-\epsilon^2\omega^{(2)\,2} &$\dots$&0&0\\
             \vdots&\vdots&\ddots            &\vdots&\vdots\\
             0&0&$\dots$        &2-\epsilon^2\omega^{(N-2)\,2} &-1\\
             0&0&$\dots$        &-1 &2-\epsilon^2\omega^{(N-1)\,2}
    \endpmatrix.
  \tag\NUM.\num$$
\hfuzz=3pt
Thus we get
\plus
$$F_N=\Norm^{N\over2}\int d^{N-1}z\,
      \exp\bigg(-{m\over2\i\epsilon\hbar}z^TBz\bigg)
     =\bigg({m\over2\pi \i\epsilon\hbar\det B}\bigg)^\half.
  \tag\NUM.\num$$
Therefore
\plus
$$F(t'',t')=\bigg({m\over2\pi\i\hbar f(t'',t')}\bigg)^\half,\qquad
  \hbox{where}\qquad f(t'',t')=\lim_{N\to\infty}\epsilon\det B.
  \tag\NUM.\num$$
Our final task is to determine $\det B$. Let us consider the $j\times
j$ matrix
\plus
$$B^{(j)}=\pmatrix 2-\epsilon^2\omega^{(1)\,2} &-1 &$\dots$&0&0\\
             -1 &2-\epsilon^2\omega^{(2)\,2} &$\dots$&0&0\\
             \vdots&\vdots&\ddots            &\vdots&\vdots\\
             0&0&$\dots$        &2-\epsilon^2\omega^{(j-1)\,2} &-1\\
             0&0&$\dots$        &-1 &2-\epsilon^2\omega^{(j)\,2}
    \endpmatrix
  \tag\NUM.\num$$
One can show that the following recursion relations holds:
$$\det B^{(j+1)}=(2-\epsilon^2\omega^{(j+1)\,2})\det B^{(j)}
  -\det B^{(j-1)}$$
with $\det B^{(1)}=2-\epsilon^2\omega^{(1)\,2}$ and $\det B^{(0)}=1$.
Let us define $g^{(j)}=\epsilon\det B^{(j)}$, then we have
$$g^{(j+1)}-2g^{(j)}+g^{(j-1)}=-\epsilon^2\omega^{(j+1)\,2}g^{(j)}.$$
Turning to a continuous notation we find for the function
$g(t)=f(t,t')$ a differential equation:
\plus
$$\ddot g(t)+\omega^2(t)g(t)=0,
  \quad\hbox{with}\quad g(t')=0,\quad\dot g(t')=1.
  \tag\NUM.\num$$
The two last equation follow from
$g(t') =g_0 =\lim_{\epsilon\to0} \epsilon \det B^{(0)}$ and
$\dot g(t') =\lim_{\epsilon\to 0}$\newline
$\big[g(t'+\epsilon)-g(t')\big]/\epsilon
 =\lim_{\epsilon\to 0} (\det B^{(1)}-\det B^{(0)}) =1$.
Finally we have to insert $g(t'')=f(t'',t')$ into equation (2.10).

\noindent
At once we recover the free particle with $g=t''-t'=T$.
\newline
The case of the usual harmonic oscillator with $\omega(t)=\omega$
(time-independent) is given by
$$g(t)={1\over\omega}\sin\omega T.$$
To get the path integral solution for the harmonic oscillator
we must calculate its classical action. It is a straightforward
calculation to show that it is given by:
\plus
$$S[x_{Cl}(t''),x_{Cl}(t')]={m\omega\over2\sin\omega T}
  \Big[({x'}_{Cl}^2
  +{x''}_{Cl}^2)\cos\omega T-2x_{Cl}'x_{Cl}''\Big].
  \tag\NUM.\num$$
and we have for the Feynman kernel:
\plus
$$K(x'',x';T)=\bigg({m\omega\over2\pi\i\hbar\sin\omega T}\bigg)^\half
  \exp\bigg\{
  -{m\omega\over2\i\hbar}\bigg[({x'}^2+{x'}^2)\cot\omega T
     -2{x'x''\over\sin\omega T}\bigg]\bigg\}.
  \tag\NUM.\num$$
By the use of the  Mehler-formula [\EMOTa,\ Vol.III,\ p.272]:
\plus
$$\e^{-(x^2+y^2)/2}
  \sum_{n=0}^\infty{1\over n!}\bigg({z\over2}\bigg)^nH_n(x)H_n(y)
 ={1\over\sqrt{1-z^2}}
  \exp\bigg[{4xyz-(x^2+y^2)(1+z^2)\over2(1-z^2)}\bigg],
  \tag\NUM.\num$$
where $H_n$ denote the Hermite-polynomials, we can expand the Feynman
kernel according to (identify $x\equiv\sqrt{m\omega/\hbar}\,x'$,
$y\equiv\sqrt{m\omega/\hbar}\,x''$ and $z=\e^{-\i\omega T}$):
\plus
$$K(x'',x';T)
  =\sum_{n=0}^\infty \e^{-\i TE_n/\hbar}\Psi_n^*(x')\Psi_n(x'')
  \tag\NUM.\num$$
with energy-spectrum and wave-functions:
$$\allowdisplaybreaks\align
  E_n      &=\hbar\omega(n+\bhalf),
  \tag\NUM.\num\\    \global\plus
  \Psi_n(x)&=\bigg({m\omega\over2^{2n}\pi\hbar n!^2}\bigg)^{1\over4}
            H_n\bigg(\sqrt{m\omega\over\hbar}\, x\bigg)
            \exp\bigg(-{m\omega\over2\hbar}x^2\bigg).
  \tag\NUM.\num
  \endalign$$
Equation (2.14) is as it stands only valid for $0<\omega T<\pi$.
Let us investigate it for larger times. Let
\plus
$$T={n\pi\over\omega}+\tau, \qquad\hbox{with}\qquad
  n\in\N_0;\quad 0<\tau<\pi/\omega
  \tag\NUM.\num$$
then we have $\sin\omega T=\e^{\i\pi n}\sin\omega\tau$,
$\cos\omega T=\e^{\i\pi n}\cos\omega\tau$ and equation (2.14) becomes
\plus
$$\multline
  K(x'',x';T)=
  \bigg({m\omega\over2\pi\hbar\sin\omega\tau}\bigg)^\half
  \\  \times
  \exp\bigg\{-{i\pi\over2}\bigg(\half+n\bigg)
  +{\i m\omega\over2\hbar\sin\omega T}
   \Big[({x'}^2+{x'}^2)\cos\omega T-2x'x''\Big]\bigg\}
  \endmultline
  \tag\NUM.\num$$
which is the formula for the propagator with the Maslov correction
(note $\sin\omega\tau=\vert\sin\omega T\vert $). Now let $\tau\to0$,
i.e.\ we consider the propagator at caustics. We obtain
\plus
$$\allowdisplaybreaks\align
  K\bigg(x'',x';{n\pi\over\omega}\bigg)
  &=\lim_{\tau\to0}\bigg({m\over2\pi\i\hbar\tau}\bigg)^\half
  \exp\bigg\{-{\i\pi n\over2}
  +{\i m\over2\hbar\tau}\Big[({x'}^2+{x'}^2)
      -2\e^{-\i\pi n}x'x''\Big]\bigg\}
  \\    &
  =\lim_{a\to0} {1\over\sqrt{\pi}\,a}
  \exp\bigg[-{1\over a^2}(x'-(-1)^nx'')^2-\i{\pi n\over2}\bigg]
  \\    &
  =\exp\bigg(-{\i\over2}\pi n\bigg)\delta(x'-(-1)^nx'').
  \tag\NUM.\num
  \endalign$$

Let us finally determine the energy dependent Green function. We use the
integral representation [\GRA, p.729], $a_1>a_2$,
$\Re(\half+\mu-\nu)>0$):
\plus
$$\multline
  \int_0^\infty\coth^{2\nu}{x\over2}\exp\bigg[
                               -{a_1+a_2\over2}t\cosh x\bigg]
  I_{2\mu}(t\sqrt{a_1a_2}\sinh x)dx
  \\
  ={\Gamma(\half+\mu-\nu)\over t\sqrt{a_1a_2}\,\Gamma(1+2\mu)}
  W_{\nu,\mu}(a_1t)M_{\nu,\mu}(a_2t).
  \endmultline
  \tag\NUM.\num$$
\edef\numCBxa{\NUM.\num}%
Here $W_{\nu,\mu}(z)$ and $M_{\nu,\mu}(z)$ denote Whittaker-functions.
We reexpress $\e^x=\sqrt{\pi x/2}$\linebreak
$[I_\half(x)+I_{-\half}(x)]$ apply equation (\numCBxa) for
$\nu=E/2\hbar\omega$ and use some relations between Whittaker- and
parabolic cylinder-functions to obtain
\plus
$$\multline
  \!\!\!\!
  G(x'',x';E)=-\half\sqrt{m\over\pi\hbar\omega}\,
     \Gamma\bigg(\half-{E\over\hbar\omega}\bigg)
  \\  \times
   D_{-\half+{E\over\hbar\omega}}\left[\sqrt{2m\omega\over\hbar}\,
   (x'+x''+\vert x''-x'\vert )\right]
   D_{-\half+{E\over\hbar\omega}}\left[-\sqrt{2m\omega\over\hbar}\,
   (x'+x''-\vert x''-x'\vert )\right].
  \\ \quad \endmultline
  \tag\NUM.\num$$
{}From this representation we can recover by means of
the expansion for the $\Gamma$-function
$$\Gamma(z)=\sum_{n=0}^\infty{(-1)^n\over n!(z+n)}
  +\int_1^\infty \e^{-t}t^{z-1}dt$$
and some relations of the parabolic cylinder-functions and the Hermite
polynomials, the wave-functions of equation (2.18).

Let us note that we can use equations (2.14, 2.23) to obtain
recursion relations for the Feynman kernel and the Green function,
respectively, of the harmonic oscillator in $D$ dimensions.
Let $\vec x\in\R^n$ and define $\mu={\vec{x'}}^2+{\vec{x''}}^2$,
$\nu=\vec x'\cdot \vec x''$, then
\plus
$$\allowdisplaybreaks\align
  K^{(D)}(\vec x'',\vec x';T)
  &={1\over2\pi}{\partial\over\partial\nu}
   K^{(D-2)}(\vec x'',\vec x';T)
  \\
  &={1\over2\pi}\e^{-\i\omega T}
   \bigg({m\omega\over\hbar}-2{\partial\over\partial\mu}\bigg)
   K^{(D-2)}(\vec x'',\vec x';T).
  \tag\NUM.\num
  \endalign$$
For the Green functions this gives
\advance\glno by -1
$$\allowdisplaybreaks\align
  G^{(D)}(\vec x'',\vec x';E)
  &\equiv\tilde G^{(D)}(\mu\vert\nu;E)
  \tag\NUM.\num\\    \global\plus
  \tilde G^{(D)}(\mu\vert\nu;E)
  &=
  {1\over2\pi}{\partial\over\partial\nu}\tilde G^{(D-2)}(\mu\vert\nu;E)
  \tag\NUM.\num\\    \global\plus
  \tilde G^{(D)}(\mu\vert\nu;E)
  &={1\over2\pi}\bigg({m\omega\over\hbar}
    -2{\partial\over\partial\mu}\bigg)
  \tilde G^{(D-2)}(\mu\vert\nu;E-\omega\hbar)
  \tag\NUM.\num
  \endalign$$
Introducing furthermore
$\xi=\half(\vert\vec x'+\vec x''\vert +\vert\vec x''-\vec x'\vert )$,
$\eta=\half(\vert\vec x'+\vec x''\vert
   -\vert\vec x''-\vec x'\vert )$ we obtain for
$D=1,3,5\dots$.
\plus
$$\multline
  G^{(D)}(\vec x'',\vec x';E)
  =-\half\sqrt{m\over\pi\hbar\omega}\,
     \Gamma\bigg(\half-{E\over\hbar\omega}\bigg)
  \bigg({1\over2\pi}\bigg)^{D-1\over2}
  \\  \times
  \bigg[{1\over\eta^2-\xi^2}
  \bigg(\eta{\partial\over\partial\xi}-\xi{\partial\over\partial\eta}
  \bigg)\bigg]^{D-1\over2}
  D_{-\half+{E\over\hbar\omega}}\left(\sqrt{2m\omega\over\hbar}\,
   \xi\right)
   D_{-\half+{E\over\hbar\omega}}\left(-\sqrt{2m\omega\over\hbar}\,
   \eta\right),
  \\ \quad \endmultline
  \tag\NUM.\num$$
respectively,
\plus
$$\multline
  G^{(D)}(\vec x'',\vec x';E)
  =-\half\sqrt{m\over\pi\hbar\omega}\,
     \Gamma\bigg({D\over2}-{E\over\hbar\omega}\bigg)
  \bigg({1\over2\pi}\bigg)^{D-1\over2}
  \\  \times
  \bigg[{1\over\eta^2-\xi^2}
  \bigg(\xi{\partial\over\partial\xi}-\eta{\partial\over\partial\eta}
  \bigg)+{m\omega\over\hbar}\bigg]^{D-1\over2}
  D_{-{D\over2}+{E\over\hbar\omega}}\left(\sqrt{2m\omega\over\hbar}\,
   \xi\right)
   D_{-{D\over2}+{E\over\hbar\omega}}\left(-\sqrt{2m\omega\over\hbar}\,
   \eta\right).
  \\ \quad \endmultline
  \tag\NUM.\num$$
Let us note that the most general solution for the general quadratic
Lagrangian is due to Grosjean and Goovaerts [\GROS, \GROGOb].

\newpage
\bigskip\bigskip
\glno=0               
\advance\chapno by 1  

\section{The Radial Path Integral}
\subsection{The General Radial Path Integral}
Radial path integrals have been first discussed by Edwards and
Gulyaev [\EG] and Arthurs [\ARTb].
Edwards and Gulyaev discussed the two- and three-dimensional cases,
Arthurs concentrated on $D=2$.
Further progress have been made by Peak and Inomata [\PI] who
calculated the path integral for the radial harmonic oscillator
including some simple applications.
See also [\GRSb] for the ordering
problematics which give rise to the various quantum potentials
in the formulation of the radial path integral.
In this section we derive the path integral for $D$-dimensional
polar coordinates. We follow in our line of reasoning references
[\GRSb] and [\STEc], where these features have been first discussed in
their full detail. Similar topics can be also found in B\"ohm and
Junker [\BJb] from a group theoretic approach. We will get an expansion
in the angular momentum $l$, where the angle dependent part can be
integrated out and a radial dependent part is left over: the radial
path integral. We discuss some properties of the radial path integral
and show that it possible to get from the short time kernel the
radial Schr\"odinger equation.

The path integral in $D$-dimensions has the form:
\plus
$$\multline
  K^{(D)}(x'',x';T)
  \\    \qquad
  =\int\limits_{x(t')=x'}^{x(t'')=x''} \CD x(t)\exp\left\{
  \ih\int^{t''}_{t'}\bigg[{m\over2}\dot x^2-V(x)\bigg]dt\right\}
  \hfill\\     \qquad
  =\lim_{N\to\infty}\Norm^{ND/2}\int dx^{(1)}\dots\int dx^{(N-1)}
  \hfill\\   \qquad\qquad\times\exp\left\{\ih\sum^N_{j=1}\bigg[
  {m\over2\epsilon\hbar}(x^{(j)}-x^{(j-1)})^2-V(x^{(j)})\bigg]\right\}.
  \hfill\endmultline
  \tag\NUM.\num$$
Now let $V(x)=V(\vert x\vert )$
and introduce $D$-dimensional polar coordinates
[\EMOTa,\ Vol.II, Chapter IX]:
\plus
$$\left.\aligned
  x_1&=r\cos\theta_1\\
  x_2&=r\sin\theta_1\cos\theta_2\\
  x_3&=r\sin\theta_1\sin\theta_2\cos\theta_3\\
     &\hbox{$\dots$}  \\
  x_{D-1}&=r\sin\theta_1\sin\theta_2\dots\sin\theta_{D-2}\cos\phi\\
  x_D&=r\sin\theta_1\sin\theta_2\dots\sin\theta_{D-2}\sin\phi
  \endaligned\qquad\qquad\qquad\right\}
  \tag\NUM.\num$$
where $0\leq \theta_\nu\leq \pi\ (\nu=1,\dots,D-2)$,
$0\leq\phi\leq 2\pi$, $r=\bigg(\sum_{\nu=1}^D x_\nu^2\bigg)^{1/2}$.
Therefore $V(x)=V(r)$.
We have to use the addition theorem:
\plus
$$\multline
  \cos\psi^{(1,2)}
  =\cos\theta_1^{(1)}\cos\theta_1^{(2)}
  \\
  +\sum_{m=1}^{D-2}\cos\theta^{(1)}_{m+1}\cos\theta_{m+1}^{(2)}
  \prod_{n=1}^m\sin\theta^{(1)}_n\sin\theta^{(2)}_n+
  \prod_{n=1}^{D-1}\sin\theta^{(1)}_n\sin\theta^{(2)}_n,
  \endmultline
  \tag\NUM.\num$$
where $\psi^{(1,2)}$ is the angle between two $D$-dimensional vectors
$x^{(1)}$ and $x^{(2)}$ so that $x^{(1)}\cdot x^{(2)}=r^{(1)}r^{(2)}
\cos\psi^{(1,2)}$.
The metric tensor in polar coordinates is
\plus
$$(g_{ab})=\diag(1,r^2,r^2\sin^2\theta_1,\dots,
                 r^2\sin^2\theta_1\dots\sin^2\theta_{D-2}).
  \tag\NUM.\num$$
If $D=3$ equation (\NUM.3) reduces to:
$$\cos\psi^{(1,2)}=\cos\theta^{(1)}\cos\theta^{(2)}
      +\sin\theta^{(1)}\sin\theta^{(2)}\cos(\phi^{(1)}-\phi^{(2)}).$$
The $D$-dimensional measure $dx$ expressed in polar coordinates is
\plus
$$\left.\aligned
  dx&=r^{D-1}drd\Omega=r^{D-1}\prod_{k=1}^{D-1}
                      (\sin\theta_k)^{D-1-k}drd\theta_k
  \\
  d\Omega
  &=\prod_{k=1}^{D-1}(\sin\theta_k)^{D-1-k}d\theta_k.
  \endaligned\qquad\qquad\right\}
  \tag\NUM.\num$$
$d\Omega$ denotes the $(D-1)$-dimensional surface element on the unit
sphere $S^{D-1}$ and $\Omega(D)=2\pi^{D/2}/\Gamma(D/2)$ is the
volume of the D-dimensional unit ($S^{D-1}$-) sphere.
The determinant of the metric tensor is given by
\plus
$$g=\det(g_{ab})=\bigg(r^{D-1}\prod_{k=1}^{D-1}
     (\sin\theta_k)^{D-1-k}\bigg)^2
  \tag\NUM.\num$$
With all this the path integral equation (\NUM.1) yields:
\plus
$$\multline
  K^{(D)}(r'',\{\theta''\},r',\{\theta'\};T)=
  \\   \qquad=
  \lim_{N\to\infty}\Norm^{ND/2}
  \int_0^\infty r^{D-1}_{(1)}dr_{(1)}\int d\Omega_{(1)}\dots
  \int_0^\infty r^{D-1}_{(N-1)}dr_{(N-1)}\int d\Omega_{(N-1)}
  \hfill\\ \qquad\qquad
  \times\prod^N_{j=1}\exp\left\{
  {\i m\over2\epsilon\hbar}\big[ r_{(j)}^2+r_{(j-1)}^2
  -2r_{(j)}r_{(j-1)}\cos\psi^{(j,j-1)}\big]
  -{\i\epsilon\over\hbar} V(r_{(j)})\right\}.
  \endmultline
  \tag\NUM.\num$$
$\{\theta\}$ denotes the set of the angular variables.
For further calculations we need the formula [\GRA,\ p.980]:
\plus
$$\e^{z\cos\psi}=\bigg({z\over2}\bigg)^{-\nu}\Gamma(\nu)
  \sum_{l=0}^\infty(l+\nu)I_{l+\nu}(z)C^\nu_l(\cos\psi),
  \tag\NUM.\num$$
(for some $\nu\neq 0,-1,-2,\dots$), where $C_l^\nu$ are Gegenbauer
polynomials and $I_\mu$ modified Bessel functions. Equation (\NUM.8) is
a generalization of the well known expansion in three dimension where
$\nu=\half$ (remember $C^{\half}_l=P_l$, [\GRA,\ p.980]):
\plus
$$\e^{z\cos\psi}=\sqrt{\pi\over 2 z}\sum_{l=0}^\infty (2l+1)
  I_{l+{1\over2}}(z)P_l(\cos\psi)
  \tag\NUM.\num$$
Note: It is in some sense possible to include the case $D=2$,
i.e.\ $\nu=0$ if one uses
\newline
$l\lim_{\lambda\to 0}\Gamma(\lambda)C^\lambda_l=2\cos l\psi$ [\GRA,\
p.1030], yielding finally [\GRA,\ p.970]:
\plus
$$\e^{z\cos\psi}=\sum_{k=-\infty}^\infty I_k(z)\e^{\i k\psi}.
  \tag\NUM.\num$$
The addition theorem for the real surface (or hyperspherical) harmonics
$S^\mu_l$ on the $S^{D-1}$-sphere has the form [\EMOTa,\ Vol.II,
Chapter IX]:
\plus
$$\sum_{\mu=1}^M S^\mu_l(\Omega^{(1)})S_l^\mu(\Omega^{(2)})=
  {1\over\Omega(D)}{2l+D-2\over D-2}C_l^{D-2\over2}(\cos\psi^{(1,2)})
  \tag\NUM.\num$$
with $\Omega=x/r$ unit vector in $\R^d$ and the $M$ linearly
independent $S^\mu_l$ of degree $l$ with $M=(2l+D-2)(l+D-3)!/l!(D-3)!$.
The orthonormality relation is
\plus
$$\int d\Omega S^\mu_l(\Omega)S_{l'}^{\mu'}(\Omega)
         =\delta_{ll'}\delta_{\mu\mu'}.
  \tag\NUM.\num$$
Combining equations (\NUM.8) and (\NUM.11) we get the expansion formula
\plus
$$\e^{z(\Omega^{(1)}\cdot\Omega^{(2)})}=\e^{z\cos\psi^{(1,2)}}=2\pi
  \bigg({2\pi\over z}\bigg)^{D-2\over2}
  \sum_{l=0}^\infty\sum_{\mu=1}^M
  S^\mu_l(\Omega^{(1)})S^\mu_l(\Omega^{(2)})
  I_{l+{D-2\over2}}(z).
  \tag\NUM.\num$$
Let $\nu=(D-2)/2>0$ in equation (\NUM.8), then the jth term in
equation (\NUM.7) becomes
\plus
$$\multline
  R_j=\exp\left\{ {\i m\over2\epsilon\hbar} (r_{(j)}^2+r_{(j-1)}^2)
  -{\i\epsilon\over\hbar} V(r_{(j)}) \right\}
  \exp\bigg[{m\over2\i\epsilon\hbar}r_{(j)}r_{(j-1)}
                                       \cos\psi^{(j,j-1)}\bigg]
  \hfill\\  \qquad
  =\bigg({2\i\epsilon\hbar\over mr_{(j)}r_{(j-1)}}\bigg)^{D-2\over2}
   \Gamma\bigg({D-2\over2}\bigg)
  \exp\left[{\i m\over2\epsilon\hbar} (r_{(j)}^2+r_{(j-1)}^2)
  -{\i\epsilon\over\hbar} V(r_{(j)})\right]
  \hfill\\  \qquad\qquad
  \times\sum_{l_j=0}^\infty\bigg(l_j+{D\over2}-1\bigg)
  I_{l_j+{D-2\over2}}
  \bigg({m\over \i\epsilon\hbar}r_{(j)}r_{(j-1)}\bigg)
  C^{D-2\over2}_{l_j}(\cos\psi^{(j,j-1)})
  \hfill\\  \qquad
  =2\pi\bigg({2\pi \i\epsilon\hbar\over mr_{(j)}r_{(j-1)}}
                                              \bigg)^{D-2\over2}
   \exp\left[{\i m\over2\epsilon\hbar} (r_{(j)}^2+r_{(j-1)}^2)
   -{\i\epsilon\over\hbar} V(r_{(j)})\right]
   \hfill\\ \qquad\qquad
   \times\sum_{l_j=0}^\infty I_{l_j+{D-2\over2}}
    \bigg({m\over \i\epsilon\hbar}r_{(j)}r_{(j-1)}\bigg)
    \sum_{\mu_j=1}^M S^{\mu_j}_{l_j}(\Omega_{(j)})
    S^{\mu_j}_{l_j}(\Omega_{(j-1)}).
  \hfill\endmultline
  \tag\NUM.\num$$
With the help of equations (\NUM.11) and (\NUM.14) now (\NUM.7) becomes:
\plus
$$\multline
  K^{(D)}(x'',x';T)
  \\   \qquad=
  (r'r'')^{-{D-2\over2}}\sum_{l=0}^\infty\sum_{\mu=1}^M
  S^\mu_l(\Omega')S^\mu_l(\Omega'')
  \hfill\\  \qquad\qquad\times
  \lim_{N\to\infty}\bigg({m\over \i\epsilon\hbar}\bigg)^N
  \int_0^\infty r_{(1)}dr_{(1)}\dots
  \dots\int_0^\infty r_{(N-1)}dr_{(N-1)}
  \hfill\\  \qquad\qquad\times
  \prod_{j=1}^N\exp\left[{\i m\over2\epsilon\hbar}
  (r_{(j)}^2+r_{(j-1)}^2)-{\i\epsilon\over\hbar} V(r_{(j)})\right]
  I_{l+{D-2\over2}}
  \bigg({m\over \i\epsilon\hbar}r_{(j)}r_{(j-1)}\bigg).
  \hfill\endmultline
  \tag\NUM.\num$$
Therefore we can separate the radial part of the path integral:
\plus
$$K^{(D)}(r'',\{\theta''\},r',\{\theta'\};T)=
  \Omega^{-1}(D)
  \sum_{l=0}^\infty {l+D-2\over D-2}C^{D-2\over2}_l(\cos\psi^{(','')})
  K_l(r'',r';T)
  \tag\NUM.\num$$
with
\plus
$$\multline
  K_l^{(D)}(r'',r';T)=(r'r'')^{-{D-2\over2}}\lim_{N\to\infty}
  \Norm^{N/2} \int_0^\infty dr_{(1)}\dots\int_0^\infty dr_{(N-1)}
  \hfill\\  \hfill\times
  \prod_{j=1}^N\mu_l[r_{(j)}r_{(j-1)}]\cdot
  \exp\left\{\ih\sum_{j=1}^N\bigg[{m\over2\epsilon}
  (r_{(j)}-r_{(j-1)})^2-\epsilon V(r_{(j)})\bigg]\right\}
  \endmultline
  \tag\NUM.\num$$
and the functional measure
\plus
$$\mu_l^{(D)}(z_{(j)})=
  \sqrt{2\pi z_{(j)}}\,\e^{-z_{(j)}}I_{l+{D-2\over2}}(z_{(j)}),
  \tag\NUM.\num$$
where $z_{(j)}=(m/\i\epsilon\hbar)r_{(j)}r_{(j-1)}$.

In the literature often use is been made of the asymptotic form of the
modified Bessel functions
\plus
$$I_\nu(z)\simeq(2\pi z)^{-\half}\e^{z-(\nu^2-{1\over4})/2z}
  \qquad(\vert z\vert\gg1,\quad
  \Re(z)>0).
  \tag\NUM.\num$$
Then the functional measure becomes (we ignore $\Re(z)=0$):
\plus
$$\mu_l^{(D)}(z_{(j)})\simeq
  \exp\left\{-{\i\epsilon\hbar\over2mr_{(j)}r_{(j-1)} }
  \left[\bigg(l+{D-2\over2}\bigg)^2
  -{1\over4} \right]\right\}
  \tag\NUM.\num$$
and $K_l^{(D)}$ reads:
\hfuzz=15pt
\plus
$$\multline
  K_l^{(D)}(r'',r';T)=(r'r'')^{-{D-1\over2}}\lim_{N\to\infty}
  \Norm^{N/2} \int_0^\infty dr_{(1)}\dots\int_0^\infty dr_{(N-1)}
  \hfill\\  \hfill
  \times\exp\left\{\ih\sum_{j=1}^N\left[{m\over2\epsilon}
  (r_{(j)}-r_{(j-1)})^2
  -\hbar^2{(l+{D-2\over2})^2-{1\over4}\over2mr_{(j)}r_{(j-1)}}
  -\epsilon V(r_{(j)})\right]\right\}.
  \endmultline
  \tag\NUM.\num$$
This last equation seems to suggests a Lagrangian formulation of the
radial path integral:
\plus
$$k_l^{(D)}(r'',r';T)
  {\mathop=^{?}}\,\int\limits_{r(t')=r'}^{r(t'')=r''}\CD r(t)
  \exp\left\{\ih\int_{t'}^{t''}\left[{m\over2}\dot r^2
  -\hbar^2{(l+{D-2\over2})^2-{1\over4}\over2mr^2}
  -V(r)\right]\right\}.
  \tag\NUM.\num$$
\hfuzz=0.1pt
with $k_l^{(D)}(r'',r';T)=(r'r'')^{(D-1)/2}K_l^{(D)}(r'',r';T)$.
But nevertheless, equation (\NUM.17) can be written in a similar manner
with a nontrivial functional measure:
\plus
$$k_l^{(D)}(r'',r';T)
  =\int\limits_{r(t')=r'}^{r(t'')=r''}\mu_l^{(D)}[r^2]\CD r(t)
  \exp\left\{\ih\int_{t'}^{t''}\bigg[{m\over2}\dot r^2-V(r)
   \bigg]dt\right\}
  \tag\NUM.\num$$
and the whole $l$-dependence is in $\mu_l$, as defined in equation
(\NUM.18) in the lattice formulation.
The asymptotic expansion of $I_\nu$ in equation (\NUM.19) in problematic
because it is only valid for $\Re(z)>0$, but we have $\Re(z)=0$.
The complete expansions in this case reads [\GRA,\ p.962]
\plus
$$\allowdisplaybreaks\align
  I_\nu(z)&\simeq{\e^z\over\sqrt{2\pi z}}
  \sum_{k=0}^\infty {(-1)^k\over(2z)^k}
  {\Gamma(\nu+k+\half)\over k!\Gamma(\nu-k+\half)}
  \\ &\qquad\qquad+
  {\exp[-z\pm(\nu+\half)i\pi]\over \sqrt{2\pi z} }
  \sum_{k=0}^\infty {1\over(2z)^k}
  {\Gamma(\nu+k+\half)\over k!\Gamma(\nu-k+\half)}
  \\
  &\simeq{1\over\sqrt{2\pi z}}\bigg\{
  \exp\bigg[ z-{\nu^2-{1\over4}\over2z}\bigg]+
  \exp\bigg[
  -z+{\nu^2-{1\over4}\over2z}\pm\i\pi\bigg(\nu+\half\bigg)\bigg]\bigg\}.
  \tag\NUM.\num\endalign$$
(``+'' for $-\pi/2<\arg(z)<3\pi/2$, ``-'' for $-3\pi/2<\arg(z)<\pi/2$).
Langguth and Inomata [\LAI] argued that the inserting of the expansion
(\NUM.20) in the path integral (\NUM.17) can be justified (adopting an
argument due to Nelson [\NELb]), but things are not such as easy. Our
arguments are as follows:
\item{1)} The Schr\"odinger equation
\plus
$$\bigg[-{\hbar^2\over2m}\bigg({d^2\over dr^2}
   +{D-1\over2r}{\d\over\d r}\bigg)
  +\hbar^2{l(l+D-2)\over2mr^2}+V(r)\bigg]\Psi(r;t)
  =\i\hbar{\partial\over\partial t} \Psi(r;t)
  \tag\NUM.\num$$
can be derived from the short time kernel of the path integral
(\NUM.17).
\item{2)} For $r',r''\to 0$ equation (\NUM.24) has the right boundary
conditions at the origin. We have for $z\to0$:
$$\sqrt{2\pi z}\e^{-z}I_\lambda(z)\simeq
  {2\over(2\lambda)!!}z^{\lambda+\half}\big[1+O(z^2)\big]$$
and it follows:
\plus
$$\mu_l^{(D)}(z_{(j)})\simeq\bigg\{
  \aligned
  &r^{\prime l+(D-1)/2}\qquad (r'\to0,\ r''>0)\\
  &r^{\prime\prime l+(D-1)/2}\qquad (r''\to0,\ r'>0)
  \endaligned\bigg.
  \tag\NUM.\num$$
for all $j$.
If $R^{(D)}_{n,l}$ and $\e^{(D)}_{n,l}$ denote the (normalized) radial
state functions and energy levels, respectively, of the $D$-dimensional
problem, we have
\plus
$$K_l^{(D)}(r'',r';T)=\sum_{n=0}^\infty
  R^{(D)}_{n,l}(r')
  R^{(D)}_{n,l}(r'')\e^{-\i T \e^{(D)/\hbar}_{n,l}/\hbar}
  \tag\NUM.\num$$
and thus for any but $s$-states - including the factor $r^{-(D-1)/2}$
in equation (\NUM.27) - we have:
\plus
$$R^{(D)}_{n,l}(r)\simeq r^{l+(D-1)/2}r^{-(D-1)/2}=r^l
  \tag\NUM.\num$$
which is the correct boundary condition for $R^{(D)}_{n,l}$.

\noindent
Both points are not true for the path integral (\NUM.22), because
\newline
1)  we did not find any obvious method for deriving the full
Schr\"oding er equation (\NUM.25), the angular dependence comes out
wrong.
\newline
2) In the ``Euclidean region'' $T\to -\i\tau(\tau>0)$
the functional measure is vanishing like
\plus
$$\multline
  \exp\left\{-\hbar{\tau\over N}\left[
  {(l+{D-2\over2})^2-{1\over4}\over2mr_1r'}+
  {(l+{D-2\over2})^2-{1\over4}\over2mr_{(N-1)}r''}\right]\right\}
  \hfill\\  \hfill\times
 \exp\left\{-\hbar{\tau\over N}
 \sum_{j=1}^{N-1}{(l+{D-2\over2})^2-{1\over4}
              \over2mr_{(j)}r_{(j-1)}}\right\}
  \endmultline
  \tag\NUM.\num$$
which is also vanishing for $r',r''\to0$, but not in the right manner.
The points $r'=r''=0$ are essential singularities, where as the
correct vanishing is powerlike.

\noindent 3)
Equation (\NUM.22) seems very suggestive, but it is quit useless in
explicit calculations. Even in the simplest case $D=3,l=0,V=0$, i.e.
\plus
$$\multline
K_l^{(D)}(r'',r';T)=(r'r'')^{-{D-1\over2}}\lim_{N\to\infty}
  \Norm^{N/2}
  \hfill\\  \hfill\times
  \int_0^\infty dr_{(1)}\dots\int_0^\infty dr_{(N-1)}
  \exp\left\{ {\i m\over2\epsilon\hbar}\sum_{j=1}^N
  (r_{(j)}-r_{(j-1)})^2\right\}
  \endmultline
  \tag\NUM.\num$$
cannot be calculated.
The integrals turn out to be repeated integrals over errorfunctions
which are not tractable. Also the method of Arthurs [\ARTb] fails.
In this method one assumes that the integrations in the limit
$\epsilon\to0$ are effectively from $-\infty$ to $+\infty$.
The path integral (\NUM.30) then becomes
$$K_l^{(D)}(r'',r';T)=(r'r'')^{-{D-1\over2}}
  \exp\bigg[{\i m\over2\epsilon\hbar}(r'-r'')^2\bigg]$$
which is wrong.
The ``mirror'' term $K_l^{(D)}(r'',-r';T)$ is missing.

\bigskip
Let us discuss some properties of the path integral (\NUM.22)
\plus
$$k_l^{(D)}(r'',r';T)
  =\int\limits_{r(t')=r'}^{r(t'')=r''}\mu^{(D)}_l[r^2]\CD r(t)
  \exp\left\{\ih\int_{t'}^{t''}\bigg[ {m\over2}
  \dot r^2-V(r)\bigg]dt\right\}.
  \tag\NUM.\num$$
\item{1)} The whole $l$- and $D$-dependence is contained
in $\mu^{(D)}_l$, as defined in (\NUM.18):
\plus
$$\mu_l^{(D)}[z_{(j)}]=
  \sqrt{2\pi z_{(j)}}\,\e^{-z_{(j)}}I_{l+{D-2\over2}}(z_{(j)}).
  \tag\NUM.\num$$
Therefore we can conclude:
\plus
$$\mu^{(D)}_l[r^2]=\mu^{(3)}_{l+{D-3\over2}}[r^2]
  \tag\NUM.\num$$
and all dimensional dependence of the path integral (\NUM.31) can be
deduced from the three dimensional case:
\plus
$$k^{(D)}_l(r'',r';T)=k^{(3)}_{l+{D-3\over2}}(r'',r',T).
  \tag\NUM.\num$$
{}From now on we will denote $k_l(T)=k^{(3)}_l(T)$.
\item{2)} The boundary conditions we have already discussed;
we have for the radial wave functions $u^{(D)}_l$ and
$R^{(D)}_l$ respectively:
\plus
$$u^{(D)}_l(r)\to r^{l+{D-1\over2}},\qquad
  R^{(D)}_l(r)\to r^l\qquad(r\to0).
  \tag\NUM.\num$$
We therefore state:
The radial path integral $k_l(T)$ can be written as a superposition of
two one dimensional path integrals:
\plus
$$k_l(r'',r';T)=k_l^{(1)}(r'',r';T)-(-1)^l k_l^{(1)}(r'',-r';T)
  \tag\NUM.\num$$
with
\hfuzz=15pt
\plus
$$\multline
  k^{(1)}_l(r'',r';T)=\lim_{N\to\infty}\Norm^{N\over2}
   \int_{-\infty}^\infty dr_{(1)}\dots\int_{-\infty}^\infty dr_{(N-1)}
  \hfill\\  \hfill
  \times\prod_{j=1}^N \mu_l^{(D)}[r_{(j)}r_{(j-1)}]
  \exp\left\{\ih\sum_{j=1}^N\bigg[{m\over2\epsilon\hbar}
  \Delta^2r_{(j)}-\epsilon V(\vert r_{(j)}\vert )\bigg]\right\}.
  \endmultline
  \tag\NUM.\num$$

\noindent 4)
{}From equation (\NUM.36) for the case $l=V=0$ we have two free particle
path integrals the solution being:
\plus
$$\aligned
  k_0^{V=0}(r'',r';T)&=
 \bigg({m\over2\pi\i\hbar T}\bigg)^\half
  \bigg[\exp\bigg(-{\i m\over2T\hbar}(r''-r')^2\bigg)
       -\exp\bigg({\i m\over2T\hbar}(r''+r')^2\bigg)\bigg]\\
  &=\bigg({m\over2\pi\i\hbar T}\bigg)^\half
  \exp\bigg(-{\i m\over2T\hbar}(r''-r')^2\bigg)
  \bigg[ 1-\exp\bigg(-{\i m\over\hbar T}r'r''\bigg)\bigg],
  \endaligned
  \tag\NUM.\num$$
the second term is the so called ``mirror'' term. It is not allowed to
drop it, even for very small $T\to0$, since this violates the boundary
condition which demands that $k_0^{V=0}$ vanishes for $r',r''\to0$.
These two parts of the Bessel functions express nothing else than the
famous mirror principle. Actually, in the above Euclidean form
$k_0^{V=0}$ is nothing else  but the heat kernel on the half line
$r\geq0$, which has the representation:
\plus
$$K^{(3)}(r'',r';T)=K^{(1)}(r'',r';T)-K^{(1)}(r'',-r';T)
  \tag\NUM.\num$$
where $K^{(1)}$ denotes the one-dimensional heat kernel in {\bf R}. In
the language of diffusion processes: $r=0$ is an absorbing boundary and
the job is done by the functional measure $\mu_l^{(D)}$.

Let us discuss the radial path integral in the language of
the Weyl-ordering and product-ordering rule [\GRSb].
We consider the Schr\"odinger equation in $D$-dimensions with a
potential $V(\vert x\vert )=V(r)$:
\plus
$$\i\hbar{\partial\over\partial t}\Psi(r,\{\theta\};t)=\bigg
 [-{\hbar^2\over2m}\Delta_{LB}
  +V(r)\bigg]\Psi(r,\{\theta\};t)
  \tag\NUM.\num$$
with $\Delta_{LB}$ the Laplace Beltrami operator.
Introducing the $D$-dimensional Legendre operator:
\plus
$$\multline
  L^2_{(D)}=\bigg[
    {\partial^2\over\partial\theta_1^2}+(D-2)\cot\theta_1
    {\partial\over\partial\theta_1}\bigg]
    +{1\over\sin^2\theta_1}\bigg[
    {\partial^2\over\partial\theta_2^2}+(D-3)\cot\theta_2
    {\partial\over\partial\theta_2}\bigg]
    +\hbox{$\dots$}\hfill\\ \hfill
    +{1\over\sin^2\theta_1\dots\sin^2\theta_{D-3}}\bigg[
    {\partial^2\over\partial\theta_{D-2}^2}+\cot\theta_{D-2}
    {\partial\over\partial\theta_{D-2}}\bigg]
    +{1\over\sin^2\theta_1\dots\sin^2\theta_{D-2}}
    {\partial^2\over\partial\phi^2}.
  \\
  \endmultline
  \tag\NUM.\num$$
\hfuzz=3pt
the Laplace operator is:
\plus
$$\Delta_{LB}={\partial^2\over\partial r^2}+{D-1\over r}
         {\partial\over\partial r}+{1\over r^2}L^2_{(D)}.
  \tag\NUM.\num$$
Rewriting the Hamiltonian $H=-{\hbar^2\over2m}\Delta_{LB}$ yields:
\plus
$$\multline
  H(p_r,r,\{p_\theta,\theta\})={p_r^2\over2m}
  \\
  +{1\over 2mr^2}\bigg[
  p^2_{\theta_1}+{1\over\sin^2\theta_1}p^2_{\theta_2}+\hbox{$\dots$}
  +{1\over\sin^2\theta_1\dots\sin^2\theta_{D-2}}
  p^2_\phi\bigg]+\Delta V_{Weyl}(\{\theta\})
  \endmultline
  \tag\NUM.\num$$
with
\plus
$$\Delta V_{Weyl}(r,\{\theta\})=-{\hbar^2\over 8mr^2}
  \bigg[1+{1\over\sin^2\theta_1}+\hbox{$\dots$}+
  {1\over\sin^2\theta_1\dots
         \sin^2\theta_{D-2}}\bigg].
  \tag\NUM.\num$$
and the hermitean momenta
\plus
$$\left.\aligned
  p_r&={\hbar\over\i}\bigg({\partial\over\partial r}
        +{D-1\over2r}\bigg)\\
  p_{\theta_\nu}&={\hbar\over\i}\bigg({\partial\over\partial\theta_\nu}
        +{D-1-\nu\over2}\cot\theta_\nu\bigg)\\
  p_\phi&={\hbar\over\i}{\partial\over\partial\phi}.
  \endaligned\qquad\qquad\right\}
  \tag\NUM.\num$$
\edef\numCDxb{\NUM.\num}%
Because of the special nature of $g_{ab}$, the product ordering gives
the same quantum potential as the Weyl-ordering, i.e.\ we have
$\Delta V_{Weyl}=\Delta V_{prod}$. In particular we have the equivalence
\plus
$$\multline
  \prod_{j=1}^{N-1}\sqrt{g^{(j)}}\cdot
  \exp\left\{{\i\epsilon\over\hbar}\sum_{j=1}^N
  \CL^N_{Cl}(r^{(j)},\{\theta^{(j)}\},r^{(j-1)},
                \{\theta^{(j-1)}\})\right\}
  \hfill\\ \hfill\dot=
  (g'g'')^{-{1\over4}}\prod_{j=1}^N\sqrt{\bar{g}^{(j)}}\cdot
  \exp\left\{{\i\epsilon\over\hbar}
  \CL^N_{Cl}(\bar r^{(j)},\{\bar{\theta}^{(j)}\})\right\}
  \endmultline
  \tag\NUM.\num$$
where
\plus
$$\multline
  \CL^N_{Cl}(r^{(j)},\{\theta^{(j)}\},r^{(j-1)},\{\theta^{(j-1)}\})
  \\   \qquad
  ={m\over2\epsilon^2}\bigg\{(r^{(j)}-r^{(j-1)})^2+
  {\widehat{r^{(j)\,2}}}
  \bigg[(\theta_1^{(j)}-\theta_1^{(j-1)})^2
  +{\widehat{\sin^2}}\theta_1^{(j)}
   (\theta^{(j)}_2-\theta^{(j-1)}_2)^2+\hbox{$\dots$}
  \hfill\\
  \hbox{$\dots$}+({\widehat{\sin^2}}\theta_1^{(j)}\dots
  {\widehat{\sin^2}}\theta^{(j-1)}_{D-2})
  (\phi^{(j)}-\phi^{(j-1)})^2\bigg]\bigg\}
  \endmultline
  \tag\NUM.\num$$
denotes a ``classical Lagrangian'' on the lattice. $\CL^N_{Cl}(\bar
r^{(j)},\{{\bar\theta}^{(j)}\})$  is similar defined as (\NUM.47),
except that one has to take all trigonometrics at mid-points. The
features are, of course, direct consequences of the product form
definition as described in section II.3. With the correct Hamiltonian
(\NUM.43) we can consider the Hamiltonian path integral for the Feynman
kernel K:
\plus
$$\multline
  K^{(D)}(r'',\{\theta''\},r',\{\theta'\};T)
  \\  \quad
  =(g'g'')^{-{1\over4}}
  \int\limits_{\scriptstyle r(t')=r'\atop
                \scriptstyle\{\theta(t')=\theta'\}}
              ^{\scriptstyle r(t'')=r''\atop
                \scriptstyle\{\theta(t'')=\theta''\}}
  \CD r(t)\CD p_r(t)\CD\{\theta\}(t)\CD\{p_\theta\}(t)
  \hfill\\  \qquad\times
  \exp\left\{\ih\int^{t''}_{t'}\left[p_r\cdot\dot r+
  \sum_{\nu=1}^{D-1}p_{\theta_\nu}\cdot
  \dot\theta_\nu-\CH_{eff}(p_r,r,\{p_\theta\},\{\theta\})
                                               \right]dt\right\}
  \hfill\\  \quad=
  (g'g'')^{-{1\over4}}
  \lim_{N\to\infty}\prod_{j=1}^{N-1}
  \int dr_{(j)}d\{\theta_{(j)}\}
  \prod_{j=1}^N\int{dp_{r_{(j)}}d\{p_{\theta_{(j)}}\}\over(2\pi\hbar)^D}
  \hfill\\    \qquad\times
  \exp\left\{\ih\left[\sum_{j=1}^N
  p_{r_{(j)}}\Delta r_{(j)}+\{p_{\theta_{(j)}}\}\Delta\{\theta_{(j)}\}
  -\epsilon\CH_{eff}(p_{r_{(j)}},r_{(j)},
         \{p_{\theta_{(j)}}\},\{\theta_{(j)}\})\right]\right\}
  \hfill\\  \quad\endmultline
  \tag\NUM.\num$$
with the effective Hamiltonian
\plus
$$\multline
  \!\!\!\!
  \CH_{eff}(p_{r_{(j)}},r_{(j)},
           \{p_{\theta_{(j)}},\theta_{(j)}\})
  ={p_{r_{(j)}}^2\over2m}
  \\
  +{1\over 2m{\widehat{r^2_{(j)}}}}\left[
  p^2_{\theta_1^{(j)}}
  +{1\over{\widehat{\sin^2}}\theta_1^{(j)}}p^2_{\theta_2}+\dots
  +{1\over{\widehat{\sin^2}}\theta_1^{(j)}
    \dots{\widehat{\sin^2}}
    \theta_{D-2}^{(j)}}p^2_{\phi_{(j)}}\right]
  +\Delta V_{Prod}(\{\theta_{(j)}\})
  \\  \quad\endmultline
  \tag\NUM.\num$$
After the integration over all momenta we get ($\Delta V\equiv
\Delta V_{Weyl}=\Delta V_{prod}$):
\plus
$$\multline
  K^{(D)}(r'',r',\{\theta''\}, \{\theta'\}; T)
  \hfill\\  \qquad=
  \int\limits_{\scriptstyle r(t')=r'\atop
                \scriptstyle\{\theta(t')=\theta'\}}
              ^{\scriptstyle r(t'')=r''\atop
                \scriptstyle\{\theta(t'')=\theta''\}}
   \CD r(t)\,D\Omega(t)\exp\left\{\ih\int_{t'}^{t''}\big[
  \CL^N_{Cl}(r,\dot r,\{\theta,\dot\theta\})
  -\epsilon\Delta V(r,\{\theta\})\big]dt \right\},
  \hfill\\   \qquad=
  \lim_{N\to\infty}\Norm^{ND\over2}
  \prod_{j=1}^{N-1}\int_0^\infty dr^{D-1}_{(j)}dr_{(j)}
  \int d\Omega_{(j)}
  \hfill\\  \qquad\qquad\times
  \exp\left\{\ih\left[\sum_{j=1}^N
  \CL^N_{eff}(r_{(j)},\{\theta_{(j)}\},r_{(j-1)},\{\theta_{(j-1)}\})
  -\epsilon\Delta V(r_{(j)},\{\theta_{(j)}\})\right]\right\}
  \hfill\\ \ \endmultline
  \tag\NUM.\num$$
with the effective Lagrangian
\plus
$$\multline
  \CL_{eff}(r,\dot r,\{\theta,\dot\theta\})\hfill\\
  \hfill    ={m\over2}[\dot r^2+r^2
  \dot\theta_1^2+r^2\sin^2\theta_1\dot\theta^2_2+\dots
  +r^2(\sin^2\theta_1\dots\sin^2\theta_{D-2})\dot\phi^2]
   -V(r,\{\theta\})
  \\  \ \endmultline
  \tag\NUM.\num$$
defined in the same way as $\CL^N_{Cl}$ in equation (\NUM.47).

The path integral (\NUM.50) with the Lagrangian given by equation
(\NUM.47) is too complicated for explicit calculations. We therefore
try to replace equation (\NUM.47) under the path integral (\NUM.50) by
the following expression:
\plus
$$\CL_{Cl}(r,\{\theta\},\dot r,\{\dot\theta\})\to
\tilde\CL_{Cl}(r,\{\theta\},\dot r,\{\dot\theta\}):=
{m\over2}R^2\dot\Omega^2-V_c(\{\theta\})
  \tag\NUM.\num$$
where $V_c$ has to be determined and
$\Omega$ denotes the $D$-dimensional unit vector on the $S^{D-1}$-%
sphere.
Thus we try to replace $\CL_{Cl}$ by a simpler expression and hope
that $V_c+\Delta V_{Weyl}$ is simple enough. We have
\plus
$$(\Omega^{(1)}-\Omega^{(2)})^2=2(1-\cos\psi^{(1,2)})
  \tag\NUM.\num$$
with the addition theorem (\NUM.3). We shall use equation (\NUM.53) to
justify the replacement (\NUM.52) and thereby derive an expression for
$V_c$. We start with the kinetic term $(x^{(j)}-x^{(j-1)})^2$ expressed
in the polar coordinates (\NUM.2), $R=r$ (not fixed), and expand it in
terms of $\Delta r$ and $\Delta\theta_\nu$.
After some tedious calculations we obtain the following identity
\plus
$$\multline
  \!\!\!\!
  \exp\left[{\i\epsilon\over\hbar}\CL^N_{Cl}
  (r^{(j)},\{\theta^{(j)}\},r^{(j-1)},\{\theta^{(j-1)}\})\right]
  \\
  \dot=\exp\bigg[{\i m\over\epsilon\hbar}
  \big(r^{(j)\,2}+r^{(j-1)\,2}-2r^{(j)}r^{(j-1)}\cos\psi^{(j,j-1)}\big)
  -\i\epsilon V_c(r^{(j)},\{\theta^{(j)}\})\bigg]
  \endmultline
  \tag\NUM.\num$$
with
\plus
$$V_c(r^{(j)},\{\theta^{(j)}\})
  ={\hbar^2\over 8mr^{(j)\,2}}
  \bigg[1+{1\over\sin^2\theta_1^{(j)}}+\dots+
  {1\over\sin^2\theta^{(j)}_1\dots\sin^2\theta^{(j)}_{D-2}}\bigg]
  \tag\NUM.\num$$
($V_c$ is the same whether or not $\Delta r^{(j)}=0$).
The result is that the potential $V_c$ generated by these steps
cancels exactly $\Delta V_{Weyl}(r,\{\theta\})$!
Therefore we get:
\plus
$$K^{(D)}(r'',r',\{\theta''\}, \{\theta'\}; T)=
  \int\limits_{\scriptstyle r(t')=r'\atop
                \scriptstyle\{\theta(t')=\theta\}}
              ^{\scriptstyle r(t'')=r''\atop
                \scriptstyle\{\theta(t'')=\theta''\}}
  \CD r(t)\,\CD\Omega(t)\exp\left\{\ih\int_{t'}^{t''}\bigg[
  {m\over2}\dot x^2-V(r)\bigg]dt\right\},
  \tag\NUM.\num$$
where  $\dot x^2$ has to be expressed in polar coordinates.
In the lattice formulation $\dot x^2$ reads
\plus
$$\dot x^2\to[r^2_{(j)}+r^2_{(j-1)}
  -2r_{(j)}r_{(j-1)}\cos\psi_{(j,j-1)}]/\epsilon^2.
  \tag\NUM.\num$$
Therefore we arrive at equation (\NUM.52) and both approaches are
equivalent as it should be.

Let us note that equation (\NUM.54) as derived in [\GRSb]  is
effectively a regularization of the highly singular terms in $\Delta V$.
This corresponds to the $1/r^2$-terms in the radial path integral
which have been regularized by the functional measure $\mu_l[r^2]$.
However, the functional measure method gives a quite practical tool
to regularize such singular terms.

\bigskip\bigskip

\subsection{The Radial Harmonic Oscillator}
Let us now discuss the most important application of equation (\NUM.17),
namely the harmonic oscillator with $V(r)=\half m\omega^2r^2$.
The calculation has first been performed by Peak and Inomata [\PI].
However, we present the more general case with
time-dependent coefficients following Goovaerts [\GOOb].
This example will be of great virtue in the solution of
various path integral problems.

\noindent
We have to study
\plus
$$\multline
   K_l(r'',r';\tau)
  \\    \quad=
  (r'r'')^{1-D\over2}\lim_{N\to\infty}\Norm^{N/2}
  \int_0^\infty dr_{(1)}\dots\int_0^\infty dr_{(N-1)}
  \hfill\\  \qquad\times
  \prod_{j=1}^N\left\{\mu_l^{(D)}[r_{(j)}r_{(j-1)}]\cdot
  \exp\left[{\i m\over2\epsilon\hbar}(r_{(j)}-r_{(j-1)})^2
  -{\i\epsilon\over2\hbar}
  m\omega^2(t^{(j)})r_{(j)}^2\right]\right\}
  \hfill\\    \quad=
  (r'r'')^{2-D\over2}\lim_{N\to\infty}
  \bigg({m\over \i\epsilon\hbar}\bigg)^{N/2}
  \int_0^\infty r_{(1)}r_{(1)}\dots\int_0^\infty r_{(N-1)}dr_{(N-1)}
  \hfill\\  \qquad\times
  \prod_{j=1}^N\left\{\exp\left[{\i m\over2\epsilon\hbar}
  (r_{(j)}^2+r_{(j-1)}^2)-{\i\epsilon\over2\hbar}
  m\omega^2(t^{(j)})r_{(j)}^2\right]
  \cdot I_{l+{D-2\over2}}\bigg({m\over \i\epsilon\hbar}
  r_{(j)}r_{(j-1)}\bigg)\right\}
  \hfill\\   \quad=
  (r'r'')^{2-D\over2}\lim_{N\to\infty}
  \bigg({\alpha\over\i}\bigg)^{N/2}\e^{\i\alpha(r^{'2}+r^{''2})/2}
  \int_0^\infty r_{(1)}dr_{(1)}\dots\int_0^\infty r_{(N-1)}dr_{(N-1)}
  \hfill\\  \qquad\times
  \exp\Big[i(\beta_{(1)}r_{(1)}^2
  +\beta_{(2)}r_{(2)}^2+\dots+\beta_{(N-1)}r_{(N-1)}^2)\Big]
  \hfill\\  \qquad\times
  \bigg[I_{l+{D-2\over2}}(-\i\alpha r_{(0)}r_{(1)})\dots
  I_{l+{D-2\over2}}(-\i\alpha r_{(N-1)}r_{(N)})\bigg]
  \hfill\\    \quad=
  (r'r'')^{2-D\over2}\lim_{N\to\infty}K_l^N(T)
  \hfill\endmultline
  \tag\NUM.\num$$
where $K_l^N(T)$ is defined by the iterated integrals.
Furthermore we have set $\alpha=m/\epsilon\hbar$ and
$\beta_{(j)}=\alpha[1-\epsilon^2m\omega^2(t^{(j)})/2]$.
We are now using [\GRA,\ p.718]
\plus
$$\int_0^\infty x\e^{-\gamma x^2}
               J_\nu(\alpha x)J_\nu(\beta x) dx
  ={1\over2\gamma}\e^{-(\alpha^2+\beta^2)/4\gamma}
                  I_\nu\bigg({\alpha\beta\over2\gamma}\bigg)
  \tag\NUM.\num$$
which is valid for $\Re(\nu)>-1$, $\vert\arg\sqrt{\gamma}\vert <\pi/4$
and $\alpha,\beta>0$. With analytic continuation [\PI] one can show that
\plus
$$\int_0^\infty r\e^{\i\alpha r^2}
         I_\nu(-\i ar)I_\nu(-\i br)dr
         ={\i\over2\alpha}\e^{(a^2+b^2)/4\alpha\i}
          I_\nu\bigg({ab\over2\alpha\i}\bigg)
  \tag\NUM.\num$$
is valid for $\nu>-1$ and $\Re(\alpha)>0$.
By means of equation (\NUM.60) we obtain for $K_l^N(T)$:
\plus
$$\multline
  K_l^N(T)=\bigg({\alpha\over\i}\bigg)^{N\over2}
  \exp\bigg({\i\beta\over2}({r'}^2+{r'}^2)\bigg)
  \int_0^\infty r_{(1)}dr_{(1)}\dots\int_0^\infty r_{(N-1)}dr_{(N-1)}
  \hfill\\  \qquad\qquad\times
  \exp\Big[i(\beta_{(1)}r_{(1)}^2
  +\beta_{(2)}r_{(2)}^2+\dots+\beta_{(N-1)}r_{(N-1)}^2)\Big]
  \hfill\\  \qquad\qquad\times
  \bigg[I_{l+{D-2\over2}}(-\i\alpha r_{(0)}r_{(1)})\dots
  I_{l+{D-2\over2}}(-\i\alpha r_{(N-1)}r_{(N)})\bigg]
  \\  \hphantom{K_N}=
  {\alpha_N\over\i}
  \exp\big(\i p_N{r'}^2+\i q_N{r'}^2\big)
  I_{l+{D-2\over2}}(-\i\alpha_N r'r'')
  \hfill\endmultline
  \tag\NUM.\num$$
where the coefficients $\alpha_N,p_N$ and $q_N$ are given by
\plus
$$\left.\aligned
  \alpha_N&=\alpha\prod_{k=1}^{N-1}{\alpha\over2\gamma_k}
  \\
  p_N     &={\alpha\over2}-\sum_{k=1}^{N-1}{\alpha_k^2\over4\gamma_k}
  \\
  q_N     &={\alpha\over2}-{\alpha^2\over4\gamma_{N-1}}
  \\
  \alpha_1&=\alpha,\qquad
  \alpha_{k+1}=\alpha\prod_{j=1}^k{\alpha\over2\gamma_k}
  \qquad(k\geq1)
  \\
  \gamma_1&=\beta_1,\qquad
  \gamma_{k+1}=\beta_{k+1}-{\alpha^2\over4\gamma_k}.
  \endaligned\qquad\qquad\qquad\right\}
  \tag\NUM.\num$$
We must now determine these quantities.
Let us start with the evaluation of $\gamma_k$. Putting
\plus
$${2\gamma_k\over\alpha}={y_{k+1}\over y_k}
  \tag\NUM.\num$$
we obtain
\plus
$$\gamma_{k+1}=\beta_{k+1}-{\alpha^2\over4\gamma_k}
  \qquad\iff\qquad
  {y_{k+1}-2y_k+y_{k-1}\over\epsilon^2}+\omega_k^2y_k=0.
  \tag\NUM.\num$$
In the limit $N\to\infty$ this gives a differential equation for $y$
$$\ddot y+\omega^2(t)y=0$$
with solution $y=\eta(t'+t)\equiv\eta(t)$. Since on one hand side
$$\aligned
  y_1&\simeq y_0+\epsilon \dot y_0\to y_0\qquad(\epsilon\to0),
  \\
  y_1&\simeq {2\over\alpha}y_0\gamma_1\to2y_0,\qquad(\epsilon\to0),
  \endaligned$$
and on the other
$$\dot y_0={y_1-y_0\over\epsilon}={1\over\epsilon}
  \bigg({2\over\alpha}\gamma_1-1\bigg)y_0\to1,\qquad(\epsilon\to1),
$$
we have the boundary conditions
\plus
$$\eta(0)=0,\qquad\dot\eta(0)=1.
  \tag\NUM.\num$$
Now let us expand $y_{k+1}\simeq\eta[
    (k+1)\epsilon]+O(\epsilon^3)$, we
observe that
$$y_{k+1}-2y_k+y_{k-1}+\epsilon^2\omega^2(t'+k\epsilon)y_k=0$$
is satisfied up to second order in $\epsilon$.
Therefore expanding (\NUM.64)
$$\gamma_{k+1}={\alpha\over2}{y_{k+2}\over y_{k+1}}
  ={\alpha\over2}{\eta[(k+2)\epsilon]+O(\epsilon^3)\over
   \eta[(k+1)\epsilon]+O(\epsilon^3)}.$$
Consequently
\plus
$$\allowdisplaybreaks\align
  \lim_{N\to\infty}\alpha_N
  &=\lim_{N\to\infty}{m\over\epsilon\hbar}\prod_{j=1}^{N-1}
  {\eta(j\epsilon)+O(\epsilon^3)\over
   \eta[(j+1)\epsilon]+O(\epsilon^3)}
  \\   &
  =\lim_{N\to\infty}{m\over\epsilon\hbar}
   {\eta(0)\over\eta(\epsilon N)}={m\over\hbar\eta(T)}.
  \tag\NUM.\num\endalign$$
Similarly
\plus
$$\allowdisplaybreaks\align
  \lim_{N\to\infty} q_N
  &=\lim_{N\to\infty}{m\over2\epsilon\hbar}
  \left(1-{\eta[(N-1)\epsilon]+O(\epsilon^3)\over
           \eta[\epsilon N]+O(\epsilon^3)}\right)
  \\   &
  =\lim_{N\to\infty}{m\over2\hbar}
  {\eta[N\epsilon]
  -\eta[t'+(N-1)\epsilon]\over
    \epsilon\eta[N\epsilon]}
  ={m\dot\eta(T)\over2\hbar\eta(T)}.
  \tag\NUM.\num\endalign$$
Finally we must calculates $p_N$. We have
\plus
$$\allowdisplaybreaks\align
  \lim_{N\to\infty}p_N
  &=\lim_{N\to\infty}
  \bigg({\alpha\over2}-\sum_{k=1}^{N-1}{\alpha_k^2\over4\gamma_k}\bigg)
  \\   &
  ={m\over2\hbar} \lim_{N\to\infty}
  \bigg({1\over\epsilon}-\sum_{k=1}^{N-1}
  {\epsilon\over\eta^2[(k+1)\epsilon]}\bigg)
  \\   &
  ={m\over2\hbar} \lim_{N\to\infty}
  \bigg({1\over\epsilon}-\int_\epsilon^T{dt\over\eta^2(t)}\bigg)
  \\   &
  ={m\over2\hbar\eta(T)} \lim_{N\to\infty}
  \bigg[{\eta(T)\over\epsilon}
   -\eta(T)\int_\epsilon^T{dt\over\eta^2(t)}\bigg]
  \equiv{m\over2\hbar\eta(T)}\xi(T).
  \tag\NUM.\num\endalign$$
Furthermore we find
$$\aligned
  \lim_{N\to\infty}\xi(\epsilon)
  &=\lim_{N\to\infty}{\eta(\epsilon)\over\epsilon}-\eta(\epsilon)
   \int_\epsilon^\epsilon{dt\over\eta^2(t)}
  \\
  \lim_{\epsilon\to0}{\eta(0)+\epsilon\dot\eta(0)\over\epsilon}
  &=\dot\eta(0)=1
  \\
  \lim_{\epsilon\to0}\dot\xi(\epsilon)
  &=\lim_{\epsilon\to0}\bigg({\dot\eta(\epsilon)\over\epsilon}
    -{1\over\eta(\epsilon)}\bigg)
  \\     &
  =\lim_{\epsilon\to0}
  {\dot\eta(0)[\eta(0)+\epsilon\dot\eta(0)]-\epsilon\over
   \epsilon[\eta(0)+\epsilon\dot\eta(0)]}=0,
  \endaligned$$
and therefore $\xi(t)$ is satisfying the boundary conditions
\plus
$$\xi(0)=1,\qquad\dot\xi(0)=0.
  \tag\NUM.\num$$
$\xi(t)$ is found to satisfy the differential equation
\plus
$$\ddot\xi+\omega^2(t)\xi=0.
  \tag\NUM.\num$$
Therefore we have
\plus
$$P(T)=\lim_{N\to\infty}p_N={m\over2\hbar}{\xi(T)\over\eta(T)}
  \tag\NUM.\num$$
and we have finally for the path integral of the radial harmonic
oscillator with time-dependent frequency:
\plus
$$\multline
  K_l(r'',r';T)
  \\
  =(r'r'')^{2-D\over2}{m\over\i\hbar\eta(T)}
  \exp\bigg[{\i m\over2\hbar} \bigg({\xi(T)\over\eta(T)}{r'}^2
  +{\dot\eta(T)\over\eta(T)}{r'}^2\bigg)\bigg]
  I_{l+{D-2\over2}}\bigg({mr'r''\over\i\hbar\eta(T)}\bigg).
  \endmultline
  \tag\NUM.\num$$
In particular for $\omega(t)=\omega=const.$:
$$\aligned
  \eta(t)&={1\over\omega}\sin\omega(t-t'),
  \qquad \dot\eta(t)=\cos\omega(t-t')
  \\
  \xi(t)&=\cos\omega(t-t')
  \endaligned$$
which yields the radial path integral solution for the radial
harmonic oscillator with time-independent frequency
\plus
$$\multline
  K_l(r'',r';T)
  \\
  =(r'r'')^{2-D\over2}{m\omega\over\i\hbar\sin\omega T}
  \exp\bigg[
  {\i m\omega\over2\hbar}({r'}^2+{r'}^2)\cot\omega T\bigg]
  I_{l+{D-2\over2}}\bigg({m\omega r'r''\over\i\hbar\sin\omega T}\bigg).
  \endmultline
  \tag\NUM.\num$$
It is possible to get the same result, if one starts with the
$D$-dimensional path integral in Cartesian coordinates:
\plus
$$K(x'',x';\tau)=\int\limits_{x(t')=x'}^{x(t'')=x''} \CD x(t)
  \exp\left\{{\i m\over2\hbar}\int_{t'}^{t''}
  \big[\dot x^2-\omega^2 x^2\big]dt\right\}
  \tag\NUM.\num$$
insert for every dimension the solution of the one-dimensional
oscillator
\plus
$$K(x''_k,x'_k;T)=\sqrt{m\omega\over2\pi i\hbar\sin\omega T}
  \exp\bigg\{
  {\i m\omega\over2\hbar}\bigg[({x'_k}^2+{x_k'}^2)\cot\omega T-
      {2x_k'x_k''\over\sin\omega T}\bigg]\bigg\}
  \tag\NUM.\num$$
and uses equations (\NUM.8) and (\NUM.11).

The next step is to calculate with the help of equation (\NUM.73)
the energy-levels and state-functions.
For this purpose we use the Hille-Hardy-formula [\GRA, p.1038]
\plus
$${t^{-\alpha/2}\over 1-t}
  \exp\bigg[ -{1\over2}(x+y){1+t\over 1-t}\bigg]
  I_\alpha\bigg({2\sqrt{xyt}\over 1-t}\bigg)
  =\sum_{n=0}^\infty
  {t^n n! \e^{-\half(x+y)}\over\Gamma(n+\alpha+1)}
  (xy)^{\alpha/2}L_n^{(\alpha)}(x) L_n^{(\alpha)}(y).
  \tag\NUM.\num$$
With the substitution $t=\e^{-2\i\omega T}$, $x=m\omega{r'}^2/\hbar$ and
$y=m\omega {r'}^2/\hbar$ in equation (\NUM.73) we get finally:
\advance\glno by -1
$$\allowdisplaybreaks\align
  \!\!\!\!
  &K_l(r'',r';T)=\sum_{N=0}^\infty
                  \e^{-\i TE_N/\hbar}R_N^l(r') R_N^l(r'')
  \tag\NUM.\num\\   \global\plus
  &E_N=\omega\hbar\bigg(N+{D\over2}\bigg)
  \tag\NUM.\num\\   \global\plus
  &R_N^l(r)=\sqrt{ {2m\omega\over\hbar r^{D-2}}\cdot
    {\Gamma({N-l\over2}+1)\over\Gamma({N+l+D\over2})}}\,
    \bigg({m\omega\over\hbar} r^2\bigg)^{l+{D-2\over2}}
    \exp\bigg(-{m\omega\over\hbar} r^2\bigg)
    L_{N-l\over2}^{(l+{D-2\over2})}\bigg({m\omega\over\hbar} r^2\bigg).
  \\    &\quad
  \tag\NUM.\num\endalign$$
\hfuzz=3pt
\goodbreak

The path integral for the harmonic oscillator suggests a
generalization in the index $l$.
This will be very important in further applications.
For this purpose we consider equation (\NUM.17) with the nontrivial
functional measure $\mu_l^{(D)}(r^2)\equiv\mu_l[r^2]$:
\plus
$$K_l(r'',r';T)
  ={1\over r'r''}\int\limits_{r(t')=r'}^{r(t'')=r''}
   \CD r(t)\mu_l[r^2]
   \exp\left\{\ih\int_{t'}^{t''}\bigg[{m\over2}\dot r^2-V(r)
   \bigg]dt\right\}
  \tag\NUM.\num$$
The functional measure corresponds to a potential
$V_l=\hbar^2{l(l+1)\over2mr^2}$ in the Schr\"odinger equation.
Assuming that we can analytically continue in $l\to\lambda$ with
$\Re(\lambda)>-1$, then we get for an arbitrary potential
$V_\lambda(r)=\hbar^2{\lambda^2-{1\over4}\over2mr^2}$
na\"\ii vely inserted into the radial path integral
\plus
$$\int\limits_{r(t')=r'}^{r(t'')=r''}
  \CD r(t)\exp\left\{\ih\int_{t'}^{t''}\left[
  {m\over2}\dot r^2-\hbar^2{\lambda^2-{1\over4}\over2mr^2}
              -{m\over2}\omega^2r^2\right]dt\right\}.
  \tag\NUM.\num$$
This path integral must now be interpreted in terms of the functional
measure in equation (\NUM.18).
Define ($z\equiv mr_{(j-1)}r_{(j)}/\i\epsilon\hbar$):
\plus
$$\mu_\lambda[r^2]=\lim_{N\to\infty}\prod_{j=1}^N
                   \sqrt{2\pi z_{(j)}}\,\e^{-z_{(j)}}I_\lambda(z_{(j)}).
  \tag\NUM.\num$$
Then equation (\NUM.81) {\bf must be interpreted} with the help of
equation (\NUM.82) as
\plus
$$\multline
  \int\limits_{r(t')=r'}^{r(t'')=r''}
  \CD r(t)\exp\left\{\ih\int_{t'}^{t''}\left[
  {m\over2}\dot r^2-\hbar^2{\lambda^2-{1\over4}\over2mr^2}
              -{m\over2}\omega^2r^2\right]dt\right\}
  \\        \qquad
  :=\int\limits_{r(t')=r'}^{r(t'')=r''} \CD r(t)\mu_\lambda[r^2]
  \exp\left[{\i m\over2\hbar}\int_{t'}^{t''}
            \big(\dot r^2-\omega^2r^2\big)dt\right]
  \hfill\\  \qquad
  ={\sqrt{r'r''} m\omega\over\i\hbar\sin\omega T}
   \exp\bigg[{\i m\omega\over2\hbar}
             \big({r'}^2+{r'}^2\big)\cot\omega T\bigg]
   I_\lambda\bigg({m\omega r'r''\over\i\hbar\sin\omega T}\bigg).
  \hfill\endmultline
  \tag\NUM.\num$$
This important equation has been derived by Peak and Inomata [\PI]
with the help of the three-dimensional radial harmonic oscillator,
and by Duru [\DURd] with the corresponding two-dimensional case.
Duru considered the radial potential problem
\plus
$$V(r)={a\over r^2}+br^2
  \tag\NUM.\num$$
with some numbers $a$ and $b$. Identifying $a=(\hbar^2/2m)(\lambda^2
-{1\over4})$ and $b=\half m\omega^2$ it is simple calculation to express
equation (\NUM.83) in terms of $a$ and $b$, which is omitted here.
However, these authors did not discuss this path integral identity in
terms of the functional measure language. That such a procedure is
actually legitimate is beyond the scope of these notes. It was
justified by Fischer, Leschke and M\"uller [\FLM] for the radial path
integral and for the P\"oschl-Teller path integral as well (see below),
where one also has to be careful with the appropriate Besselian
functional measure. In the functional measure interpretation equation
(\NUM.83) can be found in references [\GRSb] and [\STEc].
Equation (\NUM.83) is very important in numerous applications.

Let us note the free particle case. In the limit $\omega\to0$ we
obtain in equation (\NUM.73)
\plus
$$\multline
  K_l^{(\omega=0)}(r'',r';T)
  \\    \qquad=
  (r'r'')^{2-D\over2}{m\over\i\hbar T}
  \exp\bigg[{\i m\over2\hbar T}({r'}^2+{r'}^2)\bigg]
  I_{l+{D-2\over2}}\bigg({mr'r''\over\i\hbar T}\bigg)
  \hfill\\   \qquad=
  (r'r'')^{2-D\over2}\int_0^\infty dp\,p
  \exp\bigg(-{\i\hbar Tp^2\over2m}\bigg)
  J_{l+{D-2\over2}}(pr') J_{l+{D-2\over2}}(pr'')
  \hfill\endmultline
  \tag\NUM.\num$$
with wave-functions and energy spectrum
\plus
$$\Psi_p(r)=r^{2-D\over2}\sqrt{p}\,J_{l+{D-2\over2}}(pr)
  ,\qquad E_p={\hbar^2p^2\over2m}.
  \tag\NUM.\num$$
The one-dimensional case gives (i.e.\ motion on the half-line)
\plus
$$\Psi_p(r)=r\sqrt{p}\,J_\half(pr)=\sqrt{2\over\pi r}\sin pr
  ,\qquad E_p={\hbar^2p^2\over2m}.
  \tag\NUM.\num$$
However, there is an ambiguity in the boundary condition for $r=0$ for
$D=1$. The present case here, i.e.\ $\Psi_p(0)=0$, corresponds
to a specific self adjoint extension of the Hamiltonian
$H=-\hbar^2 d^2/dx^2$ for functions
$\Psi\in L^2([0,\infty))$ on the
half-line.

Finally we calculate the energy dependent Green function for the
radial harmonic oscillator. Making use of the integral representation
of the previous section we obtain
\plus
$$\multline
  G_l(r'',r';E)
  \\  \qquad
  =\i\int_0^\infty \e^{\i ET/\hbar}K_l(r'',r';T)dt
  \hfill\\  \qquad
  ={\Gamma[\half(l+{D\over2}-{E\over\hbar\omega})]\over
           \omega(r'r'')^{D/2}\Gamma(l+{D\over2})}
  W_{{E\over2\hbar\omega},\half(l+{D-2\over2})}
                     \bigg({m\omega\over\hbar}r^2_>\bigg)
  M_{{E\over2\hbar\omega},\half(l+{D-2\over2})}
           \bigg({m\omega\over\hbar}r^2_<\bigg).\qquad
  \hfill\endmultline
  \tag\NUM.\num$$
{}From this representation we can recover by means of the expansion for
the $\Gamma$-function the wave-functions of equation (\NUM.79).

The corresponding Green function for the free particle has the form
$$\allowdisplaybreaks\align
  &G_l^{(\omega=0)}(r'',r';E)
  \\
  &={2m\over\hbar(r'r'')^{D-2\over2}}
  \\  &\qquad\times
  I_{l+{D-2\over2}}\bigg[{1\over\i}\sqrt{mE\over2\hbar^2}\,
    (r'+r''-\vert r''-r'\vert )\bigg]
  K_{l+{D-2\over2}}\bigg[{1\over\i}\sqrt{mE\over2\hbar^2}\,
    (r'+r''+\vert r''-r'\vert )\bigg]
  \\   &\quad
  \tag\NUM.\num\\   \global\plus
  &={\i\pi m\over\hbar(r'r'')^{D-2\over2}}
  \\  &\qquad\times
  J_{l+{D-2\over2}}\bigg[\sqrt{mE\over2\hbar^2}\,
    (r'+r''-\vert r''-r'\vert )\bigg]
  H^{(1)}_{l+{D-2\over2}}\bigg[\sqrt{mE\over2\hbar^2}\,
    (r'+r''+\vert r''-r'\vert )\bigg],
  \\   &\quad
  \tag\NUM.\num\endalign$$
\hfuzz=3pt
where use has been made of the integral representation
\plus
$$\int_0^\infty{dx\over x}
  \exp\bigg(-px-{a+b\over2x}\bigg)
  I_\nu\bigg({a-b\over2x}\bigg)
  =2I_\nu\Big[\sqrt{p}\,\big(\sqrt{a}-\sqrt{b}\,\big)\Big]
    K_\nu\Big[\sqrt{p}\,\big(\sqrt{a}+\sqrt{b}\,\big)\Big].
  \tag\NUM.\num$$

\glno=0               
\advance\chapno by 1  

\section{Other Elementary Path Integrals}
There are two further path integral solutions based on the $\SU(2)$
[\BJb, \DURb, \INOWI] and $\SU(1,1)$ [\BJb, \BJa] group path
integration, respectively. The first yields the path integral identity
for the solution of the P\"oschl-Teller potential according to
\plus
$$\multline
  K^{(PT)}(x'',x';T)
  \\    \qquad=
  \int\limits_{x(t')=x'}^{x(t'')=x''}\CD x(t)
  \exp\left\{\ih\int_{t'}^{t''}\left[{m\over2}\dot x^2
        -{\hbar^2\over2m}\bigg(
  {\alpha^2-{1\over4}\over\sin^2x}+{\beta^2-{1\over4}\over\cos^2x}\bigg)
  \right]dt\right\}
  \hfill\\   \qquad=
  \sum_{l=0}^\infty\exp\bigg[-{\i\hbar
  T\over2m}(\alpha+\beta+2l+1)^2\bigg]
  \Psi^{(\alpha,\beta)\,*}_l(x')\Psi^{(\alpha,\beta)}_l(x'')
  \hfill\\   \qquad=
  \sqrt{\sin2x'\sin2x''}\sum_{J=0,\half}^\infty
  \exp\bigg[-{\i\hbar T\over2m}(2J+1)^2\bigg]
  \hfill\\  \qquad\qquad\times
  (2J+1)
  D^{J\,*}_{{\alpha+\beta\over2},{\beta-\alpha\over2}}(\cos2x')
  D^J_{{\alpha+\beta\over2},{\beta-\alpha\over2}}(\cos2x'')
  \endmultline
  \tag\NUM.\num$$
with the wave-functions given by
\plus
$$\allowdisplaybreaks\align
  \Psi_n^{(\alpha,\beta)}(x)
  &=
  \sqrt{(2J+1)\sin2x}\,
  D^J_{{\alpha+\beta\over2},{\beta-\alpha\over2}}(\cos2x)
  \tag\NUM.\num a\\
  &=\bigg[2(\alpha+\beta+2l+1)
  {l!\Gamma(\alpha+\beta+l+1)\over\Gamma(\alpha+l+1)\Gamma(\beta+l+1)}
  \bigg]^\half
  \\
  &\qquad\qquad\times\phantom{\bigg]}
  (\sin x)^{\alpha+\half}(\cos x)^{\beta+\half}
  P_n^{(\alpha,\beta)}(\cos2x).
  \tag\NUM.\num b\\
  \phantom{\Psi_n^{(\alpha,\beta)}(x)}
  &=\bigg[{\alpha+\beta+2l+1\over2^{\alpha+\beta+1}}
  {l!\Gamma(\alpha+\beta+l+1)\over\Gamma(\alpha+l+1)\Gamma(\beta+l+1)}
  \bigg]^\half
  \\
  &\qquad\qquad\times\phantom{\bigg]}
  \sqrt{2\sin2x}\, (1-\cos2x)^{\alpha\over2} (1+\cos2x)^{\beta\over2}
  P_n^{(\alpha,\beta)}(\cos2x).
  \tag\NUM.\num c
  \endalign$$
Here, of course we can analytically continue from integer values of $m$
and $n$ to, say, real numbers $\alpha$ and $\beta$, respectively .

Similarly we can state a path integral identity for the
modified P\"osch-Teller potential which is defined as
\plus
$$V^{(\eta,\nu)}(r)={\hbar^2\over2m}
   \bigg({\eta^2-{1\over4}\over\sinh^2r}
   -{\nu^2-{1\over4}\over\cosh^2r}\bigg).
  \tag\NUM.\num$$
This can be achieved by means of the path integration of the $\SU(1,1)$
group manifold. One gets
\plus
$$\allowdisplaybreaks\align
  K^{(MPT)}(r'',r';T)
  &=\sum_{n=0}^{N_M}\Phi_n^{(\eta,\nu)\,*}(r')
  \Phi_n^{(\eta,\nu)}(r'')
  \exp\bigg\{
  -{\i\hbar T\over2m}\Big[2(k_1-k_2-n)-1\Big]^2\bigg\}
  \\
  &+\int_0^\infty dp\,\Phi_p^{(\eta,\nu)\,*}(r')
  \Phi_p^{(\eta,\nu)}(r'') \exp\bigg(-{\i\hbar T\over2m}p^2\bigg).
  \tag\NUM.\num
  \endalign$$
\goodbreak\noindent
Introduce the numbers $k_1,k_2$ defined by:
$k_1=\half(1\pm\nu)$, $k_2=\half(1\pm\eta)$, where the correct sign
depends on the boundary conditions for $r\to0$ and $r\to\infty$,
respectively. In particular for $\eta^2={1\over4}$,
i.e.\ $k_2={1\over4},{3\over4}$, we obtain wave-functions with
even and odd  parity, respectively.
The number $N_M$ denotes the maximal number of states with
$0,1,\dots,N_M<k_1-k_2-\half$.
The bound state wave-functions read as:
\plus
$$\left.\aligned
  \Phi_n^{(k_1,k_2)}(r)
  &=N_n^{(k_1,k_2)}(\sinh r)^{2k_2-\half}
                    (\cosh r)^{-2k_1+{3\over2 }}
  \\  &\qquad\times
  {_2}F_1(-k_1+k_2+\kappa,-k_1+k_2-\kappa+1;2k_2;-\sinh^2r)
  \\
  N_n^{(k_1,k_2)}
  &={1\over\Gamma(2k_2)}
  \bigg[{2(2\kappa-1)\Gamma(k_1+k_2-\kappa)
                     \Gamma(k_1+k_2+\kappa-1)\over
    \Gamma(k_1-k_2+\kappa)\Gamma(k_1-k_2-\kappa+1)}\bigg]^\half
  \endaligned\qquad\right\}
  \tag\NUM.\num$$
($\kappa=k_1-k_2-n$), (there is a factor ``$2$'' missing in
reference [\FW]).
Note the equivalent formulation
\plus
$$\tilde\Phi_n^{(\alpha,\beta)}(r)
  =\bigg[{\beta n!\Gamma(n+\alpha+\beta+1)\over
  2^{\alpha+\beta}\Gamma(n+\alpha+1)\Gamma(n+\beta+1)}\bigg]^\half
  (1-x)^{\alpha\over2}(1+x)^{\beta-1\over2}
  P_n^{(\alpha,\beta)}(x),
  \tag\NUM.\num$$
with the substitutions $\alpha=2k_2-1$, $\beta=2(k_1-k_2-n)-1
=2\kappa-1$, $x=2/\cosh^2r-1$ with the incorporation of the
appropriate measure term, i.e.\ $dr=\left[(1+x)
\sqrt{2(1-x)}\right]^{-1}dx$.
The scattering states are given by:
\plus
$$\left.\aligned
  \Phi_p^{(k_1,k_2)}(r)
  &=N_p^{(k_1,k_2)}(\cosh r)^{2k_1-\half}(\sinh r)^{2k_2-\half}
  \\   &\qquad\qquad\times
  {_2}F_1(k_1+k_2-\kappa,k_1+k_2+\kappa-1;2k_2;-\sinh^2r)
  \\
  N_p^{(k_1,k_2)}
  &={1\over\Gamma(2k_2)}\sqrt{p\sinh\pi p\over2\pi^2}
  \Big[\Gamma(k_1+k_2-\kappa)\Gamma(-k_1+k_2+\kappa)
  \\   &\qquad\qquad\times
  \Gamma(k_1+k_2+\kappa-1)\Gamma(-k_1+k_2-\kappa+1)\Big]^\half,
  \endaligned\qquad\right\}
  \tag\NUM.\num$$
[$\kappa=\half(1+ip)$].

It is possible to state closed expressions for the (energy dependent)
Green functions for the P\"oschl-Teller and modified P\"oschl-Teller
potential, respectively. For the P\"oschl-Teller potential it has the
form (Kleinert and Mustapic [\KLEMUS])
\plus
$$\allowdisplaybreaks\align
  G&(x'',x';E)={m\over \i\hbar}\sqrt{\sin2x'\sin2x''}
  {\Gamma(m_1-L_E)\Gamma(L_E+m_1+1)\over
   \Gamma(m_1+m_2+1)\Gamma(m_1-m_2+1)}
  \\   &\times
  \bigg({1-\cos2x'\over2}\bigg)^{(m_1-m_2)/2}
  \bigg({1+\cos2x'\over2}\bigg)^{(m_1+m_2)/2}
  \\   &\times
  \bigg({1-\cos2x''\over2}\bigg)^{(m_1-m_2)/2}
  \bigg({1+\cos2x''\over2}\bigg)^{(m_1+m_2)/2}
  \\   &\times
  {_2}F_1\bigg(-L_E+m_1;L_E+m_1+1;m_1-m_2+1;{1-\cos2x'\over2}\bigg)
  \\   &\times
  {_2}F_1\bigg(-L_E+m_1;L_E+m_1+1;m_1+m_2+1;{1+\cos2x''\over2}\bigg)
  \tag\NUM.\num\endalign$$
with $m_{1/2}=\half(\lambda\pm\kappa)$, $L_E=-\half+\half \sqrt{2mE}\,
/\hbar$ and $x''\geq x'$. A similar expression is valid for the
modified P\"oschl-Teller-potential Green function (see [\KLEMUS])
which is, however, omitted here.


\glno=0               
\advance\chapno by 1  

\section{The Coulomb Potential}
The hydrogen atom is, of course, one of the most interesting
subjects in quantum mechanics. In the beginnings of quantum mechanics
and with the atom model of Rutherford it was a riddle how to tract a
system which is from the classical physics point of view unstable and
doomed to vanish into pure radiation. Its was Bohr's genius,
postulating the famous rules that only a countable number of orbits are
allowed satisfying the quantum condition $\oint pdq=nh$ ($n\in\N$).
However, this ``old'' quantum mechanics was not sufficient, because it
fails e.g.\ in the case of the helium-atom, and it was Pauli who solved
the hydrogen problem in the terms of the ``new'' quantum mechanics
developed by Schr\"odinger and Heisenberg. It is surprising, that Pauli
did not use a differential equation and solves the corresponding
Eigenvalue problem as Schr\"odinger did later, instead he exploited in
fact the ``hidden'' $\SO(4)$ symmetry of the Kepler-Coulomb-problem.
This symmetry gives classically rise to an conserved quantity,
called Lenz-Runge vector. This additional symmetry allows also a
separation of the Coulomb problem in parabolic coordinates.
This vector points from the focal point of the orbit the perihel.

Ever since the success of quantum mechanics the hydrogen atom was {\bf
the} model to test the theory, may it be Dirac's relativistic quantum
mechanics, where first the fine-structure
constant $\alpha=\e^2/\hbar c$ arises.

It was for a long time a really nuisance that this important
physical system could not be treated by path integrals. Calculating
wave functions and energy levels remains more or less a simple task in
the operator language, but even the construction of the Green function
(resolvent kernel) was impossible for a long time. It takes as long as
1979 as Duru and Kleinert [\DKa, \DKb] finally applied a long-known
transformation in astronomy (Kustaanheimo-Stiefel transformation
[\KUST]) in the path integral of the Coulomb problem and were
successful. Even more, their idea of transforming simultaneously
coordinates and the time-slicing opens new possibilities in solving the
huge amount of unsolved path integral problems. In particular,
Coulomb-related potentials, in the sense, that all these problems can
be reformulated in terms of confluent hypergeoemtric functions, like
the Morse-potential could be successfully treated. However, the
original attempt of Duru and Kleinert was done in a more or less formal
manner, and it does not takes a long time when it was refined by
Inomata [\INOb] and Duru and Kleinert [\DKb].
\newline
Closely related to the original Coulomb problem is, of course,
the $1/r$-potential discussion in $\R^D$. The $D=2$ case in discussed
in this subsection, whereas the original Coulomb problem in the next.

As it turns out its Schr\"odinger equation for the Coulomb potential
is separable in four coordinate systems:
\item{1)} Spherical coordinates:
\plus
$$\aligned
  x&=r\sin\theta\cos\phi\\
  y&=r\sin\theta\sin\phi\\
  z&=r\cos\theta
  \endaligned\qquad
  (r\geq0,0\leq\theta\leq\pi,0\leq\phi\leq2\pi).
  \tag\NUM.\num$$\plus%
Separation of variables in this coordinate system is, of course, not a
specific feature of the Coulomb problem, but a common property of all
radial potentials.
\item{2)} Parabolic coordinates:
$$\aligned
  x&=\xi\eta\cos\phi\\
  y&=\xi\eta\sin\phi\\
  z&=\half(\xi^2-\eta^2)
  \endaligned\qquad
  (\xi,\eta\geq0,0\leq\phi\leq2\pi).
  \tag\NUM.\num$$\plus%
Here electric and magnetic fields can be introduced without spoiling
separability.
\item{3)} Spheroidal coordinates:
$$\aligned
  x&=\sinh\xi\sin\eta\sin\phi\\
  y&=\sinh\xi\sin\eta\cos\phi\\
  z&=\cosh\xi\cos\eta+1
  \endaligned\qquad
  (\xi\geq0,0\leq\eta\leq\pi,0\leq\phi\leq2\pi).
  \tag\NUM.\num$$\plus%
Here a further charge can be introduced and therefore these coordinates
are suitable for the study of the hydrogen ion $H_2^+$.
\item{4)} Spheroconical coordinates:
$$\aligned
  x&=r\sn(\alpha,k)\dn(\beta,k')\\
  y&=r\cn(\alpha,k)\cn(\beta,k')\\
  z&=r\dn(\alpha,k)\sn(\beta,k')
  \endaligned\qquad\aligned
  & r\geq0,\vert\alpha\vert\leq K,\vert\beta\vert\leq2K',\\
  & 0\leq k,k'\leq1,\quad k^2+{k'}^2=1.
  \endaligned
  \tag\NUM.\num$$\plus%
$\sn(\alpha,k),\cn(\alpha,k)$ and $\dn(\alpha,k)$ denote Jacobi
elliptic functions (e.g.\ [\GRA,\ pp.910]) of
modulus $k$ and with real
and imaginary periods $4K$ and $4iK'$, respectively. Here the
corresponding wave functions in $\alpha$ and $\beta$ can be identified
with the wave functions of a quantum mechanically asymmetric top.

We will discuss the path integral for the Coulomb system in the first
two of these coordinate systems. For the usual polar- and parabolic
coordinate coordinate system the path integration can be exactly
performed. In the remaining two, however, the theory of special
functions of these coordinate is poorly developed and no solution seems
up to now available. Nevertheless it is possible to formulate the
Coulomb-problem in these coordinate systems and point out some
relations to other problems connected to these coordinates [\GROm].

It is surprising that the Coulomb path integral was first solved in
Cartesian coordinates (where the Kustaanheimo-Stiefel transformation
works) and not in polar coordinates. Looking at the appropriate
formul\ae\ we see that even in the one-dimensional case (in polar
coordinates) we need the one-dimensional realization of the
Kustaanheimo-Stiefel transformation and a space-time transformation.

\subsection{The $1/r$-Potential in $\R^2$ [\DKb, \INOa]}
We consider the Euclidean two-dimensional space with the singular
potential $V(r)=-Z\e^2/r$ ($r=\vert x\vert $, $x\in\R^2$).
Here, $\e^2$ denotes the square of an electric charged $Z$ its
multiplicity (including sign). However, as already noted, this
potential is not the potential of a point charge in $\R^2$. The
classical Lagrangian now has the form
$$\CL(x,\dot x)={m\over2}\dot x^2+{Z\e^2\over r}.
  \tag\NUM.\num$$\plus%
Of course, bound and continuous states can exist, depending on the
sign of $Z$.
The path integral is given by
$$K(x'',x';T)=\int\limits_{x(t')=x'}^{x(t'')=x''} \CD x(t)
  \exp\left[\ih\int_{t'}^{t''}
            \bigg({m\over2}\dot x^2+{Z\e^2\over r}\bigg)dt\right].
  \tag\NUM.\num$$\edef\numFHbs{\NUM.\num}\plus%
Discussion of this path integral are due to Duru and Kleinert
[\DKa,\DKb] and Inomata [\INOb]. However, the lattice-formulation is not
trivial for the $1/r$-term. In fact, it is too singular for a path
integral, respectively, a stochastic process and some regularization
must be found. This is known for some time, and it turns out that the
Kustaanheimo-Stiefel transformation does the job.

As it turns out one must perform a space-time transformation in order
to solve the path integral (\numFHbs). We perform first the
transformation
$$x_1=\xi^2-\eta^2,\qquad
  x_2=2\xi\eta
  ,\qquad
  u\equiv\pmatrix \xi\\  \eta\endpmatrix\in\R^2,
  \tag\NUM.\num$$\plus%
which casts the original Lagrangian into the form
$$\CL(u,\dot u)=2m(\xi^2+\eta^2)(\dot\xi^2+\dot\eta^2)
     +{Z\e^2\over\xi^2+\eta^2}.
  \tag\NUM.\num$$\plus%
The metric tensor $(g_{ab})$ and it's inverse $(g^{ab})$ are given by
$$(g_{ab})=4(\xi^2+\eta^2)\pmatrix 1  & 0\\ 0 &1\endpmatrix,\qquad
  (g^{ab})={1\over4(\xi^2+\eta^2)}\pmatrix 1  & 0\\ 0 &1\endpmatrix
  \tag\NUM.\num$$\plus%
with determinant $g=\det(g_{ab})=16(\xi^2+\eta^2)^2$ and
$dV=dx_1dx_2=4(\xi^2+\eta^2)d\xi d\eta$.
The hermitean momenta corresponding to the scalar product
$$(\Psi_1,\Psi_2)
  =4\int_{-\infty}^\infty\int_{-\infty}^\infty d\xi d\eta\,
   (\xi^2+\eta^2)\Psi_1(\xi,\eta)\Psi_2^*(\xi,\eta)
  \tag\NUM.\num$$\plus%
are given by
$$p_\xi={\hbar\over\i}\bigg({\partial\over\partial\xi}
        +{\xi\over\xi^2+\eta^2}\bigg),\qquad
  p_\eta={\hbar\over\i}\bigg({\partial\over\partial\eta}
        +{\eta\over\xi^2+\eta^2}\bigg).
  \tag\NUM.\num$$\plus%
Following our general theory of Chapter II.5 we start by considering
the Legendre transformed Hamiltonian
$$H_E=-{\hbar^2\over2m}\bigg({\partial^2\over\partial x_1^2}
                            +{\partial^2\over\partial x_2^2}\bigg)
      -{Z\e^2\over r}-E.
  \tag\NUM.\num$$\plus%
which gives
$$\hat H_E=-{\hbar^2\over2m}{1\over4(\xi^2+\eta^2)}
          \bigg({\partial^2\over\partial\xi^2}
               +{\partial^2\over\partial\eta^2}\bigg)
          -{Z\e^2\over\xi^2+\eta^2}-E.
  \tag\NUM.\num$$\plus%
Therefore the time-transformation is given by
$\epsilon=f(\xi,\eta)\delta$ with $f(\xi,\eta)=4(\xi^2+\eta^2)$
and the space-time transformed Hamiltonian has the form
$$\aligned
  \tilde H&=-{\hbar^2\over2m}
          \bigg({\partial^2\over\partial\xi^2}
               +{\partial^2\over\partial\eta^2}\bigg)
          -4Z\e^2-4E(\xi^2+\eta^2)\\
  &={1\over2m}(p_\xi^2+p_\eta^2)-4Z\e^2-4E(\xi^2+\eta^2)
  \endaligned
  \tag\NUM.\num$$\plus%
with the momentum operators and vanishing quantum potential $\Delta V$:
$$p_\xi={\hbar\over\i}{\partial\over\partial\xi},\qquad
  p_\eta={\hbar\over\i}{\partial\over\partial\eta},\qquad
  \Delta V=0.
  \tag\NUM.\num$$\plus%
This two-dimensional transformation can be interpreted as a
two-dimensional Kustaan\-hei\-mo-Stiefel transformation [\KUST]. It has
the general feature that it performs a change of variables to
``square-root'' coordinates, note $r=\xi^2+\eta^2$. The general
properties of transformations like this states the following problem
(see e.g.\ [\DKb, \GROm] for some review of the relevant literature):
For which values of $D$ does exist a formula
$$(x_1^2+\dots+x_D^2)(y_1^2+\dots+y_D^2)=
   z_1^2+\dots+z_D^2,
  \tag\NUM.\num$$
where the $z_i$ are homogeneous bilinear forms in $x$ and $y$. Following
 a theorem of Hurwitz and Lam this type of transformation can only be
realized in the  space-dimensions $D=1,2,4,8$. The assumption about $z$
now implies that
$$\allowdisplaybreaks\align
  z&=B(x)\cdot y
  \tag\NUM.\num\\  \global\plus
  z_i&=\sum_{k,l}(B^k)^l_i\,x_ky_l=
       \sum_l B(x)^l_iy_i.
  \tag\NUM.\num\endalign$$\plus%
For the special case $x=y$ one has
$$\sum_i z_i^2=\bigg(\sum_j x_j^2\bigg)^2
  \tag\NUM.\num$$\plus%
and the matrix $B$ satisfies the condition
$${^t}B(x)B(x)=\vert x\vert ^2.
  \tag\NUM.\num$$\plus%
The one-dimensional case is ``trivial'', but we shall use it in the
calculation of the path integral for the $1/r$-potential in
polar coordinates. The $D=2$ case we have just encountered.
The four-dimensional variation of this transformation has been used a
long time ago in astronomy for the purpose of regularizing the
Kepler problem. The $D=8$ case is degenerate.
Here the matrix $B$ has the form:
$$B(x)=\pmatrix
  x_1 & x_2 & x_3 & x_4 & x_5 & x_6 & x_7 & x_8    \\
 -x_2 & x_1 & x_4 &-x_3 & x_6 &-x_5 & x_8 &-x_7    \\
 -x_3 &-x_4 & x_1 & x_2 & x_7 &-x_8 &-x_5 & x_6    \\
 -x_4 & x_3 &-x_2 & x_1 &-x_8 &-x_7 & x_6 & x_5    \\
 -x_5 &-x_6 &-x_7 & x_8 & x_1 & x_2 & x_3 &-x_4    \\
 -x_6 & x_5 & x_8 & x_7 &-x_2 & x_1 &-x_4 &-x_3    \\
 -x_7 &-x_8 & x_5 &-x_6 &-x_3 & x_4 & x_1 & x_2    \\
 -x_8 & x_7 &-x_6 &-x_5 & x_4 & x_3 &-x_2 & x_1
  \endpmatrix
  \tag\NUM.\num$$\plus%
and maps $\R^8\to\R$ and is thus of no further use.

In the next Section (hydrogen-atom) we need the $D=4$ case which is
more involved and not one-to-one.
\newline
To incorporate the time transformation
$$s(t)=\int_{t'}^t{d\sigma\over4r(\sigma)},\qquad
  s''=s(t'')
  \tag\NUM.\num$$\edef\numFHxd{\NUM.\num}\plus%
and its lattice definition $\epsilon\to4\bar r^{(j)}\Delta s^{(j)}
=4\bujq\delta^{(j)}$ into the path integral (\numFHbs) we now use from
the set of equations (II.\numBExa), namely equation (II.\numBExa a).
Observing that in the path integral (\numFHbs) the measure changes
according to
$$\prod_{j=1}^N\bigg({m\over2\pi\i\hbar\Delta t^{(j)}}\bigg)
  \times
  \prod_{j=1}^{N-1}dx_1^{(j)}dx^{(j)}_2
  ={1\over r''}
  \prod_{j=1}^N\bigg({m\over2\pi\i\hbar\delta^{(j)}}\bigg)
  \times
  \prod_{j=1}^{N-1}d\xi^{(j)}d\eta^{(j)}
  \tag\NUM.\num$$
we arrive at the path integral transformations equations ($u\in\R^2)$:
$$\gather
  K(x'',x';T)={1\over2\pi\i\hbar}\int_{-\infty}^\infty dE\,
              \e^{-\i TE/\hbar}G(x'',x';E)
  \tag\NUM.\num\\  \global\plus
  G(x'',x';E)=\i\int_0^\infty ds''
  \bigg[\tilde K(u'',u';s'')+\tilde K(-u'',u';s'')\bigg]
  \tag\NUM.\num\endgather$$\edef\numFHaa{\NUM.\num}\plus%
where the space-time transformed path integral $\tilde K$ is given by
$$\multline
  \tilde K(u'',u';s'')=\e^{4\i Z\e^2s''/\hbar}
  \int \CD\xi(s) \int \CD\eta(s)
  \\   \times
  \exp\left\{\ih\int_0^{s''}\bigg[
       {m\over2}(\dot\xi^2+\dot\eta^2)
  +4E(\xi^2+\eta^2)\bigg]ds\right\}.
  \endmultline
  \tag\NUM.\num$$\edef\numFHab{\NUM.\num}\plus%
Note that the factor $1/r''$ has been exactly canceled by means of
equation (II.\numBExa a). This is a specific feature of this potential
problem. Furthermore we have taken into account that our mapping is of
the ``square-root'' type which gives rise to a sign ambiguity. ``Thus,
if one considers all paths in the complex $x=x_1+ix_2$-plane  from $x'$
to $x''$, they will be mapped into two different classes of
paths in the $u$-plane: Those which go from $u'$ to $u''$ and those
going from $u'$ to $-u''$. In the cut complex $x$-plane for the
function $u=\sqrt{x^2}$ these are the paths passing an even or odd
number of times through the square root from $x=0$ and $x=-\infty$. We
may choose the $u'$ corresponding to the initial $x'$ to lie on the
first sheet (i.e.\ in the right half $u$-plane). The final $u''$ can be
in the right as well as the left half-plane and all paths on the
$x$-plane go over into paths from $u'$ to $u''$ and those from $u'$ to
$-u''$ [\DKb]''.  Thus the two contributions arise in equation
(\numFHaa). Equation (\numFHab) is interpreted as the path integral of a
two-dimensional isotropic harmonic oscillator with frequency
$\omega=\sqrt{-8E/m}$. Therefore we get
$$\tilde K(u'',u';s'')=
  {m\omega\over2\i\hbar\sin\omega s''}
  \exp\bigg\{{4\i Z\e^2s''\over\hbar}
   -{m\omega\over2\pi\i\hbar}\bigg[({u'}^2+{u'}^2)\cot\omega s''
  -2{u'\cdot u''\over\sin\omega s''}\bigg]\bigg\}.
  \tag\NUM.\num$$\plus%
Introducing now two-dimensional polar-coordinates
$$\xi=\sqrt{r}\,\cos{\phi\over2},\qquad
  \eta=\sqrt{r}\,\sin{\phi\over2},\qquad
  (r>0,\phi\in[0,2\pi])
  \tag\NUM.\num$$\plus%
so that $u'\cdot u''=\sqrt{r'r''}\cos(\phi''-\phi)/2$.
Thus
$$\multline
  G(x'',x';E)
  ={m\omega\over\pi\hbar}\int_0^\infty{ds''\over\sin\omega s''}
  \\  \times
  \exp\bigg[
  {4\i Z\e^2\over\hbar}s''-{m\omega\over2\i\hbar}(r'+r'')\bigg]
  \cosh\bigg({m\omega\sqrt{r'r''}\over \i\hbar\sin\omega s''}
    \cos{\phi''-\phi'\over2}\bigg)
  \endmultline
  \tag\NUM.\num$$\edef\numFHae{\NUM.\num}\plus%
Using the expansion
$$\cos\bigg(z\cos{\phi\over2}\bigg)
  =\sum_{l=-\infty}^\infty \e^{\i l\phi}I_{2l}(z)
  \tag\NUM.\num$$\plus%
we get for the kernel $G(x'',x';E)$:
$$G(x'',x';E)={1\over2\pi}
               \sum_{l=-\infty}^\infty \e^{\i l(\phi''-\phi') }
              \,G_l(r'',r';E),
  \tag\NUM.\num$$\plus%
where the radial kernel $G_l(E)$ is given by
$$\multline
  G_l(r'',r';E)
  \\  \qquad
  ={2m\omega\over\hbar}\int_0^\infty{ds''\over\sin\omega s''}
  \exp\bigg[{4\i Z\e^2s''\over\hbar}
              -{m\omega\over2\i\hbar}(r'+r'')\cot\omega s''\bigg]
  I_{2l}\bigg({m\omega\sqrt{r'r''}\over \i\hbar\sin\omega s''}\bigg)
  \hfill\\
  \hbox{(Set $\omega={2\i p\hbar\over m}$, substitution
        $u={2p\hbar s''\over m}$ and Wick-rotation)}
  \hfill\\   \qquad
  ={2m\over\hbar}\int_0^\infty{du\over\sinh u}
  \exp\bigg[{2\i\over ap}u+\i p(r'+r'')\coth u\bigg]
  I_{2l}\bigg({2p\sqrt{r'r''}\over\i\sinh u}\bigg)
  \hfill\\
  \hbox{(Substitution $\sinh u=1/\sinh v$)}
  \hfill\\   \qquad
  ={2m\over\hbar}\int_0^\infty
  \bigg(\coth{v\over2}\bigg)^{2\over ap}
  \exp\big[\i p(r'+r'')\cosh v\big]
  I_{2l}\big(-2\i p\sqrt{r'r''}\sinh v\big)dv
  \hfill\\
  \hbox{(Reinsert $\hbar p=\sqrt{2mE}$)}
  \hfill\\   \qquad
  ={1\over\sqrt{r'r''}}\sqrt{-{m\over2E}}
   {1\over(2l)!}
   \Gamma\bigg(\half+l-{Z\e^2\over\hbar}\sqrt{-{m\over2E}}\,\bigg)
   \hfill\\    \hfill\times
   W_{{Z\e^2\over\hbar}\sqrt{-{m\over2E}},l }
                        \bigg(\sqrt{-{8mE\over\hbar^2}}\,r_>\bigg)
   M_{{Z\e^2\over\hbar}\sqrt{-{m\over2E}},l }
                        \bigg(\sqrt{-{8mE\over\hbar^2}}\,r_<\bigg).
  \endmultline
  \tag\NUM.\num$$\edef\numFHac{\NUM.\num}%
In the last step we have used (\numCBxa) and
$r_>,r_<$ denotes the larger (smaller) of $r',r''$, respectively.
The representation (\numFHac) shows that $G_l(E)$ has in the complex
energy-plane poles which are given at the negative integers
$n_l=0,-1,-2,\dots$ at the argument of the $\Gamma$-function and a
cut on the real axis with a branch point at $E=0$.
Thus we get a discrete and continuous spectrum, respectively:
$$\allowdisplaybreaks\align
  E_N&=-{mZ^2\e^4\over2\hbar^2(N-\half)^2},\qquad(N=1,2,\dots)
  \tag\NUM.\num\\    \global\plus
  E_p&={\hbar^2p^2\over2m},\qquad(p\in\R).
  \tag\NUM.\num\endalign$$
\advance\glno by -1
\edef\numFHah{\NUM.\num}\plus%
\edef\numFHai{\NUM.\num}\plus%
Here we have introduced the principle quantum number $N:=n_l+l+1$.
This  is the well-known result.

To determine the discrete spectrum we consider the first line in
equation (\numFHac), identify
in the Hille-Hardy formula $t=\e^{-2u}$, $x=m\omega r'$
and $y=m\omega r''$ and thus get
$$\multline
  G_l(r'',r';E)={m\over\hbar}\sum_{n=0}^\infty
  {(-2\i p\sqrt{r'r''})^{2l}\over n+l+\half-{\i\over ap}}
  \\  \times
  {n!\over(2l+n)!}\exp[-\i p(r'+r'')]
  L_n^{(2l)}(-2\i pr')L_n^{(2l)}(-2\i pr'').
  \endmultline
  \tag\NUM.\num$$\edef\numFHad{\NUM.\num}\plus%
Taking equation (\numFHad) and the $n^{th}$ residuum gives the energy
levels and the wave functions. Inserting into equation (\numFHae) thus
yields the Green functions for the discrete levels for the
two-dimensional $1/r$-potential
$$G(x'',x';E)=\hbar\sum_{N=1}^\infty\sum_{l=-\infty}^\infty
  {\Psi_{N,l}(r',\phi')\Psi^*_{N,l}(r'',\phi'')\over E_N-E}
  \tag\NUM.\num$$\plus%
and the wave functions of the discrete spectrum are given by:
$$\multline
  \Psi_{N,l}(r,\phi)=
 \bigg[{(N-l-1)!\over\pi a^2(N-\half)^3(N+l-1)!}\bigg]^\half
  \bigg({2r\over a(N-\half)}\bigg)^l
  \\   \times
  \exp\bigg[-{r\over a(N-\half)}-\i l\phi\bigg]
  L_{N-l-1}^{(2l)}\bigg({2r\over a(N-\half)}\bigg).
  \endmultline
  \tag\NUM.\num$$\edef\numFHaf{\NUM.\num}\plus%
Furthermore we have used the abbreviation $a=\hbar^2/(mZ\e^2)$ (the
first ``Bohr-radius''). The correct normalization to unity can be seen
from the relation
$$\int_0^\infty x^{2l+\lambda+2}\,\e^{-x}
  \Big[L_{N-l-1}^{(2l+\lambda+1)}(x)\Big]^2dx
  ={2(N+{\lambda\over2})(N+l+\lambda)!\over(N-l-1)!},
  \tag\NUM.\num a$$
respectively
$$\int_0^\infty x^{\mu+a}\e^{-x}\Big[L_n^{(\mu)}(x)\Big]^2dx
  ={(2n+\mu+1)\Gamma(n+\mu+1)\over n!}.
  \tag\NUM.\num b$$\plus%
In order to determine the continuous wave functions we use the
dispersion relation [\GRA, p.987]
$$\multline
  \int_{-\infty}^\infty dx\,e^{-2ix\rho}
  \Gamma\big(\bhalf+\nu+ix\big)\Gamma\big(\bhalf+\nu-ix\big)
  M_{ix,\nu}(\alpha)M_{ix,\nu}(\beta)
  \\
  ={2\pi\sqrt{\alpha\beta}\over\cosh\rho}
  \exp\big[-(\alpha+\beta)\tanh\rho\big]
  J_{2\nu}\bigg({2\sqrt{\alpha\beta}\over\cosh\rho}\bigg).
  \endmultline
  \tag\NUM.\num$$\plus%
and get ($\hbar\tilde p=\sqrt{2mE}\,$)
$$\multline
  G_l(r'',r';E)
  \\  \qquad
  =2\i\tilde p\int_0^\infty{ds''\over\sin\omega s''}
  \exp\bigg[{4\i Z\e^2\over\hbar}s''
  -\tilde p(r'+r'')\cot\omega s''\bigg]
  I_{2l}\bigg({2\tilde p\sqrt{r'r''}\over\sin\omega s''}\bigg)
  \hfill\\  \qquad
  ={\i\over\pi\sqrt{r'r''}}
  \int_0^\infty ds''\exp\bigg({4\i Z\e^2\over\hbar}s''\bigg)
  \hfill\\  \qquad\qquad\times
  \int_0^\infty
  {\vert\Gamma(\half+l+ip)\vert ^2\over\Gamma^2(2l+1)}
  \exp\bigg(-{4\i p\tilde p\hbar s''\over m}+\pi p\bigg)
  M_{\i p,l}(-2\i\tilde pr')M_{-\i p,l}(2\i\tilde pr'')dp
  \hfill\\  \qquad=
  \hbar\int_0^\infty{dp\over\hbar^2p^2/2m-E}
  \exp\bigg({\pi\over ap}\bigg)
  \hfill\\  \qquad\qquad\times
  {\Gamma\big(\half+l+{\i\over ap}\big)
   \Gamma\big(\half+l-{\i\over ap}\big)\over
        2\pi\sqrt{r'r''}(2l)!}
  M_{-{\i\over ap},l}(2\i pr')M_{ {\i\over ap},l}(-2\i pr'').
  \hfill\endmultline
  \tag\NUM.\num$$\plus%
Here again the residuum at $E=\hbar^2p^2/2m$ has been taken in
the integral.  Thus the wave functions of the continuous spectrum
are given by ($\hbar p=\sqrt{2mE}$):
$$\Psi_{p,l}(r,\phi)=\sqrt{1\over4\pi^2r}{1\over(2l)!}
  \Gamma\bigg(\half+l+{\i\over ap}\bigg)
  \exp\bigg({\pi\over2ap}-\i l\phi\bigg)
  M_{{\i\over ap},l}(-2\i pr).
  \tag\NUM.\num$$\edef\numFHag{\NUM.\num}\plus%
Note that
$$ \bigg\vert\Gamma\bigg(\half+l+{\i\over ap}\bigg)\bigg\vert
  M_{{\i\over ap},l}(-2\i pr)
  =\bigg\vert\Gamma\bigg(\half+l+{\i\over ap}\bigg)\bigg\vert
  M_{-{\i\over ap},l}(2\i pr)
  \tag\NUM.\num$$\plus%
and the wave-functions are therefore basically real.
These results of the $1/r$ potential for the discrete and continuous
spectrum, respectively, are equivalent with the results of the operator
approach. The complete Feynman kernel therefore reads
\hfuzz=20pt
$$\multline
  K(x'',x';T)=\sum_{l=0}^\infty\sum_{N=1}^\infty
  \e^{-\i TE_N/\hbar}\Psi_{N,l}^*(r',\phi')\Psi_{N,l}(r'',\phi'')
  \\
  +\sum_{l=0}^\infty\int_0^\infty dp
  \e^{-\i TE_p/\hbar}\Psi_{p,l}^*(r',\phi')\Psi_{p,l}(r'',\phi'')
  \endmultline
  \tag\NUM.\num$$\plus%
\hfuzz=3pt
with wave functions as given in equations (\numFHaf) and (\numFHag)
and energy spectrum equations (\numFHah, \numFHai).

\eject

\subsection{The $1/r$-Potential in $\R^3$ - The Hydrogen Atom
  [\DKa, \DKb, \GMV, \HOI, \INOa, \INOb]}
We consider the Euclidean three-dimensional space with the singular
potential $V(r)\!=\!-Z\e^2/r$ ($r=\vert x\vert $, $x\in\R^3$) with $Z$
and $\e^2$ as in the previous Section. This potential corresponds in
the space $\R^3$, of course, to the potential of a point charge and is
thus the kernel of the Poisson equation in $\R^3$. The classical
Lagrangian has the form
$$\CL(x,\dot x)={m\over2}\dot x^2+{Z\e^2\over r}.
  \tag\NUM.\num$$\plus%
The path integral is given by
$$K(x'',x';T)=\int\limits_{x(t')=x'}^{x(t'')=x''}\CD x(t)
  \exp\left[\ih\int_{t'}^{t''}
  \bigg({m\over2}\dot x^2+{Z\e^2\over r}\bigg)dt\right].
  \tag\NUM.\num$$\edef\numFHaj{\NUM.\num}\plus%
This path integral was first successfully solved (actually the Green
function or resolvent kernel) by Duru and Kleinert [\DKa],
followed by further contributions of Inomata [\INOb] and
Duru and Kleinert [\DKb], Ho and Inomata [\HOI] Grinberg, Mara\~non and
Vucetich [\GMV,\GMVc] and Kleinert [\KLEh].

The transformation in the two-dimensional case corresponds
to the two-dimensional realization of the Kustaanheimo-Stiefel
transformation and we must look for a generalization.
However, we have a the relevant space $\R^3$ whereas the relevant
Kustaanheimo-Stiefel transformation maps $\R^4\to\R^3$ which
means that the transformation in question is not one-to-one.
We have namely [\DKb]
$$\pmatrix x_1\\  x_2\\  x_3\endpmatrix=
  \pmatrix u_3  & u_4  & u_1  & u_2\\
          -u_2  &-u_1  & u_4  & u_3\\
          -u_1  & u_2  & u_3  &-u_4\\
           u_4  &-u_3  & u_2  &-u_1\endpmatrix
  \cdot\pmatrix u_1 \\ u_2 \\ u_3 \\ u_4\endpmatrix,
  \tag\NUM.\num$$
or in matrix notation $x=A\cdot u$ with $x\in\R^3$ and $u\in\R^4$,
respectively.
Several modifications of the matrix $A$ are used in the
literature, e.g.\ [\CBBI,\CBI]
$$A=\pmatrix u_3  & u_4  & u_1  & u_2\\
          -u_4  & u_3  & u_2  &-u_1\\
          -u_1  &-u_2  & u_3  & u_4\\
          -u_2  & u_1  &-u_4  &-u_3\endpmatrix,
  \tag\NUM.\num b$$
and see [\KUST] as well.
The transformation has the property
$A^tA=u_1^2+u_2^2+u_3^2+u_4^2=r=\vert x\vert $ and no Jacobean exist.
However, a fourth coordinate can be described by
\plus
$$x_4(s)=2\int_{s'}^s(u_4u'_1-u_3u'_2+u_2u'_3-u_1u'_4)d\sigma.
  \tag\NUM.\num$$\edef\numFHbt{\NUM.\num}\plus%
Therefore we must circumvent this problem. The idea [\DKa,\INOb] goes
at follows: We consider the lattice formulation of equation (\numFHaj)
and introduce a factor ``one'' by means of
$$\multline
  1=\sqrt{m\over2\pi\i\hbar T}\int_{-\infty}^\infty d\xi''
     \exp\bigg[-{m\over2\i\hbar T}(\xi''-\xi')^2\bigg]
  \\
  =\lim_{N\to\infty}\Norm^{N\over2}
  \prod_{j=1}^N\int_{-\infty}^\infty d\xi^{(j)}
  \exp\bigg[{\i m\over2\hbar\epsilon}\Delta^2\xi^{(j)}\bigg]
  \endmultline
  \tag\NUM.\num$$\plus%
which gives for equation (\numFHaj)
$$\multline
  K(x'',x';T)=\int_{-\infty}^\infty d\xi''
  \lim_{N\to\infty}\Norm^{2N}
  \prod_{j=1}^{N-1}\int d^3x^{(j)}\int_{-\infty}^\infty d\xi^{(j)}
  \\   \times
  \exp\left\{\ih\sum_{j=1}^N\bigg[{m\over2\epsilon}
   (\Delta^2x^{(j)}+\Delta^2\xi^{(j)})+\epsilon{Z\e^2\over\bar r^{(j)}}
   \bigg]\right\}
  \endmultline
  \tag\NUM.\num$$\edef\numFHbi{\NUM.\num}\plus%
with the to-be-defined quantity $\bar r^{(j)}$. We now realize the
transformation $A(u)$ on midpoints [\INOb]
$\buj_a=\half(u^{(j)}_a
+u^{(j-1)}_a)$ (we use in the following the
abbreviation $x_4\equiv\xi$):
$$\Delta x^{(j)}_a=2\sum_{b=1}^4 A^{ab}(\buj)\Delta u^{(j)}_b
  \tag\NUM.\num$$\plus%
We have
$$A^t(\buj)\cdot A(\buj)=\bujq_1+\bujq_2+\bujq_3+\bujq_4
  =:\bar r^{(j)}
  \tag\NUM.\num$$\plus%
which defines $\bar r^{(j)}$. Furthermore
$$\multline
  \Delta^2x^{(j)}_1+\Delta^2x^{(j)}_2+\Delta^2x^{(j)}\sb 3
  +\Delta^2x^{(j)}_4
  \\
  =4\bar r^{(j)}\big[\Delta^2u^{(j)}_1
  +\Delta^2u^{(j)}_2+\Delta^2u^{(j)}_3
                          +\Delta^2u^{(j)}_4\big]
  =:4\bar r^{(j)}\Delta^2u^{(j)}.
  \endmultline
  \tag\NUM.\num$$\plus%
Infinitesimal this  has the form $dx_a^{(j)}=2A^{ab}(\buj)du_a^{(j)}$
(sums over repeated indices understood) and the Jacobean of this
transformation exists and is given by
$${\partial(x_1,x_2,x_3,\xi)\over\partial(u_1,u_2,u_3,u_4)}
  =2^4\bar r^{(j)\,2}.
  \tag\NUM.\num$$\plus%
Thus we arrive at the transformed classical Lagrangian
$$\CL(u,\dot u)=2m u^2\dot u^2+{Z\e^2\over u^2}.
  \tag\NUM.\num$$\plus%
We now repeat the reasoning of the previous Section by observing
that under the time-transformation equation (\numFHxd)
$$s(t)=\int_{t'}^t{d\sigma\over4r(\sigma)},\qquad
  s''=s(t'')$$
and its lattice definition $\epsilon\to4\bar r^{(j)}\Delta s^{(j)}
=4\bujq\delta^{(j)}$ the measure in the path integral
transforms according to
$$\prod_{j=1}^N\Norm^2\times
  \prod_{j=1}^{N-1}(2\buj)^4d^4u^{(j)}={1\over(4r'')^2}
  \prod_{j=1}^N\bigg({m\over2\pi\i\hbar\delta^{(j)}}\bigg)^2
  \times\prod_{j=1}^{N-1}d^4u^{(j)}.
  \tag\NUM.\num$$\plus%
Respecting again equation (II.\numBExa a) which produces a factor $4r''$
we get the path integral transformation
$$\gathered
  K(x'',x';T)={1\over2\pi\i\hbar}\int_{-\infty}^\infty dE\,
              \e^{-\i TE/\hbar}G(x'',x';E)
  \\
  G(x'',x';E)=\i\int_0^\infty ds''
     \e^{4\i Z\e^2s''/\hbar}\tilde K(u'',u';s'' )
  \endgathered
  \tag\NUM.\num$$\plus%
where the space-time transformed path integral $\tilde K$ is given by
$$\multline
  \!\!\!\!
  \tilde K(u'',u';s'')
  \\
  ={1\over4r''}\int_{-\infty}^\infty d\xi''\lim_{N\to\infty}
  \bigg({m\over2\pi\i\hbar\delta}\bigg)^{2N}
  \hfill\\  \qquad\times
  \prod_{j=1}^{N-1}\int d^4u^{(j)}
  \exp\left[\ih\sum_{j=1}^N\bigg(
  {m\over2\delta}\Delta^2u^{(j)}+4\delta E\bujq\bigg)\right]
  \hfill\\
  =\bigg({m\omega\over2\pi\i\hbar\sin\omega s''}\bigg)^2
  \int_{-\infty}^\infty{d\xi''\over4r''}
  \exp\bigg\{-{m\omega\over2\i\hbar}\bigg[
  ({u'}^2+{u'}^2)\cot\omega s''-2{u'\cdot u''\over\sin\omega s''}
    \bigg]\bigg\}.
  \hfill\endmultline
  \tag\NUM.\num$$\edef\numFHba{\NUM.\num}\plus%
Here we have used once again the known solution for the harmonic
oscillator, where we have actually a four-dimensional isotropic
harmonic oscillator with frequency $\omega=\sqrt{-8E/m}$.
Note that the factor $1/r''$ has not been canceled as in the
two-dimensional case.

Up to now we have not discussed the problem of an eventually
quantum correction appearing in the transformation procedure.
As usual we consider the Legendre transformed Hamiltonian
$$H_E=-{\hbar^2\over2m}\Delta_{(3)}-{Z\e^2\over r}-E.
  \tag\NUM.\num$$\plus%
Let us write $H_E$ in the coordinates:
$$\left.\aligned
  u_1&=u\cos\alpha\cos\beta\\
  u_2&=u\cos\alpha\sin\beta\\
  u_3&=u\sin\alpha\cos\gamma\\
  u_4&=u\sin\alpha\sin\gamma\endaligned
          \qquad
  \aligned
  (u=\vert u\vert =\sqrt{r}&,\alpha=\hbox{${\theta\over2}$},
  \beta\pm\gamma=\phi)\\
  (0\leq\theta\leq\pi&,0\leq\phi\leq2\pi)\endaligned\qquad\right\}
  \tag\NUM.\num$$\edef\numFHbw{\NUM.\num}\plus%
Then the Schr\"odinger equation in polar coordinates
$$\bigg[-{\hbar^2\over2}\bigg({\partial^2\over\partial r^2}
   +{2\over r}{\partial\over\partial r}
   +{1\over r^2}L^2\bigg)-{Z\e^2\over r}-E\bigg]\Psi(r,\theta,\phi)
   =0
  \tag\NUM.\num$$\plus%
is transformed into
$$\bigg[-{\hbar^2\over2}{1\over 4s^2}
  \bigg({\partial^2\over\partial s^2}
   +{3\over s}{\partial\over\partial s}
   -4K^2\bigg)-{Z\e^2\over s^2}-E\bigg]\Psi(u)=0
  \tag\NUM.\num$$\plus%
with
$$K^2=-{\partial^2\over\partial\alpha^2}
      -(\tan\alpha-\cot\alpha){\partial\over\partial\alpha}
      -{1\over\sin^2\alpha}{\partial^2\over\partial\beta^2}
      -\cos^2\alpha{\partial^2\over\partial\gamma^2}
  \tag\NUM.\num$$\plus%
is the Casimir-operator of the group $\SO(4)$.
Writing back into $u_1,\dots,u_4$
we get for $H_E$ in the coordinates $q\in\R^4$:
$$\tilde H=4u^2\hat H_E=-{1\over2m}\Delta_4-4Z\e^2-4Eu^2$$
and with $p_{u_k}=-\i\hbar\partial/\partial u_k$:
$$H_{eff}(p_u,q)={1\over2m}\sum_{k=1}^4 p_{u_
 k}^2-4Z\e^2-4Eu^2.
  \tag\NUM.\num$$\plus%
and no quantum correction appears.
Therefore equation (\numFHba) is correct.
\newline
To perform the $\xi''$-integration in equation (\numFHba) we introduce
polar coordinates
$$\aligned
  u_1&=\sqrt{r}\sin{\theta\over2}\cos{\alpha+\phi\over2}\\
  u_2&=\sqrt{r}\sin{\theta\over2}\sin{\alpha+\phi\over2}\\
  u_3&=\sqrt{r}\cos{\theta\over2}\cos{\alpha-\phi\over2}\\
  u_4&=\sqrt{r}\cos{\theta\over2}\sin{\alpha-\phi\over2}
  \endaligned\qquad\left(\aligned
  &0\leq\theta\leq\pi\\
  &0\leq\phi\leq2\pi\\
  &0\leq\alpha\leq4\pi\endaligned\right).
  \tag\NUM.\num$$\plus%
We have with equation (\numFHbt)
$$\xi(s)=-\int_0^s(\alpha'-\cos\theta\,\phi')r(\sigma)d\sigma.
  \tag\NUM.\num$$\edef\numFHbu{\NUM.\num}\plus%
The polar coordinates give the expansion
$$\multline
  u'\cdot u''=\sqrt{r'r''}
  \\   \times
  \bigg[\sin{\theta'\over2}\sin{\theta''\over2}
  \cos\bigg({\alpha''-\alpha'+\phi''-\phi'\over2}\bigg)
  +\cos{\theta'\over2}\cos{\theta''\over2}
  \cos\bigg({\alpha'-\alpha''+\phi''-\phi'\over2}\bigg)\bigg]
  \\  \quad
  \endmultline
  \tag\NUM.\num$$\plus%
Using the expansion
$$\e^{z\cos\phi}=\sum_{\nu=-\infty}^\infty \e^{\i\nu\phi}I_\nu(z)
  \tag\NUM.\num$$\plus%
twice, respecting due to the explicit form of $\xi(s)$ from equation
(\numFHbu) above
$$\int_{-\infty}^\infty{d\xi(s'')\over r''}=\int_0^{4\pi}d\alpha
  \tag\NUM.\num$$\plus%
and that the integration over $\alpha''$ yields
$4\pi\delta_{\nu_1\nu_2}$ we get for the resolvent kernel:
$$\multline
  G(x'',x';E)={\i\over\pi}\int_0^\infty ds''
  \bigg({m\omega\over2\i\hbar\sin\omega s''}\bigg)^2
  \exp\bigg[{4\i Z\e^2s''\over\hbar}
           -{m\omega\over2\i\hbar}(r'+r'')\cot\omega s''\bigg]
  \\    \times
  \sum_{\nu=-\infty}^\infty \e^{\i\nu(\phi''-\phi')}
  I_\nu\bigg({m\omega\sqrt{r'r''}\over \i\hbar\sin\omega s''}
                        \sin{\theta'\over2}\sin{\theta''\over2}\bigg)
  I_\nu\bigg({m\omega\sqrt{r'r''}\over \i\hbar\sin\omega s''}
                        \cos{\theta'\over2}\cos{\theta''\over2}\bigg).
  \endmultline
  \tag\NUM.\num$$\edef\numFHaz{\NUM.\num}\plus%
The last sum can be exactly performed with the help of
$$\bigg({z_1-z_2\,\e^{-\i\phi}
 \over z_1-z_2\,\e^{\i\phi}}\bigg)^{\nu\over2}
  I_\nu\left(\sqrt{z_1^2+z_2^2-2z_1z_2\cos\phi}\,\right)
  =\sum_{n=-\infty}^\infty(-1)^n
  I_n(z_2)I_{\nu+n}(z_1)\e^{\i n\phi},
  \tag\NUM.\num$$\plus%
in particular
$$\sum_{\nu=-\infty}^\infty \e^{\i\nu\phi}I_\nu(z)I_\nu(z')=
       I_0\left(\sqrt{z^2+{z'}^2+2zz'\cos\phi}\,\right),
  \tag\NUM.\num$$\plus%
and the trigonometric expressions
$\sin^2{\alpha\over2}=\half(1-\cos\alpha)$ and
$\cos^2{\alpha\over2}=\half(1+\cos\alpha)$. Thus
$$\multline
  G(x'',x';E)
  ={\i\over\pi}\int_0^\infty ds''
  \bigg({m\omega\over2\i\hbar\sin\omega s''}\bigg)^2
  \\  \times
  \exp\bigg[{4\i Z\e^2s''\over\hbar}
           -{m\omega\over2\i\hbar}(r'+r'')\cot\omega s''\bigg]
  I_0\bigg({m\omega\sqrt{r'r''}\over \i\hbar\sin\omega s''}
           \cos{\gamma\over2}\bigg),
  \endmultline
  \tag\NUM.\num$$\plus%
where $\cos\gamma=\cos\theta'\cos\theta''
+\sin\theta'\sin\theta''\cos(\phi''-\phi')$.
The $I_0$-Bessel function can be expanded in terms of
Legendre-polynomials:
$$I_0\bigg(z\cos{\gamma\over2}\bigg)={2\over z}
  \sum_{l=0}^\infty(2l+1)P_l(\cos\gamma)I_{2l+1}(z).
  \tag\NUM.\num$$\edef\numFHau{\NUM.\num}\plus%
This relation can be derived from the general expansion
$$\bigg({kz\over2}\bigg)^{\mu-\nu}I_\nu(kz)
  =k^\mu\sum_{l=0}^\infty{\Gamma(\mu+l)\over l!\Gamma(1+\nu)}
   (-1)^l(2l+\mu){_2}F_1(-l,l+\mu;1+\nu;k^2)I_{2l+\mu}(z).
  \tag\NUM.\num$$\edef\numFHat{\NUM.\num}%
Equation (\numFHau) is recovered with the identifications $\mu=1$,
$\nu=0$, $k=\cos{\gamma\over2}$ and
$P_l(z) = (-1)^l {_2}F_1(-l,l+1;1;{z+1\over2})$.
Thus the resolvent kernel can be expanded according to
$$\allowdisplaybreaks\align
  G(x'',x';E)&=G(r'',\theta'',\phi'',r',\theta',\phi';E)
  \\
  &={1\over4\pi}\sum_{l=0}^\infty(2l+1)P_l(\cos\gamma)G_l(r'',r';E)
  \tag\NUM.\num\\
  &=\sum_{l=0}^\infty\sum_{n=-l}^l
   Y^{n\,*}_l(\theta',\phi')Y^{n}_l(\theta'',\phi'')G_l(r'',r';E),
  \qquad\qquad
  \global\plus
  \tag\NUM.\num\endalign$$\plus%
where the radial Green function $G_l(E)$ is given by
$$\multline
  G_l(r'',r';E)
  \\  \qquad
  ={2m\omega\over\hbar\sqrt{r'r''}}
  \int_0^\infty{ds''\over\sin\omega s''}
  \exp\bigg[{4\i Z\e^2s''\over\hbar}
           -{m\omega\over2\i\hbar}(r'+r'')\cot\omega s''\bigg]
  I_{2l+1}\bigg({m\omega\sqrt{r'r''}\over \i\hbar\sin\omega s''}\bigg)
  \hfill\\ \qquad
  ={1\over r'r''}\sqrt{-{m\over2E}}
   {1\over(2l+1)!}
   \Gamma\bigg(1+l+{\i Z\e^2\over\hbar}\sqrt{m\over2E}\,\bigg)
  \hfill\\  \hfill\times
   W_{{Z\e^2\over\hbar}\sqrt{-{m\over2E}},l+\half}
                          \left(\sqrt{-8mE\over\hbar^2}\,r_>\right)
   M_{{Z\e^2\over\hbar}\sqrt{-{m\over2E}},l+\half}
                          \left(\sqrt{-8mE\over\hbar^2}\,r_<\right).
  \endmultline
  \tag\NUM.\num$$\plus%
{}From the poles of the $\Gamma$-function  at $z=-n_r=0,-1,-2,\dots$
we read off the bound state energy levels which are given by
$$E_N=-{mZ^2\e^4\over2\hbar^2N^2},\qquad(N=n_r+l+1=1,2,3,\dots)
  \tag\NUM.\num$$\edef\numFHam{\NUM.\num}\plus%
which are the well-known Balmer levels.

The determination of the wave functions is analogous to the previous
two-dimensional case and thus we only state the result for the
{\bf bound state wave functions of the hydrogen atom}:
$$\Psi_{N,l,n}(r,\theta,\phi)
  ={2\over N^2}
  \bigg[{(N-l-1)!\over a^3(N+l)!}\bigg]^\half
  \exp\bigg(-{r\over aN}\bigg)\bigg({2r\over aN}\bigg)^l
  L_{N-l-1}^{(2l+1)}\bigg({2r\over aN}\bigg)Y_l^n(\theta,\phi).
  \tag\NUM.\num$$\edef\numFHak{\NUM.\num}\plus%
A quite complicated attempt was made by Grinberg, Mara\~non and
Vucetich [\GMV] to determine these functions from the wave
 functions
of the corresponding four-dimensional oscillator.
The {\bf continuous state wave functions of the hydrogen atom read} as:
$$\Psi_{p,l,n}(r,\theta,\phi)
  ={1\over\sqrt{2\pi}(2l+1)!r}
   \Gamma\bigg(1+l+{\i\over ap}\bigg)
  \exp\bigg({\pi \over2ap}\bigg)
  M_{{\i\over ap},l+\half}(-2\i pr)Y_l^n(\theta,\phi).
  \tag\NUM.\num$$\edef\numFHal{\NUM.\num}\plus%
Of course, these wave functions form a complete set.
The spectrum of the continuous states is, of course, given by
$$E_p={\hbar^2p^2\over2m},\qquad(p>0).
  \tag\NUM.\num$$\edef\numFHan{\NUM.\num}\plus%
These results coincides with the operator approach.
The complete Feynman kernel therefore reads
$$\multline
  K(x'',x';T)=\sum_{l=0}^\infty\sum_{N=1}^\infty
  \e^{-\i TE_N/\hbar}\Psi_{N,l}^*(r',\phi')\Psi_{N,l}(r'',\phi'')
  \\
  +\sum_{l=0}^\infty\int_0^\infty dp
  \,\e^{-\i TE_p/\hbar}\Psi_{p,l}^*(r',\phi')\Psi_{p,l}(r'',\phi'')
  \endmultline
  \tag\NUM.\num$$\plus%
with wave functions as given in equations (\numFHak) and (\numFHal)
and energy spectrum equations (\numFHam) and (\numFHan), respectively.

\subsection{Coulomb Potential and $1/r$-Potential in $D$ Dimensions
[\CHc, \HOI, \STEb]}
In this Section we discuss the $1/r$-potential in $D$ dimensions. For
$D=3$ we have, of course, the Coulomb problem. The whole  calculation
is quite similar as in the previous cases, we just have to use a
Kustaanheimo-Stiefel transformation in one dimension.

The $D$-dimensional path integral problem was first discussed by
Chetouani and Hammann [\CHc], whereas the radial problem
by Ho and Inomata [\HOI] and  [\STEb].

We start with the $D$-dimensional path integral with  the singular
potential $V(r)=-q_1q_2/r$, where $r=\vert x\vert $ ($x\in\R^D$):
$$K(x'',x';T)=\int\limits_{x(t')=x'}^{x(t'')=x''} \CD x(t)
  \exp\left[\ih\int_{t'}^{t''}\bigg({m\over2}\dot x^2
                     +{q_1q_2\over\vert x\vert }\bigg)dt\right].
  \tag\NUM.\num$$\plus%
We denote the coupling by $q_1\times q_2$ to emphasize the possibility
that two ``charges'' $q_1$ and $q_2$ can interact which each other.
As we have studied very explicitly in Section III.3.1, the angular
variables can be integrated out yielding
$$K(x'',x';T)=\sum_{l=0}^\infty
  S_l^{(D)}(\{\theta''\})S_l^{(D)\,*}(\{\theta'\}) K_l(r'',r';T),
  \tag\NUM.\num$$\plus%
where the radial kernel $K_l(T)$ is given by
$$\multline
  K_l(r'',r';T)
  \\  \qquad=
  (r'r'')^{1-D\over2}\int\limits_{r(t')=r'}^{r(t'')=r''}\CD r(t)
  \exp\left[\ih\int_{t'}^{t''}\bigg({m\over2}\dot r^2
     -\hbar^2{(l+{D\over2}-1)^2-{1\over4}
                   \over2mr^2}+{q_1q_2\over r}\bigg)dt\right]
  \hfill\\ \qquad\equiv
  (r'r'')^{1-D\over2}\int\limits_{r(t')=r'}^{r(t'')=r''}\CD r(t)
  \mu_{l+{D\over2}-1}[r^2]
  \exp\left[\ih\int_{t'}^{t''}\bigg({m\over2}\dot r^2
                              +{q_1q_2\over r}\bigg)dt\right].
  \hfill\endmultline
  \tag\NUM.\num$$\edef\numFHaw{\NUM.\num}\plus%
The $S_l^{(D)}(\{\theta\})$ are the hyperspherical harmonics on the
$S^{D-1}$-sphere. The corresponding Hamiltonian to the path integral
(\numFHaw) has the form
$$\aligned
  H&=-{\hbar^2\over2m}\bigg({d^2\over dr^2}+{D-1\over r}
                            {\d\over\d r}\bigg)
     +\hbar^2{l(l+D-2)\over2 mr^2}-{q_1q_2\over r}
  \\
  &={1\over2m}p_r^2+\hbar^2{(l+{D\over2}-1)^2-{1\over4}\over2mr^2}
                                                  -{q_1q_2\over r}
  \endaligned
  \tag\NUM.\num$$\plus%
with  the momentum operators equation (\numCDxb).
We now perform the space-time transformation
$$s(t)=\int_{t'}^t{d\sigma\over4r(\sigma)},\qquad
  s''=s(t''),\qquad r=F(u)=u^2
  \tag\NUM.\num$$\plus%
with $f=4u^2$, thus that ${F'}^2=f$ and we get the space-time
formed Hamiltonian
$$H={1\over2m}p_u^2+\hbar^2{(2l+D-2)^2-{1\over4}\over2mu^2}
                                                      -4Eu^2-4q_1q_2
  \tag\NUM.\num$$\plus%
with the momentum operator
$$p_u={\hbar\over\i}\bigg({\d\over\d u}+{2D-3\over2u}\bigg).
  \tag\NUM.\num$$\plus%
Here we have applied the results of space-time transformations for
one-dimensional Hamiltonians.
Thus we get the transformation formul\ae
$$\gathered
  K_l(r'',r';T)={1\over2\pi\i\hbar}\int_{-\infty}^\infty dE\,
              \e^{-\i TE/\hbar}G_l(r'',r';E)
  \\
  G_l(r'',r';E)=2\i(u'u'')^{{3\over2}-D}\int_0^\infty ds''\,
               \e^{4\i q_1q_2s''/\hbar}\tilde K_l(u'',u';s''),
  \endgathered
  \tag\NUM.\num$$\plus%
where the kernel $\tilde K_l(s'')$ is given by
$$\multline
  \tilde K_l(u'',u';s'')
  \\   \qquad
  =\int\limits_{u(t')=u'}^{u(t'')=u''}\CD u(s)\mu_{2l+D-2}[u^2]
  \exp\left[\ih\int_0^{s''}\bigg({m\over2}\dot u^2
  +4Eu^2\bigg)ds\right]
  \hfill\\  \qquad
  =\sqrt{u'u''}{m\omega\over \i\hbar\sin\omega T}
  \exp\bigg[-{m\omega\over2\i\hbar}
     ({u'}^2+{u'}^2)\cot\omega T\bigg]
  I_{2l+D-2}\bigg({m\omega u'u''\over \i\hbar\sin\omega T}\bigg),
  \hfill\endmultline
  \tag\NUM.\num$$\plus%
with $\omega=\sqrt{-8E/m}$.
The Green function we obtain in the usual way with the result
$$\multline
  G_l(r'',r';E)
  \\  \quad
  =(r'r'')^{2-D\over2}
   {2m\omega\over\hbar}\int_0^\infty{ds''\over\sin\omega s''}
  \hfill\\  \qquad\times
  \exp\bigg[{4\i q_1q_2s''\over\hbar}
               -{m\omega\over2\hbar i}(r'+r'')\cot\omega s''\bigg]
  I_{2l+D-2}\bigg({m\omega\sqrt{r'r''}\over \i\hbar\sin\omega s''}\bigg)
  \hfill\\   \quad
  =(r'r'')^{1-D\over2}\sqrt{-{m\over2E}}\,
   {\Gamma\big(l+{D-1\over2}-{q_1q_2\over\hbar}
               \sqrt{-{m\over2E}}\,\big)\over(2l+D-2)! }
   \hfill\\    \qquad\times
   W_{{q_1q_2\over\hbar}\sqrt{-{m\over2E}},l+{D-2\over2} }
                        \bigg(\sqrt{-{8mE\over\hbar^2}}\,r_>\bigg)
   M_{{q_1q_2\over\hbar}\sqrt{-{m\over2E}},l+{D-2\over2} }
                        \bigg(\sqrt{-{8mE\over\hbar^2}}\,r_<\bigg).
  \hfill\endmultline
  \tag\NUM.\num$$\edef\numFHao{\NUM.\num}%
Thus it is easy to display the wave functions and spectrum for
the discrete contribution for the $D$-dimensional $1/r$-problem
[$a=\hbar^2/(m\vert q_1q_2\vert )$, $n\in\N$]:
\edef\numFHap{\NUM.\num}%
$$\allowdisplaybreaks\align
  \Psi_{N,l}(r,\{\theta\})&=
  \bigg[{1\over2(N+{D-3\over2})}{(N-l-1)!\over(N+l+D-3)!}\bigg]^\half
  \bigg({2\over a(N+{D-3\over2})}\bigg)^{D/2}
  \bigg({2r\over a(N+{D-3\over2})}\bigg)^l
  \\   &\qquad\qquad\times
  \exp\bigg[-{r\over a(N+{D-3\over2})}\bigg]
  L_{N-l-1}^{(2l+D-2)}\bigg({2r\over a(N+{D-3\over2})}\bigg)
  S_l^{(D)}(\{\theta\})
  \\  &\quad
  \tag\NUM.\num a\\
  &=(-1)^{N-l-1}\bigg({1\over N+{D-3\over2}}\bigg)
  \bigg[{1\over a(N+l+D-3)!(N-L-1)!}\bigg]^\half
  \\   &\qquad\qquad\times
  r^{-{D-1\over2}}
  W_{N+{D-3\over2},l+{D-2\over2}}\bigg({2r\over a(N+{D-3\over2})}\bigg)
  S_l^{(D)}(\{\theta\})
  \tag\NUM.\num b\\
  E_N&=-{mq_1^2q_2^2\over2\hbar^2
  (N+{D-3\over2})^2},\qquad(N=1,2,\dots)
  \global\plus
  \tag\NUM.\num\endalign$$\edef\numFHaq{\NUM.\num}\plus%
For the continuous spectrum we get similarly
$$\multline
  \Psi_{p,l}(r,\{\theta\})=r^{1-D\over2}
  {\Gamma\big(l+{D-1\over2}+{\i\over ap}\big)
                                     \over\sqrt{2\pi}(2l+D-2)!}
  \\   \times
  \exp\bigg({\pi mq_1^2q_2^2\over2\hbar^2 p}\bigg)
  M_{{\i\over ap},l+{D-2\over2}}(-2\i pr)S_l^{(D)}(\{\theta\})
  \endmultline
  \tag\NUM.\num$$\plus%
with $E_p={\hbar^2p^2\over2m}$. The complete Green function thus has
the form
$$\multline
  G(x'',x';E)=\sum_{N=1}^\infty\sum_{l=0}^\infty
  \exp\bigg[\i T{mq_1^2q_2^2
  \over2\hbar^3(N+{D-3\over2})^2}\bigg]
  \Psi_{N,l}^*(r',\{\theta'\})\Psi_{N,l}(r'',\{\theta''\})
  \\
  +\sum_{l=0}^\infty\int_0^\infty
   \exp\bigg(-\i\hbar T{p^2\over2m}\bigg)
   \Psi_{p,l}^*(r',\{\theta'\})\Psi_{p,l}(r'',\{\theta''\}).
  \endmultline
  \tag\NUM.\num$$\plus%
Finally we can state the following path integral identity
$\big(\omega=\sqrt{-8E/m}\big)$:
$$\allowdisplaybreaks\align
  \int &\CD r(t)\mu_\lambda[r^2]\exp\left[\ih\int_{t'}^{t''}
         \bigg({m\over2}\dot r^2+{\alpha\over r}\bigg)dt\right]
  \\   &
  ={2m\omega\sqrt{r'r''}\over\i\pi\hbar}
  \int_{-\infty}^\infty dE\,\e^{-\i TE\hbar}
  \\   &\qquad\times
  \int_0^\infty{ds''\over\sin\omega s''}
  \exp\bigg[{4i\alpha s''\over\hbar}
    -{m\omega\over2\i\hbar}(r'+r'')\cot\omega s''\bigg]
  I_{2\lambda}\bigg({m\omega\sqrt{r'r}\over \i\hbar\sin\omega s''}\bigg)
  \\   &
  =\i\hbar\int_{-\infty}^\infty dE\,\e^{-\i TE/\hbar}
  \sqrt{-{m\over2E}}\,
  {\Gamma(\lambda+\half-{q_1q_2\over\hbar}\sqrt{-{m\over2E}}\,)
  \over\Gamma(2\lambda+1)}
  \\   &\qquad\times
  W_{{q_1q_2\over\hbar}\sqrt{-{m\over2E}},\lambda}
  \bigg(\sqrt{-{8mE\over\hbar^2}}r_>\bigg)
  M_{{q_1q_2\over\hbar}\sqrt{-{m\over2E}},\lambda}
  \bigg(\sqrt{-{8mE\over\hbar^2}}r_<\bigg)
  \\   &
  =\i\hbar\int_0^\infty dE\,\e^{-\i TE}
  \left\{\sum_{n=0}^\infty{1\over E_n-E}\Psi_n^*(r')\Psi_n(r'')
  +\int_{-\infty}^\infty {dp\over E_p-E}\Psi_p^*(r')\Psi_p(r'').
  \right\}
  \\  &
  \tag\NUM.\num\endalign$$
The discrete state wave functions and energy spectrum
are given by ($a=\hbar^2/mq_1q_2$)
$$\allowdisplaybreaks\align
  \Psi_n(r)
  &={1\over n+\lambda+\half}
  \bigg[{n!\over a\Gamma(n+2\lambda+1)}\bigg]^\half
  \bigg({2r\over a(n+\lambda+\half)}\bigg)^{\lambda+\half}
  \\  &\qquad\qquad\times
  \exp\bigg[-{r\over a(n+\lambda+\half)}\bigg]
  L_n^{(2\lambda)}\bigg({2r\over a(n+\lambda+\half)}\bigg)
  \tag\NUM.\num a\\
  &={(-1)^n\over n+\lambda+\half}
  \bigg[{1\over an!\Gamma(n+2\lambda+1)}\bigg]^\half
  W_{n+\lambda+\half,\lambda}\bigg({2r\over a(n+\lambda+\half)}\bigg)
  \tag\NUM.\num b\\  \global\plus
  E_n&=-{m\alpha^2\over2\hbar^2(n+\lambda+\half)^2}.
  \tag\NUM.\num\endalign$$\plus%
For the continuous states we have similarly for the wave functions and
the energy spectrum
$$\Psi_p(r)=\sqrt{1\over2\pi}
  {\Gamma(\lambda+\half-{\i\over ap})\over\Gamma(2\lambda+1)}
  \exp\bigg({\pi\over2ap}\bigg)
  M_{{\i\over ap},\lambda}(-2\i pr)
  \tag\NUM.\num$$\plus%
with $E_p={\hbar^2p^2\over2m}$. In particular this gives the identities
$$\multline
  {m\over \i\hbar k}{\Gamma(\half+\lambda-\mu)\over\Gamma(1+2\lambda)}
  W_{\mu,\lambda}(2\i kr_>)M_{\mu,\lambda}(2\i kr_<)
  \\   \qquad
  ={m\over\hbar k}\sum_n{n!\over\Gamma(n+2\lambda+1)}
  {1\over n+\lambda+\half-\mu}
  \hfill\\ \qquad\qquad\times
  \exp\Big[-k(r'+r'')\Big] (4k^2r'r'')^{\lambda+\half}
  L_n^{(2\lambda)}(2kr') L_n^{(2\lambda)}(2kr'')
  \hfill\\    \qquad
  +{1\over2\pi}{m\over\hbar k}\int_{-\infty}^\infty
  {dp\,\e^{\pi p}\over p-\i\mu}
  \hfill\\ \qquad\qquad\times
  {\Gamma(\half+\lambda+ip)\Gamma(\half+\lambda-\i p)
  \over\Gamma^2(1+2\lambda)}
  M_{\i p,\lambda}(-2\i kr'')M_{-\i p,\lambda}(2\i kr'),
  \hfill\endmultline
  \tag\NUM.\num$$\plus%
or alternatively (and explicitly, $\alpha=(q_1q_2/\hbar)\sqrt{-m/2E}$)
$$\multline
  \sqrt{-{m\over2E}}
  {\Gamma(\half+\lambda-\alpha)\over\Gamma(1+2\lambda)}
  W_{\alpha,\lambda}\bigg(\sqrt{-{8mE\over\hbar^2}}r_>\bigg)
  M_{\alpha,\lambda}\bigg(\sqrt{-{8mE\over\hbar^2}}r_<\bigg)
  \\
  =\hbar\sum_{n=1}^\infty {1\over E_n-E}\Psi_n^*(r')\Psi_n(r'')
  +\hbar\int_0^\infty {dp\over E_p-E}\Psi_p^*(r')\Psi_p(r'').
  \endmultline
  \tag\NUM.\num$$\plus%
Let us note that it is not difficult to analyze the Green function by
taking the $n^{th}$ residuum in equation (\numFHao).
Using the appropriate expansion for the
$\Gamma$-function again we get
$$\multline
  G(r'',r';E) =\sum_{n=0}^\infty {(-1)^n\over n!}
  {1\over n+\half+\lambda-{q_1q_2\over\hbar}\sqrt{-{m\over2E_n}}}
  \sqrt{-{m\over2E_n}}{1\over\Gamma(2\lambda+1)}
  \\   \times
  W_{n+\lambda+\half,\lambda}\bigg({2r_>\over a(n+\lambda+\half)}\bigg)
  M_{n+\lambda+\half,\lambda}\bigg({2r_<\over a(n+\lambda+\half)}\bigg)
  +\hbox{regular terms}
  \endmultline
  \tag\NUM.\num$$
which gives using the relations of the Whittaker-functions with the
Laguerre-poly\-no\-mials the wave functions (\numFHap) and the energy
spectrum (\numFHaq). In particular we have for $D=1$
\advance\glno by -1
$$\allowdisplaybreaks\align
  \Psi_n(x)&=\bigg({1\over an^3}\bigg)^\half
            {2r\over an}\e^{-r/an}L_n^{(1)}\bigg({2r\over an}\bigg),
  \tag\NUM.\num\\  \global\plus
  E_n&=-{mZ^2\e^4\over2\hbar^2n^2},\qquad n\in\N
  \tag\NUM.\num\\  \global\plus
  \Psi_p(x)&={1\over\sqrt{2\pi}}\Gamma\bigg(1-{\i\over ap}\bigg)
             \e^{\pi/2ap}M_{{\i\over ap},\half}(-2\i pr).
  \tag\NUM.\num\endalign$$
\advance\glno by -2
\edef\numFHar{\NUM.\num}\plus\plus%
\edef\numFHas{\NUM.\num}\plus%
Because, the domains $x<0$ and $x>0$ are separated the wave-functions
(\numFHar,\numFHas) are doubly degenerated.

\noindent
We can study several special cases (see [\CHc]):
\newline i) We first consider $D=2$ and get
\hfuzz=20pt
$$\allowdisplaybreaks\align
  G&(x'',x';E)
  \\
  &={(r'r'')^{-\half}\over2\pi}\sqrt{m\over2E}\sum_{l=-\infty}^\infty
   {\Gamma\big(\vert l\vert -{\i q_1q_2\over\hbar}\sqrt{m\over2E}\big)
  \over(2\vert l\vert )!}
  \\    &\qquad\times
  \e^{\i l(\phi''-\phi')}
   W_{ {\i q_1q_2\over\hbar}\sqrt{m\over2E},\vert l\vert }
                        \bigg(\sqrt{-{8mE\over\hbar^2}}\,r_>\bigg)
   M_{-{\i q_1q_2\over\hbar}\sqrt{m\over2E},\vert l\vert }
                        \bigg(\sqrt{-{8mE\over\hbar^2}}\,r_<\bigg)
   \\    &
   ={2m\over\hbar}\int_0^\infty{du\over\sinh u}
   \exp\bigg[{q_1q_2\over\hbar}\sqrt{-{2m\over E}} u
       -{\sqrt{-2mE}\over\hbar}(r'+r'')\coth u\bigg]
  \\    &\qquad\times
   {1\over2\pi}\sum_{l=-\infty}^\infty \e^{\i l(\phi''-\phi')}
   I_{2l}\bigg({2\sqrt{-2mE}\sqrt{r'r''}\over\hbar\sinh u}\bigg)
   \\    &
   ={m\over\pi\hbar}\int_0^\infty{du\over\sinh u}
  \cosh\bigg({\sqrt{-8mE}\over\hbar\sinh u}
                    \sqrt{r'r''}\cos{\phi''-\phi'\over2}\bigg).
  \\    &\qquad\times
   \exp\bigg[{q_1q_2\over\hbar}\sqrt{-{2m\over E}} u
  -\sqrt{-2mE\over\hbar^2}\,(r'+r'')\cosh u\bigg].
  \tag\NUM.\num\endalign$$\edef\numFHav{\NUM.\num}\plus%
\hfuzz=3pt
This is equivalent with equation (\numFHae).

\noindent ii)
We consider $D$ arbitrary. The Green function reads in the
integral representation, where we expand in addition the hyperspherical
harmonics in terms of Gegenbauer-polynomials
\hfuzz=10pt
$$\multline
  G(x'',x';E)
  ={2m\over\hbar}(r'r'')^{2-D\over2}
  \int_0^\infty{du\over\sinh u}
  \exp\bigg[{q_1q_2\over\hbar}\sqrt{-{2m\over E}} u
       -{\sqrt{-2mE}\over\hbar}(r'+r'')\coth u\bigg]
   \hfill\\  \hfill\times
   \sum_{l=0}^\infty {(2l+D-2)\Gamma({D-2\over2})\over4\pi^{D/2}}
   C_l^{D-2\over2}(\cos\psi^{(1,2)})
   I_{2l+D-2}\bigg({2\sqrt{-2mE}\sqrt{r'r''}\over\hbar\sinh u}\bigg).
   \endmultline
  \tag\NUM.\num$$\plus%
\hfuzz=3pt
The $l$-summation can be performed by means of equation (\numFHat)
and taking into account that we have for the Gegenbauer-polynomials
$$\aligned
  C_n^\lambda(\cos\theta)
  &={\Gamma(n+2\lambda)\over n!\Gamma(2\lambda)}
    {_2}F_1\bigg(2\lambda+n,-n;\lambda+\half;\sin^2{\theta\over2}\bigg)
  \\
  &=(-1)^n{\Gamma(n+2\lambda)\over n!\Gamma(2\lambda)}
  {_2}F_1\bigg(2\lambda+n,-n;\lambda+\half;\cos^2{\theta\over2}\bigg).
  \endaligned
  \tag\NUM.\num$$\plus%
Thus we get
\hfuzz=10pt
$$\multline
  G(x'',x';E)
  \\  \qquad
  ={2m\over\hbar}(r'r'')^{2-D\over2}(4\pi)^{1-D\over2}
  \int_0^\infty{du\over\sinh u}
  \exp\bigg[{q_1q_2\over\hbar}\sqrt{-{2m\over E}}u
       -{\sqrt{-2mE}\over\hbar}(r'+r'')\coth u\bigg]
  \hfill\\  \qquad\qquad\times
  \sum_{l=0}^\infty(-1)^l(2l+D-2)
  {\Gamma(l+D-2)\over l!\Gamma({D-1\over2})}
  \hfill\\   \qquad\qquad\times
  {_2}F_1\bigg(l+D-2,-l;{D-1\over2};\cos^2{\psi^{(1,2)}\over2}\bigg)
  I_{2l+d-2}\bigg({2\sqrt{-2mE}\sqrt{r'r''}\over\hbar\sinh u}\bigg)
  \hfill\\   \qquad
  ={2m\over\hbar}
  \bigg({4\pi\hbar\over\sqrt{-2mE}}\bigg)^{1-D\over2}
  \bigg(\sqrt{r'r''}\cos{\psi^{(1,2)}\over2}\bigg)^{3-D\over2}
  \hfill\\    \qquad\qquad\times
  \int_0^\infty{du\over(\sinh u)^{D+1\over2}}
  \exp\bigg[{q_1q_2\over\hbar}\sqrt{-{2m\over E}}u
       -{\sqrt{-2mE}\over\hbar}(r'+r'')\coth u\bigg]
   \hfill\\   \qquad\qquad\times
   I_{D-3\over2}\bigg({2\sqrt{-2mE}\sqrt{r'r''}\over\hbar\sinh u}
                                     \cos{\psi^{(1,2)}\over2}\bigg)
  \hfill\endmultline
  \tag\NUM.\num$$\plus%
Alternatively this can be written as
$$\multline
  \!\!\!\!
  G(x'',x';E)
  \\ \quad
  ={2m\over\hbar}
  \bigg({4\pi\hbar\over\sqrt{-2mE}}\bigg)^{1-D\over2}
  \bigg({r'r''+x'\cdot x''\over2}\bigg)^{3-D\over4}
  \int_0^\infty{du\over(\sinh u)^{D+1\over2}}
  \hfill\\  \qquad\times
  \exp\bigg[{q_1q_2\over\hbar}\sqrt{-{2m\over E}}u
       -{\sqrt{-2mE}\over\hbar}(r'+r'')\coth u\bigg]
   I_{D-3\over2}\bigg({2\sqrt{-2mE}\over\hbar\sinh u}
                                \sqrt{r'r''+x'\cdot x''\over2}\bigg)
  \hfill\\ \quad
  ={m\over\hbar}\bigg({2\pi\hbar\over\sqrt{-2mE}}\bigg)^{1-D\over2}
  (x^2-y^2)^{3-D\over4}\int_0^\infty{du\over(\sinh u)^{D+1\over2}}
  \hfill\\  \qquad\times
  \exp\bigg[{q_1q_2\over\hbar}\sqrt{-{2m\over E}}u
       -{\sqrt{-2mE}\over\hbar}x\coth u\bigg]
   I_{D-3\over2}\bigg(\sqrt{-2mE}
           {\sqrt{x^2-y^2}\over\hbar\sinh u}\bigg) ,
  \hfill\endmultline
  \tag\NUM.\num$$\plus%
\hfuzz=3pt
where $x=r'+r''$ and $y=\vert x''-x'\vert $. For $D=2$ we recover, of
course, equations (\numFHae,\numFHav) with $I_{-\half}(z)=(2/\pi
z)^\half\cosh z$.
\newline
$D=1$ yields with $\psi^{(1,2)}\to2\pi$:
$$\multline
  G(x'',x';E)
  \\  \qquad
  ={2m\over\hbar}\sqrt{x'x''}\int_0^\infty{du\over\sinh u}
  \hfill\\   \qquad\qquad\times
  \exp\bigg[{q_1q_2\over\hbar}\sqrt{-{2m\over E}}u
       -{\sqrt{-2mE}\over\hbar}(x'+x'')\coth u\bigg]
   I_1\bigg({2\sqrt{-2mE}\sqrt{x'x''}\over\hbar\sinh u}\bigg)
  \hfill\\   \qquad
  =\sqrt{-{m\over2E}}\,
  \Gamma\bigg(1-{q_1q_2\over\hbar}\sqrt{-{m\over2E}}\,\bigg)
  \hfill\\  \qquad\qquad\times
   W_{{q_1q_2\over\hbar}\sqrt{-{m\over2E}},\half}
                        \bigg(\sqrt{-{8mE\over\hbar^2}}\,r_>\bigg)
   M_{{q_1q_2\over\hbar}\sqrt{-{m\over2E}},\half}
                        \bigg(\sqrt{-{8mE\over\hbar^2}}\,r_<\bigg).
  \hfill\endmultline
  \tag\NUM.\num$$\plus%
Furthermore one can show, by using the relation:
$$\bigg({d\over zdz}\bigg)^m\bigg({I_\nu(z)\over z^\nu}\bigg)
  ={I_{\nu+m}(z)\over z^{\nu+m}}$$
that the $D$-dimensional Green function for the $1/r$-potential is
connected with the $(D-2)$-dimensional Green function by the relation
$$G_D(x\vert y;E)=-{1\over2\pi y}{\partial\over\partial y}
  G_{D-2}(x\vert y;E).
  \tag\NUM.\num$$
Therefore
$$\allowdisplaybreaks\align
  G_D(x\vert y;E)
  &=\bigg({-\partial\over2\pi y\partial y}\bigg)^{D-1\over2}
                  G_1(x\vert y;E),\qquad D=1,3,5,\dots,
  \tag\NUM.\num\\
  G_D(x\vert y;E)
  &=\bigg({-\partial\over2\pi y\partial y}\bigg)^{D-2\over2}
                  G_2(x\vert y;E),\qquad D=2,4,6,\dots.
  \global\plus
  \tag\NUM.\num\endalign$$\plus%

\newpage

\subsection{Axially Symmetric Coulomb-Like Potential
[\CARPa-\CGHc, \GROm-\GROp, \SOKb, \SOKc]}
It is possible to consider the even more complicated potential problem,
namely [\GROm]
$$\allowdisplaybreaks\align
  \CL_{Cl}
  &={m\over2}(\dot x^2+\dot y^2+\dot z^2)+{q^2\over\sqrt{x^2+y^2+z^2}}
  \\  &\qquad\qquad\qquad\qquad
     -{b\hbar^2\over2m}{1\over x^2+y^2}
     -{c\hbar^2\over2m}{z\over(x^2+y^2)\sqrt{x^2+y^2+z^2}}
  \\  &\quad
  \tag\NUM.\num a\\
  &={m\over2}(\dot r^2+r^2\dot\theta^2+r^2\sin^2\theta\dot\phi\sp2)
   +{q^2\over r}-{b\hbar^2\over2mr^2\sin^2\theta}
   -{c\hbar^2\over2m}{\cos\theta\over r^2\sin^2\theta}
  \tag\NUM.\num b\\
  &={m\over2}(\xi^2+\eta^2)(\dot\xi^2+\dot\eta^2)
    +{m\over2}\xi^2\eta^2\dot\phi^2+{2q^2\over\xi^2+\eta^2}
    -{b\hbar^2\over2m\xi^2\eta^2}
    -{c\hbar^2\over m}{\eta^2-\xi^2\over\xi^2\eta^2(\xi^2+\eta^2)}.
  \\  &\quad
  \tag\NUM.\num c\endalign$$
\edef\numFHbk{\NUM.\num a}
\edef\numFHbl{\NUM.\num b}
\edef\numFHbm{\NUM.\num c}\plus%
Here I have displayed the corresponding classical Lagrangian in
cartesian, polar and para\-bo\-lic coordinates. It belongs to the class
of potentials mentioned long ago by Makarov, Smorodinsky, Valiev and
Winternitz [\MSVW] which are separable in a specific coordinate system
and furthermore exactly solvable. This potential was discussed by
Carpio-Bernido, Bernido and Inomata [\CBBI] starting from cartesian
coordinates and using the four-dimensional realization of the
Kustaanheimo-Stiefel transformation. However, this potential is
separable in polar as well as parabolic coordinates and we shall give
the explicit calculation for all of them.

\bigskip\noindent
$\underline{\hbox{1) Cartesian coordinates}}$
\newline
In our line of reasoning we follow reference [\CBBI], however
with some simplifications because we have already discussed the
realization of the four-dimensional Kustaanheimo-Stiefel in the
discussion of the hydrogen atom. We consider the path integral
representation corresponding to equation (\numFHbl)
$$\multline
  K(x'',x';T)=\lim_{N\to\infty}\Norm^{3N\over2}
  \prod_{j=1}^{N-1}\int dx^{(j)}dy^{(j)}dz^{(j)}
  \\
  \times\exp\left\{\ih\sum_{j=1}^N\bigg[{m\over2\epsilon}
  (\Delta^2x^{(j)}+\Delta^2y^{(j)}+\Delta^2z^{(j)})
  +\epsilon{q^2\over r^{(j)}}
  \right.\\  \left.\vphantom{\sum_j^N}
  -{\epsilon b\hbar^2\over2m}{1\over r^{(j)\,2}-z^{(j)\,2}}
  -{\epsilon c\hbar^2\over2m}{z^{(j)}\over r^{(j)}
                  (r^{(j)\,2}-z^{(j)\,2})}\bigg]\right\}.
  \endmultline
  \tag\NUM.\num$$\plus%
Similarly to equation (\numFHaj) we insert a factor ``one''
$$1=\lim_{N\to\infty}\Norm^{N\over2}
  \prod_{j=1}^N\int_{-\infty}^\infty d\xi^{(j)}
  \exp\bigg[{\i m\over2\epsilon\hbar}\Delta^2\xi^{(j)}\bigg].
  \tag\NUM.\num$$\plus%
We repeat the steps from equations (\numFHbi) to (\numFHba), i.e.\ we
realise the  four-dimensional Kus\-taan\-hei\-mo-Stiefel transformation
on midpoints and arrive together with the transformation formul\ae
$$\gathered
  K(x'',x';T)={1\over2\pi\i\hbar}\int_{-\infty}^\infty
  dE\,\e^{-\i TE/\hbar}G(x'',x';E)
  \\
  G(x'',x';E)=\i\int_0^\infty ds''\,\e^{4\i q^2s''/\hbar}
  \tilde K(u'',u';s''),
  \endgathered
  \tag\NUM.\num$$\plus%
with the transformed path integral
$$\tilde K(u'',u';s'')={1\over4r''}\int_{-\infty}^\infty d\xi''
  \tilde K_1(u_1'',u_1',u_2'',u_2';s'')\times
  \tilde K_2(u_3'',u_3',u_4'',u_4';s''),
  \tag\NUM.\num$$
and the kernels $K_1(s''),K_2(s'')$, respectively, are given by
$$\allowdisplaybreaks\align
  \tilde K_1&(u_1'',u_1',u_2'',u_2';s'')
  \\
  &=\int\limits_{u_1(t')=u_1'}^{u_1(t'')=u_1''}\CD u_1(s)
  \int\limits_{u_2(t')=u_2'}^{u_2(t'')=u_2''}\CD u_2(s)
  \\  &\qquad\times
  \exp\left\{\ih\int_0^{s''}\bigg[{m\over2}(\dot u_1^2+\dot u_2^2)
    +4E(u_1^2+u_2^2)-{(b+c)\hbar^2\over2m(u_1^2+u_2^2)}\bigg]ds\right\}
  \\
  &=\int\limits_{\rho_1(t')=\rho_1'}^{\rho_1(t'')=\rho_1''} \rho_1
  \CD\rho_1(s) \
  \int\limits_{\phi_1(t')=\phi_1'}^{\phi_1(t'')=\phi_1''}\CD\phi_1(s)
  \\   &\qquad\times
  \exp\left\{\ih\int_0^{s''}
  \bigg[{m\over2}(\dot\rho_1^2+\rho_1^2\dot\phi_1^2)
  +4E\rho_1^2-\hbar^2{b+c-{1\over4}\over2m\rho_1^2}\bigg]ds\right\}
  \\   &\quad
  \tag\NUM.\num\\  \global\plus
  &={1\over2\pi\sqrt{\rho_1'\rho_1''}}
  \sum_{\nu_1=-\infty}^\infty \e^{\i\nu_1(\phi_1''-\phi_1')}
  \\
  &\qquad\times
  \int\limits_{\rho_1(t')=\rho_1'}^{\rho_1(t'')=\rho_1''}\CD\rho_1(s)
  \exp\left\{\ih\int_0^{s''} \bigg[{m\over2}\dot\rho_1^2
  +4E\rho_1^2-\hbar^2{b+c+\nu^2-{1\over4}\over2m\rho_1^2}
  \bigg]ds\right\}
  \\
  &={m\omega\over2\pi\i\hbar\sin\omega s''}
  \sum_{\nu_1=-\infty}^\infty \e^{\i\nu_1(\phi_1''-\phi_1')}
  \\  &\qquad\times
  \exp\bigg[-{m\omega\over2\i\hbar}({\rho_1'}^2+{\rho_1''}\sp2)
     \cot\omega s''\bigg]
  I_{\lambda_1}\bigg({m\omega\rho_1'\rho_1''\over \i\hbar\sin\omega s''}
    \bigg).
  \tag\NUM.\num\endalign$$\plus%
Here we have introduced two-dimensional polar coordinates $u_1
=\rho_1\cos\phi_1$, $u_2=\rho_1\sin\phi_1$, separated the $\phi_1$ and
the $\rho_1$ path integration in the usual way, applied the solution of
the radial harmonic oscillator and have used the abbreviations
$\omega=\sqrt{-8E/m}$, $\lambda_1=+\sqrt{\nu^2+b+c}$. Similarly
$$\multline
  \tilde K_2(u_3'',u_3',u_4'',u_4';s'')
  ={m\omega\over2\pi\i\hbar\sin\omega s''}
  \sum_{\nu_2=-\infty}^\infty \e^{\i\nu_2(\phi_2''-\phi_2')}
  \\  \times
  \exp\bigg[-{m\omega\over2\i\hbar}({\rho_2'}^2+{\rho_2''}\sp2)
     \cot\omega s''\bigg]
  I_{\lambda_2}\bigg({m\omega\rho_2'\rho_2''\over \i\hbar\sin\omega s''}
    \bigg)
  \endmultline
  \tag\NUM.\num$$\plus%
with $u_3=\rho_2\cos\phi_2$, $u_4=\rho_2\sin\phi_2$,
$\lambda_2=+\sqrt{\nu^2+b-c}$.
We identify variables [Cayley-Klein parameters, c.f.equation (\numFHbw)]
$$\aligned
  \rho_1&=\sqrt{r}\,\cos{\theta\over2}\\
  \phi_1&={\alpha+\phi\over2}
  \endaligned\qquad\aligned
  \rho_2&=\sqrt{r}\,\sin{\theta\over2}\\
  \phi_2&={\alpha-\phi\over2}
  \endaligned
  \tag\NUM.\num$$\plus%
($0\leq\theta\leq\pi$, $0\leq\phi\leq2\pi$, $0\leq\alpha\leq4\pi$,
$d\xi''=r''d\alpha''$). Collecting factors  we thus obtain
$$\multline
  G(x'',x';E)={\i\over\pi}\int_0^\infty ds''
  \bigg({m\omega\over2\i\hbar\sin\omega s''}\bigg)^2
  \exp\bigg[{4\i q^2s''\over\hbar}
            -{m\omega\over2\i\hbar}(r'+r'')\cot\omega s''\bigg]
  \\    \hfill\times
  \sum_{\nu=-\infty}^\infty \e^{\i\nu(\phi''-\phi')}
  I_{\lambda_1}\bigg({m\omega\sqrt{r'r''}\over \i\hbar\sin\omega s''}
                        \sin{\theta'\over2}\sin{\theta''\over2}\bigg)
  I_{\lambda_2}\bigg({m\omega\sqrt{r'r''}\over \i\hbar\sin\omega s''}
                        \cos{\theta'\over2}\cos{\theta''\over2}\bigg).
  \\ \ \endmultline
  \tag\NUM.\num$$\edef\numFHbo{\NUM.\num}\plus%
(Compare with equation (\numFHaz)).
Using the ``addition theorem '' [\EMOTa,\ Vol.II,\ p.99]:
$$\multline
  {z\over2}J_\mu(z\cos\alpha\cos\beta)J_\nu(z\sin\alpha\sin\beta)
  \\   \qquad
  =(\sin\alpha\sin\beta)^\nu(\cos\alpha\cos\beta)^\mu
  \hfill\\  \qquad\qquad\times
  \sum_{l=0}^\infty(-1)^l(\mu+\nu+2l+1)
  {\Gamma(\mu+\nu+l+1)\Gamma(\nu+l+1)\over
                            l!\,\Gamma^2(\nu+1)\Gamma(l+\mu+1)}
  J_{\mu+\nu+2l+1}(z)
  \hfill\\   \qquad\qquad\times
  {_2}F_1(-l,\mu+\nu+l+1;\nu+1;\sin^2\alpha)
  {_2}F_1(-l,\mu+\nu+l+1;\nu+1;\sin^2\beta).
  \hfill\endmultline
  \tag\NUM.\num$$\edef\numFHxc{\NUM.\num}\plus%
we get in the usual way by performing he $s''$ integration
\hfuzz=9pt
$$\allowdisplaybreaks\align
  &G(x'',x';E)
  \\  &
  ={1\over2\pi}\sum_{\nu=-\infty}^\infty \e^{\i\nu(\phi''-\phi')}
  \bigg(\sin{\theta'\over2}\sin{\theta''\over2}\bigg)^{\lambda_1}
  \bigg(\cos{\theta'\over2}\cos{\theta''\over2}\bigg)^{\lambda_2}
  \\   &\ \times
  \sum_{l=0}^\infty (\lambda_1+\lambda_2+2l+1)
  {\Gamma(\lambda_1+\lambda_2+l+1)\Gamma(\lambda_1+l+1)
      \over l!\,\Gamma^2(\lambda_1+1)\Gamma(\lambda_2+l+1)}
  \\   &\ \times
  {_2}F_1\bigg(-l,\lambda_1+\lambda_2+l+1;
            \lambda_1+1;\sin^2{\theta'\over2}\bigg)
  {_2}F_1\bigg(-l,\lambda_1+\lambda_2+l+1;
            \lambda_1+1;\sin^2{\theta''\over2}\bigg)
  \\   &\ \times
  {m\omega\over\sqrt{r'r''}\hbar}
  \int_0^\infty {ds''\over\sin\omega s''}\e^{4\i q^2s''/\hbar}
  \\   &\ \times
  \exp\bigg[-{m\omega \over2\i\hbar}
  (r'+r'')\cot\omega s''\bigg]I_{\lambda_1+\lambda_2+2l+1}
  \bigg({m\omega\sqrt{r'r''}\over \i\hbar\sin\omega s''}\bigg)
  \\  &
  ={1\over2\pi}\sum_{\nu=-\infty}^\infty \e^{\i\nu(\phi''-\phi')}
  \sum_{l=0}^\infty {(\lambda_1+\lambda_2+2l+1)\over2}
  {\Gamma(\lambda_1+\lambda_2+l+1) l!
      \over\Gamma(\lambda_1+l+1) \Gamma(\lambda_2+l+1)}
  \\   &\ \times
  \bigg(\sin{\theta'\over2}\sin{\theta''\over2}\bigg)^{\lambda_1}
  \bigg(\cos{\theta'\over2}\cos{\theta''\over2}\bigg)^{\lambda_2}
  P_l^{(\lambda_1,\lambda_2)}(\cos\theta')
  P_l^{(\lambda_1,\lambda_2)}(\cos\theta'')
  \\   &\ \times
  {1\over r'r''}\sqrt{-{m\over2E}}\,
  {\Gamma[\half(\lambda_1+\lambda_2)+l+1-p]
  \over\Gamma(\lambda_1+\lambda_2+2l+2)}
  \\   &\ \times
  W_{p,l+{1+\lambda_1+\lambda_2\over2}}(-2\i kr_>)
  M_{p,l+{1+\lambda_1+\lambda_2\over2}}(-2\i kr_<).
  \tag\NUM.\num\endalign$$\plus%
\hfuzz=3pt
with the abbreviations $p=q^2\sqrt{-m/2E\hbar^2}$,
$k=\sqrt{+2mE}/\hbar$. This is the result of reference [\CBBI].
The energy-levels are determined by the poles of the $\Gamma$-function
and are given by
$$E_n=-{mq^4\over2\hbar^2(n+l+1+{\lambda_1+\lambda_2\over2})\sp2}
  ,\qquad n\in\N_0.
  \tag\NUM.\num$$\edef\numFHbn{\NUM.\num}\plus%
The corresponding wave functions will be calculated below, where we
separate the radial from the angular path integrations directly.

\goodbreak
\bigskip\noindent
$\underline{\hbox{2) Polar coordinates}}$
\newline
We consider the path integral corresponding the Lagrangian (\numFHbk):
$$\multline
  K(x'',x';T)
  \\  \qquad
  =\int\limits_{r(t')=r'}^{r(t'')=r''}\CD r(t)
  \int\limits_{\theta(t')=\theta'}^{\theta(t'')=\theta''}
  \sin\theta\CD\theta(t)
  \int\limits_{\phi(t')=\phi'}^{\phi(t'')=\phi''}\CD\phi(t)
  \hfill\\ \qquad\qquad\times
  \exp\left\{\ih\int_{t'}^{t''}\bigg[{m\over2}
  (\dot r^2+r^2\dot\theta^2+r^2\sin^2\theta\dot\phi^2)
  \right.\hfill\\  \hfill\left.\vphantom{\int_A^B}
   +{q^2\over r}-{b\hbar^2\over2mr^2\sin^2\theta}
   -{c\hbar^2\over2m}{\cos\theta\over r^2\sin^2\theta}
   +{\hbar^2\over8m}\bigg(1+{1\over\sin^2\theta}\bigg)\bigg]dt\right\}
   \\  \qquad
   ={1\over2\pi}\sum_{\nu=-\infty}^\infty \e^{\i\nu(\phi''-\phi')}
   K_\nu(r'',r',\theta'',\theta';T)
   \hfill\endmultline
  \tag\NUM.\num$$\plus%
with the kernel $K_\nu(T)$, which in turn is also separated:
$$\allowdisplaybreaks\align
  K_\nu&(r'',r',\theta'',\theta';T)
  \\   &
  =(r'r''\sin\theta'\sin\theta'')^{-\half}
  \int\limits_{r(t')=r'}^{r(t'')=r''}\CD r(t)
  \int\limits_{\theta(t')=\theta'}^{\theta(t'')=\theta''}
  \CD\theta(t)
  \\   &\qquad\times
  \exp\left\{\ih\int_{t'}^{t''}\bigg[
   {m\over2}(\dot r^2+r^2\dot\theta^2)
  -\hbar^2{\nu^2+b+c\cos\theta-{1\over4}\over2mr^2\sin^2\theta}
  +{q^2\over r}+{\hbar^2\over8mr^2}\bigg]dt\right\}
  \\   &
  =(r'r''\sin\theta'\sin\theta'')^{-\half}
  \\   &\qquad\times
  \int\limits_{r(t')=r'}^{r(t'')=r''} r\CD r(t)
  \int\limits_{\theta(t')=\theta'}^{\theta(t'')=\theta''}\CD\theta(t)
  \exp\Bigg\{\ih\int_{t'}^{t''}\Bigg[
   {m\over2}\dot r^2+{4m\over2}r^2\bigg({\dot\theta\over2}\bigg)^2
  \\   &\qquad
  -\hbar^2{\nu^2+b+c-{1\over4}\over8mr^2\sin^2\theta}
  -\hbar^2{\nu^2+b-c-{1\over4}\over8mr^2\cos^2\theta}
  +{q^2\over r}+{\hbar^2\over8mr^2}\Bigg]dt\Bigg\}
  \\   &
  =\sum_{l=0}^\infty
  \Psi_l^{(\lambda_1,\lambda_2)}(\theta')
  \Psi_l^{(\lambda_1,\lambda_2)}(\theta'')
  K_{l\nu}(r'',r';T),
  \tag\NUM.\num\endalign$$\plus%
the (P\"oschl-Teller) wave functions
$$\allowdisplaybreaks\align
  \Psi_l^{(\lambda_2,\lambda_1)}(\theta)
  &=\bigg[{\lambda_1+\lambda_2+2l+1\over2^{1+\lambda_
 1+\lambda_2}}
  {l!\,\Gamma(\lambda_1+\lambda_2+l+1)\over
   \Gamma(\lambda_1+l+1)\Gamma(\lambda_2+l+1)}\bigg]^\half
  \\  &\qquad\times
   (1-\cos\theta)^{\lambda_1\over2}
   (1+\cos\theta)^{\lambda_2\over2}
   P_l^{(\lambda_1,\lambda_2)}(\cos\theta),
  \tag\NUM.\num\endalign$$\plus%
($\lambda_{1,2}$ as before) and the radial path integral
$$K_{l\nu}(r'',r';T)
  ={1\over r'r''}\int\limits_{r(t')=r'}^{r(t'')=r''}\CD r(t)
  \exp\left\{\ih\int_{t'}^{t''}\bigg[{m\over2}\dot r^2
         +{q^2\over r}-\hbar^2{\lambda^2-{1\over4}\over2mr^2}
  \bigg]dt\right\}
  \tag\NUM.\num$$\plus%
$\big[\lambda=\half(\lambda_1+\lambda_2+2l+1]\big]$.
This path integral is nothing but a Coulomb potential path integral,
however with generalized angular momentum. Thus we just
analytically continue the known result in $l$ and obtain
$$G_\lambda(r'',r';E)
  ={1\over r'r''}\sqrt{-{m\over2E}}\,
   {\Gamma(\lambda+1-p)\over\Gamma(2\lambda+2)}
   W_{p,\lambda+\half}(-2\i kr_>)M_{p,\lambda+\half}(-2\i kr_<).
  \tag\NUM.\num$$\plus%
The energy spectrum was already stated in equation (\numFHbn) and the
corresponding wave functions are
$$\multline
  \Psi_{\lambda}(r,\theta,\phi)={\e^{\i\nu\phi}\over\sqrt{2\pi}}
  \Psi_l^{(\lambda_1,\lambda_2)}(\theta)
  \\  \times\left\{
  \aligned
  &{2\over(n+\lambda+\half)^2}
  \bigg[{n!\over a^3(n+\lambda+\half)
  \Gamma(n+2\lambda+2)}\bigg]^\half
  \bigg({2r\over a(n+\lambda+\half)}\bigg)^\lambda
  \\   &\qquad\qquad\times
  \exp\bigg(-{r\over a(n+\lambda+\half)}\bigg)
  L_n^{(2\lambda+1)}\bigg({2r\over a(n+\lambda+\half)}\bigg)
  \\
  &\sqrt{1\over2\pi}\,
  {\Gamma(\lambda+\half+{\i\over ap})\over r\Gamma(2\lambda+2)}
  \exp\bigg({\pi\over2ap}\bigg)
  M_{{\i\over ap},\lambda+\half}(-2\i pr)
  \endaligned\right.
  \endmultline
  \tag\NUM.\num$$\plus%
and $a$ denotes the Bohr radius.

\bigskip\noindent
$\underline{\hbox{3) Parabolic coordinates}}$
\newline
The potential described in equation (\numFHbm) is also separable in
parabolic coordinates, in the operator approach as well in the path
integral formalism. We consider the path integral formulation
corresponding to the Lagrangian (\numFHbm):
$$\multline
  K(x'',x';T)\equiv
  K(\xi'',\xi',\eta'',\eta',\phi'',\phi';T)
  \\  \qquad
  =\int\limits_{\xi(t')=\xi'}^{\xi(t'')=\xi''}\CD\xi(t)
   \int\limits_{\eta(t')=\eta'}^{\eta(t'')=\eta''}\CD
  \eta(t) (\xi^2+\eta^2)\xi\eta
  \int\limits_{\phi(t')=\phi'}^{\phi(t'')=\phi''}\CD\phi(t)
  \hfill\\  \qquad\qquad\times
  \exp\left\{\ih\int_{t'}^{t''}\bigg[{m\over2}
   (\xi^2+\eta^2)(\dot\xi^2+\dot\eta^2)
    +{m\over2}\xi^2\eta^2\dot\phi^2
  \right.\hfill\\   \vphantom{\int_A^B}\left.\hfill
    +{2q^2\over\xi^2+\eta^2}
    -{\hbar^2(b-{1\over4})\over2m\xi^2\eta^2}
    -{c\hbar^2\over m}{\eta^2-\xi^2\over\xi^2\eta^2(\xi^2+\eta\sp2)}
  \bigg]dt\right\}
  \\ \qquad
  ={1\over2\pi}\sum_{\nu=-\infty}^\infty \e^{\i\nu(\phi''-\phi')}
  K_\nu(\xi'',\xi',\eta'',\eta';T)
  \hfill\endmultline
  \tag\NUM.\num$$\plus%
with the kernel $K_\nu(T)$
$$\multline
  K_\nu(\xi'',\xi',\eta'',\eta';T)
  =(\xi'\xi''\eta'\eta'')^{-\half}
  \int\limits_{\xi(t')=\xi'}^{\xi(t'')=\xi''}\CD\xi(t)
  \int\limits_{\eta(t')=\eta'}^{\eta(t'')=\eta''}\CD
  (\xi^2+\eta^2)
   \\   \qquad\times
  \exp\left\{\ih\int_{t'}^{t''}\bigg[{m\over2}
   (\xi^2+\eta^2)(\dot\xi^2+\dot\eta^2)
  \right.\hfill\\   \vphantom{\int_A^B}\left.\qquad\qquad
    +{2q^2\over\xi^2+\eta^2}
    -\hbar^2{b+\nu^2-{1\over4}\over2m\xi^2\eta^2}
    -{c\hbar^2\over m}{\eta^2-\xi^2\over\xi^2\eta^2(\xi^2+\eta\sp2)}
  \bigg]dt\right\}
  \endmultline
  \tag\NUM.\num$$\plus%
This path integral is now tracked by a time-transformation
according to
$$\epsilon=\widehat{\xi_{(j)}^2+\eta_{(j)}^2},\qquad
  s(t)=\int_{t'}^t{d\sigma\over \xi^2(\sigma)+\eta^2(\sigma)},
  \tag\NUM.\num$$\plus%
therefore
$$\gathered
  K_\nu(\xi'',\xi',\eta'',\eta';T)
  ={1\over2\pi\i\hbar}\int_{-\infty}^\infty dE\,\e^{-\i ET/\hbar}
  G_\nu(\xi'',\xi',\eta'',\eta';E)
  \\
  G_\nu(\xi'',\xi',\eta'',\eta';E)
  =\i\int_0^\infty ds''\,\e^{2\i q^2s''/\hbar}
  \tilde K_\nu(\xi'',\xi',\eta'',\eta';s'')
  \endgathered
  \tag\NUM.\num$$\plus%
with (note that this is a two-dimensional time-transformation
and the prefactor is ``one''):
$$\tilde K_\nu(\xi'',\xi',\eta'',\eta';s'')
  =\tilde K_1(\xi'',\xi';s'')
    \times\tilde K_2(\eta'',\eta';s'').
  \tag\NUM.\num$$\plus%
This gives for the kernels $\tilde K_{1,2}(s'')$
$$\allowdisplaybreaks\align
  \tilde K_1(\xi'',\xi';s'')
  &={1\over\sqrt{\xi'\xi''}}
  \int\limits_{\xi(t')=\xi'}^{\xi(t'')=\xi''}
  \mu_{\lambda_1}[\xi^2]\CD\xi(s)\exp\left\{\ih\int_0^{s''}
  \bigg[{m\over2}\dot\xi^2 +E\xi^2\bigg]ds\right\}
  \\   &
  ={m\omega\over \i\hbar\sin\omega s''}
  \exp\bigg[-{m\omega\over2\i\hbar}
               ({\xi'}^2+{\xi'}^2)\cot\omega s''\bigg]
  I_{\lambda_1}\bigg({m\omega\xi'\xi''\over \i\hbar\sin\omega s''}\bigg)
  \\   &\
  \tag\NUM.\num\endalign$$\plus%
with $\omega=\sqrt{-2E/m}$, $\lambda_1=+\sqrt{\nu^2+b+c}$.
Similarly for $\tilde K_2(s'')$
$$\allowdisplaybreaks\align
  \tilde K_2(\eta'',\eta';s'')
  &={1\over\sqrt{\eta'\eta''}}
  \int\limits_{\eta(t')=\eta'}^{\eta(t'')=\eta''}
  \mu_{\lambda_2}[\eta^2]\exp\left\{\ih\int\sb0^{s''}
  \bigg[{m\over2}\dot\eta^2+E\eta^2\bigg]ds\right\}
  \\  &
  ={m\omega\over \i\hbar\sin\omega s''}
  \exp\bigg[-{m\omega\over2\i\hbar}
                 ({\eta'}^2+{\eta'}^2)\cot\omega s''\bigg]
  I_{\lambda_2}
           \bigg({m\omega\eta'\eta''\over \i\hbar\sin\omega s''}\bigg)
  \\   &\
  \tag\NUM.\num\endalign$$\plus%
($\lambda_2=+\sqrt{\nu^2+b-c}$). Integrating over $s''$ yields
$$G^{(bound)}(x'',x';E)
  =\hbar\sum_{\nu=-\infty}^\infty\sum_{n_1,n_2=0}^\infty
  {\Psi_N^*(\xi',\eta',\phi')\Psi_N(\xi'',\eta'',\phi'')
  \over E_N-E}
  \tag\NUM.\num$$\plus%
with the energy spectrum (note $\omega=q^2/\hbar N$)
$$E_N=-{mq^4\over2\hbar^2N^2},\qquad
  N=n_1+n_2+1+{\lambda_1+\lambda_2\over2}.
  \tag\NUM.\num$$\plus%
The wave functions have the form
$$\multline
  \Psi_{\nu,n_1,n_2}(\xi,\eta,\nu)
  ={\e^{\i\nu\phi}\over\sqrt{2\pi}}
  \bigg[{2\over a^2N^3}\cdot
  {2n_1!n_2!\over\Gamma(n_1+\lambda_1+1)\Gamma(n_2+\lambda_2+1)}
  \bigg]^\half
  \hfill\\  \qquad\times
  \bigg({\xi\over aN}\bigg)^{\lambda_1}
  \bigg({\eta\over aN}\bigg)^{\lambda_2}
  \exp\bigg[-{\xi^2+\eta^2\over2aN}\bigg]
  L_{n_1}^{(\lambda_1)}\bigg({\xi^2\over aN}\bigg)
  L_{n_2}^{(\lambda_2)}\bigg({\eta^2\over aN}\bigg).
  \hfill\endmultline
  \tag\NUM.\num$$\plus%
For the continuous spectrum we obtain
$$G^{(cont.)}(x'',x';E)
  =\hbar\sum_{\nu=-\infty}^\infty
  \int_0^\infty dp\int_{-\infty}^\infty d\zeta
  {\Psi_{p,\zeta,\nu}^*(\xi',\eta',\phi')
   \Psi_{p,\zeta,\nu}(\xi'',\eta'',\phi'')\over E_p-E}
  \tag\NUM.\num$$\plus%
with
$$E_p={\hbar^2p^2\over2m}
  \tag\NUM.\num$$\plus%
and the wave functions
$$\multline
  \Psi_{p,\zeta,\nu}(\xi,\eta,\phi)
  ={\e^{\i\nu\phi}\over(2\pi)^{3\over2}}
  {\Gamma[
   {1+\lambda_1\over2}+{\i\over2}(\zeta+{\i\over ap})]
   \Gamma[
   {1+\lambda_2\over2}+{\i\over2}(\zeta-{\i\over ap})]
   \over\xi\eta\Gamma(1+\lambda_1)\Gamma(1+\lambda_2)}
  \\   \times
  M_{-{\i\over2}(\zeta+{\i\over ap}),{\lambda_1\over2}} (-\i p\xi^2)
  M_{-{\i\over2}(\zeta-{\i\over ap}),{\lambda_2\over2}} (-\i p\eta^2).
  \endmultline
  \tag\NUM.\num$$\plus%
Finally we state the relation of the Green function in parabolic
coordinates. We have
$$\multline
  G(\xi'',\xi',\eta'',\eta',\phi'',\phi';E)
  ={\i\over2\pi}\bigg({m\omega\over \i\hbar}\bigg)^2
  \sum_{\nu=-\infty}^\infty \e^{\i\nu(\phi''-\phi')}
  \hfill\\  \qquad\times
  \int_0^\infty{ds''\over\sin^2\omega s''}
  I_{\lambda_1}
       \bigg({m\omega\xi'\xi''\over \i\hbar\sin\omega s''}\bigg)
  I_{\lambda_2}
       \bigg({m\omega\eta'\eta''\over \i\hbar\sin\omega s''}\bigg)
  \hfill\\  \qquad\times
  \exp\bigg[{2\i q^2\over\hbar}s''-{m\omega\over2\i\hbar}
  ({\xi'}^2+{\xi'}^2+{\eta'}^2+{\eta'}^2)\cot\omega s''\bigg].
  \hfill\endmultline
  \tag\NUM.\num$$\plus%
Note the similarity to the $u_1,u_2,u_3,u_4$ approach.
Actually the Kustaanheimo-Stiefel approach in cartesian
coordinates produces a separation in parabolic coordinates.
We switch bach to polar coordinates by means of
$$\aligned
  \xi^2& =r+z=r(1+\cos\theta)=2r\cos^2{\theta\over2}  \\
  \eta^2&=r-z=r(1-\cos\theta)=2r\sin^2{\theta\over2}
  \endaligned
  \tag\NUM.\num$$\plus%
Using again the addition theorem (\numFHxc) with a rescaling
$s''\to2s''$, $\omega\to\omega/2$ so that $\omega=\sqrt{-8E/m}$
we recover the Green function (\numFHbo).

Let us finally note that even more complicated Coulomb-like potentials
can be exactly solved [\GROo, \GROp] which are, however, only separable
in parabolic coordinates.


\newpage
\def\Chapter{Bibliography}
\def\rightheadline{\eightpoint\eightrm\hfil Bibliography\hfil}
\centerline{\fourteenbf B\large  IBLIOGRAPHY}
\bigskip
\eightpoint
\eightrm
\baselineskip=10pt

\item{[\ALB]}
S.Albeverio:
Some Recent Developments and Applications of Path Integrals;
in:
M.C.Gutzwiller et al.(eds.):
{\it Bielefeld Encounters in Physics and Mathematics VII; Path
Integrals From meV to MeV}, 1985
({\it World Scientific}, Singapore, 1986).
\item{[\ABHKa]}
S.Albeverio, P.Blanchard and R.H\o egh-Krohn:
Some Applications of Functional Integration; in ``Mathematical Problems
in Theoretical Physics'', p.265 (Eds.: R.Schrader, R.Seider, and
D.A.Uhlenbrock) {\it Lecture Notes in Physics} {\bf 153}
({\it Springer-Verlag}, Berlin, 1982).
\item{[\ACHRS]}
S.Albeverio, Ph.Combe, R.H\o egh-Krohn, G.Rideau, M.Sirgue-Collin,
M.Sirgue and R.Stora (Eds.):
Feynman Path Integrals, {\it Lecture Notes in Physics} {\bf 106}
({\it Springer-Verlag}, Berlin, 1979).
\item{[\ARTb]}
A.M.Arthurs:
Path Integrals in Polar Coordinates;
{\it Proc.Roy.Soc.(London)} {\bf A 313} (1969) 445;
\newline
Path Integrals in Curvilinear Coordinates;
{\it Proc.Roy.Soc.(London)} {\bf A 318} (1970) 523.
\item{[\BJb]}
M.B\"ohm and G.Junker:
Path Integration Over Compact and Noncompact Rotation Groups;
{\it J.Math.Phys.}\ {\bf 28} (1987) 1978.
\item{[\CAST]}
D.P.L.Castrigiano and F.St\"ark:
New Aspects of the Path Integrational Treatment of the Coulomb
Potential;
{\it J.Math.Phys.}\ {\bf 30} (1989) 2785.
\item{[\CARPa]}
M.V.Carpio-Bernido:
Path Integral Quantization of Certain Noncentral Systems with
Dynamical Symmetries;
{\it J.Math.Phys.}\ {\bf 32} (1991) 1799.
\item{[\CARPb]}
M.V.Carpio-Bernido:
Green Function for an Axially Symmetric Potential Field: A Path
Integral Evaluation in Polar Coordinates;
{\it J.Phys.A: Math.Gen.}\ {\bf 24} (1991) 3013.
\item{[\CBB]}
M.V.Carpio-Bernido and C.C.Bernido:
An Exact Solution of a Ring-Shaped Oscillator Plus a
$c\sec^2\theta/r^2$ Potential;
{\it Phys.Lett.}\ {\bf A 134} (1989) 395.
\item{[\CBBI]}
M.V.Carpio-Bernido, C.C.Bernido and A.Inomata:
Exact Path Integral Treatment of Two Classes of Axially Symmetric
Potentials; in ``Third International Conference on
Path Integrals From meV to MeV'', 1989, p.442;
Eds.: V.Sa-yakanit et al.({\it World Scientific}, Singapore, 1989).
\item{[\CBI]}
M.V.Carpio-Bernido and A.Inomata:
Path Integral Treatment of the Hartmann Potential;
in ``Bielefeld Encounters in Physics and Mathematics VII; Path
Integrals From meV to MeV'', 1985;
eds.:M.C.Gutz\-willer et al.({\it World Scientific}, Singapore, 1986).
\item{[\CGHa]}
L.Chetouani, L.Guechi and T.F.Hammann:
Exact Path Integral for the Ring Potential;
{\it Phys. Lett.}\ {\bf A 125} (1987) 277.
\item{[\CGHc]}
L.Chetouani, L.Guechi and T.F.Hammann:
Exact Path Integral Solution of the Coulomb Plus Aharonov-Bohm
Potential;
{\it J.Math.Phys.}\ {\bf 30} (1989) 655.
\item{[\CGHd]}
L.Chetouani, L.Guechi and T.F.Hammann:
Generalized Canonical Transformations and Path Integrals;
{\it Phys.Rev.}\ {\bf A 40} (1989) 1157.
\item{[\CHc]}
L.Chetouani and T.F.Hammann:
Coulomb Green's Function, in a $n$-Dimensional Euclidean Space;
{\it J.Math.Phys.}\ {\bf 27} (1986) 2944.
\item{[\DDS]}
R.De, R.Dutt and U.Sukhatme:
Mapping of Shape Invariant Potentials Under Point Canonical
Transformations;
{\it J.Phys.A: Math.Gen.}\ {\bf 25} (1992), L 843.
\item{[\DEWb]}
B.S.DeWitt:
Dynamical Theory in Curved Spaces.\ I.\ A Review of the Classical and
Quantum Action Principles;
{\it Rev.Mod.Phys.}\ {\bf 29} (1957) 377.
\item{[\DEWMa]}
C.DeWitt-Morette:
Feynman Path Integrals: I.Linear and Affine Technique,
II.The Feynman Green Function;
{\it Commun.Math.Phys.}\ {\bf 37} (1974) 63.
\item{[\DEWMNb]}
C.DeWitt-Morette, A.Maheswari and B.Nelson:
Path Integration in Non-Relativistic Quantum Mechanics;
{\it Phys.Rep.}\ {\bf 50} (1979) 255.
\item{[\DIRd]}
P.A.M.Dirac:
The Lagrangian in Quantum Mechanics;
{\it Phys.Zeitschr.Sowjetunion} {\bf 3} (1933) 64;
reprinted in: Quantum Electrodynamics (Ed.J.Schwinger)
{\it Dover}, New York, 1958, p.312;
\item{[\DOTO]}
J.C.D'Olivio and M.Torres:
The Canonical Formalism and Path Integrals in Curved Spaces;
{\it J.Phys.A: Math.Gen.}\ {\bf 21} (1988) 3355;
The Weyl Ordering and Path Integrals in Curved Spaces;
Path Summation: Achievements and Goals, Trieste, 1987, p.481,
eds: S.Lundquist et al.\ ({\it World Scientific}, Singapore, 1988).
\item{[\DOWc]}
J.S.Dowker:
Covariant Feynman Derivation of Schr\"odinger's Equation in a
Riemannian Space;
{\it J.Phys.A: Math., Nucl.Gen.}\ {\bf 7} (1974) 1256.
\item{[\DOWd]}
J.S.Dowker:
Path Integrals and Ordering Rules:
{\it J.Math.Phys.}\ {\bf 17} (1976) 1873.
\item{[\DMa]}
J.S.Dowker and I.W.Mayes:
The Canonical Quantization of Chiral Dynamics;
{\it Nucl.Phys.}\ {\bf B 29} (1971) 259.
\item{[\DURb]}
I.H.Duru:
Path Integrals Over $\SU(2)$ Manifold and Related Potentials;
{\it Phys.Rev.}\ {\bf D 30} (1984) 2121.
\item{[\DURd]}
I.H.Duru:
On the Path Integral for the Potential $V=ar^{-2}+br^2$;
{\it Phys.Lett.}\ {\bf A 112} (1985) 421.
\item{[\DKa]}
I.H.Duru and H.Kleinert:
Solution of the Path Integral for the H-Atom;
{\it Phys.Lett.}\ {\bf B 84} (1979) 185.
\item{[\DKb]}
I.H.Duru and H.Kleinert:
Quantum Mechanics of H-Atoms From Path Integrals;
{\it Fort\-schr.Phys.}\ {\bf 30} (1982) 401.
\item{[\DKS]}
R.Dutt, A.Khare and U.P.Sukhatme:
Supersymmetry, Shape Invariance, and Exactly Solvable Potentials;
{\it Amer.J.Phys.}\ {\bf 56} (1988), 163.
\item{[\EG]}
S.F.Edwards and Y.V.Gulyaev:
Path Integrals in Polar Co-ordinates;
{\it Proc.Roy.Soc.(London)} {\bf A 279} (1964) 229.
\item{[\EMOTa]}
A.Erd\'elyi, W.Magnus, F.Oberhettinger and F.G.Tricomi (Eds.):
Higher Transcendental Functions, Vol.I-III
({\it McGraw Hill}, New York, 1955).
\item{[\FEYa]}
R.P.Feynman:
The Principle of Least Action in Quantum Mechanics;
Ph.D.Thesis, Princeton University, May 1942.
\item{[\FEYb]}
R.P.Feynman:
Space-Time Approach to Non-Relativistic Quantum Mechanics;
{\it Rev.Mod.Phys.}\ {\bf 20} (1948) 367.
\item{[\FH]}
R.P.Feynman and A.Hibbs: Quantum Mechanics and Path Integrals
({\it McGraw Hill, New York}, 1965).
\item{[\FLM]}
W.Fischer, H.Leschke and P.M\"uller:
Changing Dimension and Time: Two Well-Founded and Practical Techniques
for Path Integration in Quantum Physics;
{\it Universit\"at Erlangen-N\"urnberg preprint}, August 1991,
{\it J.Phys.A: Math.Gen.}\ {\bf 25} (1992).
\newline
Path Integration in Quantum Physics by Changing the Drift of the
Underlying Diffusion Process;
{\it Universit\"at Erlangen-N\"urnberg preprint}, February 1992.
\item{[\FW]}
A.Frank and K.B.Wolf:
Lie Algebras for Systems With Mixed Spectra.I.
The Scattering P\"oschl-Teller Potential;
{\it J.Math.Phys.}\ {\bf 25} (1985) 973.
\item{[\GY]}
I.M.Gelfand and A.M.Jaglom:
Die Integration in Funktionenr\"aumen und ihre Anwendung in der
Quantentheorie;
{\it Fortschr. Phys.}\ {\bf 5} (1957) 517;
Integration in Functional Spaces and its Applications in Quantum
Physics;
{\it J.Math.Phys.}\ {\bf 1} (1960) 48.
\item{[\GJ]}
J.-L.Gervais and A.Jevicki:
Point Canonical Transformations in the Path Integral;
{\it Nucl.Phys.}\ {\bf B 110} (1976) 93.
\item{[\GOOb]}
M.J.Goovaerts:
Path-Integral Evaluation of a Nonstationary Calogero Model;
{\it J.Math.Phys.}\ {\bf 16} (1975) 720.
\item{[\GRA]}
I.S.Gradshteyn and I.M.Ryzhik:
Table of Integrals, Series, and Products
({\it Academic Press,} New York, 1980).
\item{[\GMV]}
H.Grinberg, J.Mara\~non and H.Vucetich:
Atomic Orbitals of the Nonrelativistic Hydrogen Atom in a
Four-Dimensional Riemann Space Through the Path Integral Formalism;
{\it J.Chem.Phys.}\ {\bf 78} (1983) 839;
\newline
The Hydrogen Atom as a Projection of an Homogeneous Space;
{\it Zeitschr.Phys.}\ {\bf C 20} (1983) 147.
\item{[\GMVc]}
H.Grinberg, J.Mara\~non and H.Vucetich:
Some Remarks on Coulomb and Oscillator Problems in $\R^N$;
{\it KINAM} {\bf 5} (1983) 127.
\item{[\GROa]}
C.Grosche:
The Product Form for Path Integrals on Curved Manifolds;
{\it Phys.Lett.}\ {\bf A 128} (1988) 113.
\item{[\GROe]}
C.Grosche:
Path Integral Solution of a Class of Potentials Related to the
P\"oschl-Teller Potential;
{\it J.Phys.A: Math.Gen.}\ {\bf 22} (1989) 5073.
\item{[\GROj]}
C.Grosche:
Separation of Variables in Path Integrals and Path Integral Solution
of Two Potentials on the Poincar\'e Upper Half-Plane;
{\it J.Phys.A: Math.Gen.}\ {\bf 23} (1990) 4885.
\item{[\GROm]}
C.Grosche:
Coulomb Potentials by Path Integration;
{\it Fort\-schr.Phys.}\ {\bf 40} (1992), 695.
\item{[\GROo]}
C.Grosche:
Path Integral Solution of a Non-Isotropic Coulomb-Like Potential;
{\it Phys.Lett.}\ {\bf A 165} (1992), 185.
\item{[\GROp]}
C.Grosche:
Path Integral Solution of Two Potentials Related to the SO(2,1)
Dynamical Algebra;
{\it Trieste preprint}, SISSA/186/ 91/FM, December 1991.
\item{[\GRSb]}
C.Grosche and F.Steiner:
Path Integrals on Curved Manifolds;
{\it Zeitschr.Phys.}\ {\bf C 36} (1987) 699.
\item{[\GRSf]}
C.Grosche and F.Steiner:
Feynman Path Integrals;
to appear in {\it Lecture Notes in Physics}.
\item{[\GRSg]}
C.Grosche and F.Steiner:
A Table of Feynman Path Integrals;
to appear in {\it Springer Tracts in Modern Physics}.
\item{[\GROS]}
C.C.Grosjean:
A General Formula for the Calculation of Gaussian Path-Integrals in
Two and Three Euclidean Dimensions;
{\it J.Comput.Appl.Math.}\ {\bf 23} (1988) 199.
\item{[\GROGOb]}
C.C.Grosjean and M.J.Goovaerts:
The Analytical Evaluation of One-Dimensional Gaussian Path-Integrals;
{\it J.Comput.Appl. Math.}\ {\bf 21} (1988) 311.
\item{[\HIR]}
A.C.Hirshfeld:
Canonical and Covariant Path Integrals;
{\it Phys.Lett.}\ {\bf A 67} (1978) 5.
\item{[\HOI]}
R.Ho and A.Inomata:
Exact Path Integral Treatment of the Hydrogen Atom;
{\it Phys.Rev.Lett.}\ {\bf 48} (1982) 231.
\item{[\INH]}
L.Infeld and T.E.Hull:
The Factorization Method;
{\it Rev.Mod.Phys.}\ {\bf 23} (1951), 21.
\item{[\INOa]}
A.Inomata:
Exact Path-Integration for the Two Dimensional Coulomb Problem;
{\it Phys.Lett.}\ {\bf A 87} (1982) 387.
\item{[\INOb]}
A.Inomata:
Alternative Exact-Path-Integral-Treatment of the Hydrogen Atom;
{\it Phys.Lett.}\ {\bf A 101} (1984) 253.
\item{[\INOd]}
A.Inomata:
Remarks on the Time Transformation Technique for Path Integration; in
``Bielefeld Encounters in Physics and Mathematics VII; Path Integrals
from meV to MeV'', 1985, p.433; Eds.:
M.C.Gutzwiller et al.\ ({\it World Scientific}, Singapore, 1986);
\newline
Recent Developments of Techniques for Solving Nontrivial Path Integrals;
in ``Path Summation: Achievements and Goals'', Trieste, 1987, p.114;
Eds.:S.Lundquist et al.\ ({\it World Scientific}, Singapore, 1986);
\newline
Time Transformation Techniques in Path Integration;
in ``Path Integrals from meV to MeV'', p.112; Eds.: V.Sa-yakanit
et al.\ ({\it World Scientific}, Singapore, 1986);
\item{[\INOWI]}
A.Inomata and R.Wilson:
Path Integral Realization of a Dynamical Group;
{\it Lecture Notes in Physics} {\bf 261}, p.42
({\it Springer-Verlag},  Berlin, 1985);
Factorization-Algebraization-Path Integration and Dynamical Groups;
in ``Symmetries in Science II'', eds.: B.Gruber and R.Lenczewski, p.255
({\it Plenum Press}, New York, 1986).
\item{[\BJa]}
G.Junker and M.B\"ohm:
The $\SU(1,1)$ Propagator as a Path Integral Over Noncompact Groups;
{\it Phys.Lett.}\ {\bf A 117} (1986) 375.
\item{[\KLEh]}
H.Kleinert:
How to do the Time Sliced Path Integral for the H Atom;
{\it Phys.Lett.}\ {\bf A 120} (1987) 361.
\item{[\KLEk]}
H.Kleinert:
Quantum Mechanics and Path Integral in Spaces with Curvature and
Torsion;
{\it Mod.Phys.Lett.}\ {\bf A 4} (1990) 2329.
\item{[\KLEm]}
H.Kleinert:
Path Integrals in Quantum Mechanics, Statistics and Polymer Physics
({\it World Scientific}, Singapore, 1990).
\item{[\KLEMUS]}
H.Kleinert and I.Mustapic:
Summing the Spectral Representations of P\"oschl-Teller and
Rosen-Morse Fixed-Energy Amplitudes;
{\it J.Math.Phys.}\ {\bf 33} (1992) 643.
\item{[\KUST]}
P.Kustaanheimo and E.Stiefel:
Perturbation Theory of Kepler Motion Based on Spinor Regularization;
{\it J.Rein.Angew.Math.}\ {\bf 218} (1965), 204.
\item{[\LAI]}
W.Langguth and A.Inomata:
Remarks on the Hamiltonian Path Integral in Polar Coordinates;
{\it J.Math.Phys.}\ {\bf 20} (1979) 499.
\item{[\LRTd]}
F.Langouche, D.Roekaerts and E.Tirapegui:
Functional Integration and Semiclassical Expansion
({\it Reidel}, Dordrecht, 1982).
\item{[\LEEb]}
T.D.Lee:
Particle Physics and Introduction to Field Theory
({\it Harwood Academic Publishers}, Chur, 1981).
\item{[\MCLS]}
D.W.McLaughlin and L.S.Schulman:
Path Integrals in Curved Spaces;
{\it J.Math.Phys.}\ {\bf 12} (1971) 2520.
\item{[\MSVW]}
A.A.Makarov, J.A.Smorodinsky, Kh.Valiev and P.Winternitz:
A Systematic Search for Nonrelativistic Systems With Dynamical
Symmetries;
{\it Nuovo Cimento} {\bf A 52} (1967), 1061.
\item{[\MARa]}
M.S.Marinov:
Path Integrals in Quantum Theory: An Outlook of Basic Concepts;
{\it Phys.Rep.}\ {\bf 60} (1980) 1.
\item{[\MIZa]}
M.M.Mizrahi:
The Weyl Correspondence and Path Integrals;
{\it J.Math.Phys.}\ {\bf 16} (1975) 2201.
\item{[\MIZc]}
M.M.Mizrahi:
On the Semiclassical Expansion in Quantum Mechanics for Arbitrary
Hamiltonians;
{\it J.Math.Phys.}\ {\bf 18} (1977) 786.
\item{[\MIZd]}
M.M.Mizrahi:
Phase Space Path Integrals, Without Limiting Procedure;
{\it J.Math.Phys.}\ {\bf 19} (1978) 298.
\item{[\MORE]}
C.Morette:
On the Definition and Approximation of Feynman's Path Integrals;
{\it Phys.Rev.}\ {\bf 81} (1951) 848.
\item{[\MDEW]}
C.Morette-DeWitt:
Feynman's Path Integral: Definition Without Limiting Procedure;
{\it Commun.Math.Phys.}\ {\bf 28} (1972) 47.
\item{[\DEWMc]}
C.Morette-DeWitt:
The Semiclassical Expansion;
{\it Ann.Phys.(N.Y.)} {\bf 97} (1976) 367.
\item{[\DEWMEL]}
C.Morette-DeWitt and K.D.Elworthy:
New Stochastic Methods in Physics;
{\it Phys.Rep.}\ {\bf 77} (1981) 122.
\item{[\NELb]}
E.Nelson:
Feynman Integrals and the Schr\"odinger Equation;
{\it J.Math.Phys.}\ {\bf 5} (1964) 332.
\item{[\OMO]}
M.Omote:
Point Canonical Transformations and the Path Integral;
{\it Nucl.Phys.}\ {\bf B 120} (1977) 325.
\item{[\PAKSc]}
N.K.Pak and I.S\"okmen:
General New-Time Formalism in the Path Integral;
{\it Phys.Rev.}\ {\bf A 30} (1984) 1629.
\item{[\PAUb]}
W.Pauli: Die allgemeinen Prinzipien der Wellenmechanik,
in S.Fl\"ugge: Handbuch der Physik, Band V/1
({\it Springer-Verlag}, Berlin, 1958).
\item{[\PI]}
D.Peak and A.Inomata:
Summation Over Feynman Histories in Polar Coordinates;
{\it J.Math.Phys.}\ {\bf 10} (1969) 1422.
\item{[\RS]}
M.Reed and B.Simon:
Methods of Modern Mathematical Physics, Vol.II
({\it Academic Press, New York}, 1975).
\item{[\SCHUc]}
L.S.Schulman:
Techniques and Applications of Path Integration
({\it John Wiley \&\ Sons}, New York, 1981).
\item{[\SIMON]}
B.Simon:
Functional Integration and Quantum Physics
({\it Academic Press}, New York, 1979).
\item{[\SOKb]}
I.S\"okmen:
Exact Path-Integral Solution of the Ring-Shaped Potential;
{\it Phys.Lett.}\ {\bf A 115} (1986) 249.
\item{[\SOKc]}
I.S\"okmen:
Exact Path-Integral Solution for a Charged Particle in a Coulomb Plus
Aharonov-Bohm Potential;
{\it Phys.Lett.}\ {\bf A 132} (1988) 65.
\item{[\STEa]}
F.Steiner:
Space-Time Transformations in Radial Path Integrals;
{\it Phys.Lett.}\ {\bf A 106} (1984) 356.
\item{[\STEb]}
F.Steiner:
Exact Path Integral Treatment of the Hydrogen Atom;
{\it Phys.Lett.}\ {\bf A 106} (1984) 363.
\item{[\STEc]}
F.Steiner:
Path Integrals in Polar Coordinates From eV to GeV;
in ``Bielefeld Encounters in Physics and Mathematics VII; Path
Integrals From meV to MeV'', 1985, p.335;
Eds.: M.C.Gutzwiller et al.({\it World Scientific}, Singapore, 1986).
\item{[\TROT]}
H.F.Trotter:
On the Product of Semi-Groups of Operators;
{\it Proc.Amer.Math.Soc.}\ {\bf 10} (1959) 545.



\enddocument